\documentclass[11pt,twoside]{article}
\usepackage[makeroom]{cancel}
\usepackage{latexsym}
\usepackage{longtable}
\usepackage{epsfig}
\usepackage{graphicx,bbm,psfrag}
\usepackage{amssymb,amsmath,amsthm,booktabs,mathtools}
\usepackage{bm}% bold math
\usepackage{color}
\usepackage{dutchcal}
\usepackage{hyperref}
\hypersetup{
    colorlinks,
    citecolor=red,
    linkcolor=blue,
}

\newtheorem{theorem}{Theorem}[section]
\newtheorem{rema}{Remark}[section]

\def\doi#1{\\ DOI: \href{https://www.doi.org/#1}{#1}}

\newcommand{\bc}{\begin{center}}
\newcommand{\ec}{\end{center}}
\def\ba#1{\begin{array}{#1}\displaystyle}
\newcommand{\ea}{\end{array}}

\newcommand{\beq}{\begin{equation}}
\newcommand{\eeq}{\end{equation}}
\newcommand{\beqa}{\begin{eqnarray}}
\newcommand{\eeqa}{\end{eqnarray}}

\newcommand{\n}{\nonumber\\}
\newcommand{\bi}{\begin{itemize}}
\newcommand{\ei}{\end{itemize}}
\def\lt#1{\left#1}
\def\rt#1{\right#1}
\def\t#1{\tilde{#1}}
\def\h#1{\hat{#1}}

\def\frc#1#2{\frac{#1}{#2}}

\newcommand{\p}{\partial}

\newcommand{\bra}{\langle}
\newcommand{\ket}{\rangle}
\newcommand{\Z}{{\mathbb{Z}}}
\newcommand{\N}{{\mathbb{N}}}
\newcommand{\R}{{\mathbb{R}}}
\newcommand{\C}{{\mathbb{C}}}

\newcommand{\ep}{\epsilon}

\newcommand{\ri}{{\rm i}}
\newcommand{\re}{{\rm e}}
\newcommand{\dd}{{\rm d}}

\def\hi#1{\mathcal{#1}}
\def\whi#1{{\mathcal{#1}'}}
\def\wwhi#1{{\mathcal{#1}''}}

\newcommand{\bal}{{\rm bal}}
\newcommand{\dif}{{\rm dif}}
\newcommand{\scat}{{\rm scat}}

\def\la#1{\mathcal{#1}}

\newcommand{\halmos}{\rule{1ex}{1.4ex}}
\newcommand{\eproof}{\hspace*{\fill}\mbox{$\halmos$}}

\begin{document}

\begin{titlepage}

\begin{center}
{\Large {\bf Diffusion and superdiffusion from hydrodynamic projections}}

\vspace{1cm}

{\large Benjamin Doyon}
\vspace{0.2cm}

{\small\em
Department of Mathematics, King's College London, Strand, London WC2R 2LS, UK}

\end{center}

\vspace{1cm}
\noindent {\bf Abstract:}

Hydrodynamic projections, the projection onto conserved charges representing ballistic propagation of fluid waves, give exact transport results in many-body systems, such as the exact Drude weights. Focussing one one-dimensional systems, I show that this principle can be extended beyond the Euler scale, in particular to the diffusive and superdiffusive scales. By hydrodynamic reduction, Hilbert spaces of observables are constructed that generalise the standard space of conserved densities and describe the finer scales of hydrodynamics. The Green-Kubo formula for the Onsager matrix has a natural expression within the diffusive space. This space is associated with quadratically extensive charges, and projections onto any such charge give generic lower bounds for diffusion. In particular, bilinear expressions in linearly extensive charges lead to explicit diffusion lower bounds calculable from the thermodynamics, and applicable for instance to generic momentum-conserving one-dimensional systems. Bilinear charges are interpreted as covariant derivatives on the manifold of maximal entropy states, and represent the contribution to diffusion from scattering of ballistic waves. An analysis of fractionally extensive charges, combined with clustering properties from the superdiffusion phenomenology, gives lower bounds for superdiffusion exponents. These bounds reproduce the predictions of nonlinear fluctuating hydrodynamics, including the Kardar-Parisi-Zhang exponent $2/3$ for sound-like modes, the Levy-distribution exponent $3/5$ for heat-like modes, and the full Fibonacci sequence.
%Finally, I speculate that in many-body integrability, wave scattering states span the diffusive space, and that this parallels the notion that the number of conserved quantities coincides with the number of degrees of freedom. 

\vfill

\hfill \today

\end{titlepage}

\tableofcontents

\section{Introduction}

Finding organising principles for emergent many-body behaviours is one of the most important tasks of theoretical physics. Thermodynamics and hydrodynamics offer powerful frameworks, based on the postulates that states and dynamics emerging at large scales are described by few, long-lived degrees of freedom \cite{Spohn-book}. However, characterising these degrees of freedom and quantifying their effects are difficult problems. This is especially true beyond the Euler scale, where one would like to study the Onsager matrix, representing diffusion, and other objects in a model-independent way. In one dimension, a particularly rich phenomenology of diffusion and superdiffusion is observed. Nonlinear fluctuating hydrodynamics (NFH) \cite{SpohnNonlinear} offers flexible tools and has been very successful especially for classical, conventional fluids but also beyond, see e.g.~\cite{1742-5468-2015-3-P03007,KulkarniFluctuating2015,PopkovFibonacci2015,PopkovFibonacci2016,SchutzFibonacci2018,ChenDeGierHirikiSasamotoFluctuHydro18,BulchandaniKardar2019}. Less  phenomenological methods based on microscopics include bound on diffusion from conserved quantities \cite{Prosenquad}, and in integrable systems, exact results from generalised hydrodynamics \cite{PhysRevLett.117.207201,PhysRevX.6.041065} and Bethe-ansatz techniques, which have provided much insight \cite{dNBD,dNBD2,GHKV2018,IlievskiSuper2018,LZP19,GopalaKinetic2019,GopalaAnomalous2019,deNardisAnomalous2019,BulchandaniKardar2019}. But despite this progress, a framework for establishing general and rigorous results is still largely lacking.

In this paper, focussing on one dimension of space, I propose such a framework. I show that observables of quantum and classical many-body systems can be organised into equivalence classes with Hilbert space structures, which naturally extract their contributions at finer and finer scales of hydrodynamics. The construction is based on the available one-parameter groups of transformations, such as space and time translations, and clustering properties of local and quasi-local observables, and is similar to the Gelfand-Naimark-Segal construction in $C^*$ algebras. The Drude weight, Onsager matrix, hydrodynamic equations and hydrodynamic entropy production are naturally expressed using the inner products on Hilbert spaces associated to the ballistic and diffusive scales. In particular, the diffusive Hilbert space $\hi H_\dif$ gives rise to a formally exact projection formula for the Onsager matrix, similar to the standard hydrodynamic projection \cite{Spohn-book,SciPostPhys.3.6.039} for the Drude weights. Superdiffusive scales correspond to a family of Hilbert spaces parameterised by the superdiffusive exponent. 

I explain how these Hilbert spaces can be related to spaces of conserved charges with various extensivity properties. The linearly extensive charges, studied in greater detail in \cite{Doyon2017,DoyonProjection}, represent the ballistic scale. These are simply related to the standard conserved quantities of many-body systems, such as the total energy and momentum, and are known to bound  the Drude weights \cite{mazur69,CasZoPre95,ZoNaPre97,ProsenQuasilocal1,ProsenQuasilocal2}. I show, likewise, that the quadratically extensive charges give rise to generic lower bounds on the Onsager matrix coefficients, generalising Prosen's lower bound \cite{Prosenquad}. Further, fractionally extensive charges give generic lower bounds for superdiffusion exponents.

Bilinear expressions in conserved quantities provide examples of quadratically and fractionally extensive charges. Analysing these, I obtain a number of results providing explicit lower bounds for diffusion and superdiffusion, which are calculable purely from the thermodynamic averages of conserved densities and currents. The strength of the result depends on the assumed space-time clustering properties of three-point functions in the state of interest.

Under Lieb-Robinson-type clustering, the diagonal Onsager matrix element for the current $\la j_k$ is strictly positive, and may be infinite, if  $\bra Q_i Q_j \la j_k\ket^{\rm c} - \sum_{mn}\bra Q_i Q_j \la q_m\ket^{\rm c} \mathsf C^{mn} \bra Q_m j_k\ket^{\rm c}$ does not vanish for some $i,j$ ($\la q_i$ are conserved densities, $Q_i$ their total space integral, $\mathsf C_{ij}$ is the static covariance matrix, $\mathsf C^{ij}$ is its inverse, and $\bra\cdots\ket^{\rm c}$ are connected correlation functions in some maximal entropy state). In particular, consider a conventional, isentropic, one-dimensional Galilean gas, where the number of particles and their total momentum are the only relevant conserved charges. Its Navier-Stokes equation for the velocity field $v$ is $\p_t v + v\p_x v = \rho^{-1} (-\p_x P + \zeta \p_x^2 v)$, where $\rho$ is the gas density, $P$ the pressure and $\zeta$ the bulk viscosity. Then, by hydrodynamic projections, we find that the bulk viscosity is bounded as
\beq\label{viscosity}
	\zeta \geq \frc{\rho \chi}{2v_{\rm LR}} \big(\p_\rho v_{\rm s}\big)^2
\eeq
where $\chi = T\p \rho/\p\mu$ is the susceptibility ($T$ the temperature, $\mu$ the chemical potential),  $v_{\rm s} = \sqrt{\dd P/\dd \rho}$ is the speed of sound and $v_{\rm LR}$ plays the role of the Lieb-Robinson velocity; see Eq.~\eqref{L11bound}.

Result \eqref{viscosity} is shown from general clustering assumptions, and, I expect, can be made rigorous in a large family of models. However, it is known from NFH that, in many cases, $\zeta$ is in fact infinite. A formal linear response calculation of Euler-scale three-point functions indicates that clustering is in fact controlled by hydrodynamic velocities. This gives stronger lower bounds. In particular, it reproduces the result from NFH that the diagonal Onsager matrix elements (and in particular $\zeta$) must be infinite whenever the NFH three-point couplings are nonzero. Further assuming finer clustering properties set by power laws around ballistic trajectories, I show that the superdiffusion exponent is bounded from below as $\alpha \geq 2/3$ if the fully diagonal three-point coupling is nonzero and the scaling function has finite variance. This lower bound is the exponent of the Kardar-Parisi-Zhang universality class that is expected by NFH in this case. I obtain other lower bounds from partially diagonal three-point couplings and scaling functions with fat tails, which, if saturated, reproduce exponents for the heat mode ($3/5$) and for the Fibonacci sequence \cite{PopkovFibonacci2015,PopkovFibonacci2016,SchutzFibonacci2018} predicted by NFH and mode coupling theory \cite{SpohnNonlinear,PopkovFibonacci2015}. Thus, various aspects of superdiffusion up to now only accessible via NFH, are seen as consequences of Hilbert space structures along with assumptions of clustering, without the need for the hydrodynamic equation or the addition of phenomenological noise.

The space of linearly extensive charges is the tangent space to the manifold of maximal entropy states. I explain how covariant derivatives are obtained from bilinear charges. The Hilbert space they span, the ``wave scattering space" $\hi H_\scat\subset \hi H_\dif$, has the physical interpretation as that of two-body scattering states of ballistic waves. In particular, in this language, states formed of co-propagating (same-velocity) ballistic waves are at the source of superdiffusion.

Projection onto $\hi H_\scat$ reproduces a formula for the Onsager matrix conjectured recently \cite{MedenjakDiffusion2019} from ideas based on NFH. As anticipated in this work, this formula is therefore here shown to provide a lower bound for the Onsager matrix. It is observed in \cite{MedenjakDiffusion2019} that this formula  agrees, when specialised to the hydrodynamics of integrable systems \cite{PhysRevLett.117.207201,PhysRevX.6.041065}, with the exact Onsager matrix. The latter was obtained by a form factor expansion in \cite{dNBD,dNBD2}, and its diagonal part was derived from a linear response mechanism in \cite{GHKV2018}. Thus, the projection onto $\hi H_\scat$ saturates the Onsager matrix in integrable systems. From this perspective, the present results and techniques put in a precise framework some of the ideas in \cite{MedenjakDiffusion2019,dNBD,dNBD2,GHKV2018}.

Finally, interpreting the diffusive and wave-scattering Hilbert spaces, I propose a notion of many-body integrability that parallels the Liouville condition that the number of conservation laws must agree with that of degrees of freedom.

The paper is organised as follows. In Section \ref{secthydro} the hydrodynamic spaces are introduced, and their relation with the Drude weight and the Onsager matrix are explained. In Section \ref{sectext}, extensive charges are defined and their consequences on generic lower bounds for diffusion and superdiffusion are proven. In Section \ref{sectwave}, bilinear charges are introduced, and the explicit lower bounds discussed above are derived.  In Section \ref{sectadd}, the relation with the hydrodynamic equation, the geometric interpretation of the ballistic and wave-scattering spaces, and the proposal for many-body integrability are discussed. Concluding remarks are made in Section \ref{sectcon}. All calculations are done in a general framework for one-dimensional many-body systems under natural assumptions, and we expect the results to be applicable to one-dimensional models such as chains and gases of particles, integrable or not. Mathematically rigorous lower bounds in various families of models will be obtained in future works.

\section{Superdiffusion bounds}\label{sectoverview}

Fourier's law and diffusive hydrodynamics are irreversible large-scale equations. Yet they are expected to emerge in many reasonable isolated, extended system, whose microscopic dynamics is reversible. Connecting irreversible characteristics to microscopic quantities without phenomenological inputs such as noise or collision assumptions -- the essence of Hilbert's sixth problem -- is one of the most important challenges of theoretical and mathematical physics. In particular, given the ubiquity of diffusive effects, an important question is as to the existence of universal, model-independent principles that would establish its presence from simpler conditions. What basic properties of the reversible dynamics and of the state guarantee the emergence of diffusion?

In this section, I overview some of the main ideas that emerge from the organisation principles based on Hilbert spaces, developed in the sections that follow. I explain how to bound the strength of diffusion and superdiffusion -- dynamical quantities -- by {\em non-dynamical, equal-time correlation functions} -- static quantities --, in a universal, largely model-independent way. Non-dynamical correlation functions are more easily accessible than dynamical quantities, and thus, I hope, this is a small step towards extracting irreversible characteristics from microscopic quantities, using a limited amount phenomenological input. The sections that follow put this into a formal context, taking into account the full space of hydrodynamic modes. However, for illustration, in this section the argument is presented in its simplest form, for a system of interacting classical particles under simple assumptions, concentrating on a single fluid mode.

I explain how, in one and two dimensions of space, it is possible to establish that hydrodynamic (super)diffusion must be present whenever certain susceptibilities are nonzero. Although the rest of the paper concentrates on one-dimensional systems, it is useful and not more costly here to address higher-dimensionality. Explicit bounds are obtained on the Onsager matrix that represents diffusion, and on the dynamical exponents controlling superdiffusion. I find, generically, the Kardar-Parisi-Zhang dynamical exponent $3/2$ in one dimension, and logarithmic superdiffusion in two dimensions, in agreement with previous studies. These results are based on a certain physically plausible assumption on 3-point dynamical correlation functions, pictorially represented in Fig.~\ref{correlation}, which is difficult to prove rigorously with current techniques. In the sections that follow, it is also shown that under weaker assumptions, most likely easier to prove, the weaker result of nonzero diffusion (positive Onsager matrix) follows, in particular leading to \eqref{viscosity}.

%How to understand emergent phenomena in complex many-body systems is one of the deepest questions in theoretical physics. One such phenomenon is irreversibility. Particularly striking is its emergence in isolated, thermodynamically large systems; isolated systems, no matter how large, obey a reversible microscopic dynamics. Irreversibility appears in many related physical processes, such as the production of entropy and thermalisation, and has been the subject of much discussion since Boltzmann \cite{}. The common understanding is that irreversibility is rendered possible by properties of both the microscopic dynamics, such as molecular chaos, and the state the system is in.

%One irreversible physical process of high interest is that of hydrodynamic diffusion.

%At long wavelengths and timescales, many-body systems tend to follow emergent hydrodynamic equations. These equations describe how disturbances propagate -- the reversible ballistic dynamics -- and spread -- the irreversible diffusive dynamics. Fourier's law is a simple example, representing the purely diffusive spreading of heat. Rigorously establishing the hydrodynamic equations in any particular model is an extremely daunting task, see \cite{} for striking results in the hard-sphere gas. However, given the ubiquity of diffusive effects, an important question is as to the existence universal, model-independent principles that would establish its presence from simpler conditions. What basic properties of the reversible dynamics and of the state guarantee the emergence of diffusion?

\subsection{Diffusion and superdiffusion in one and two dimensions}

Because of the kinematic constraints of low-dimensionality, diffusive effects have been observed to be anomalous in many one- and two-dimensional systems. On the one hand, one way to express this is to discuss correlation functions of local observables. Consider an interacting system of particles, with $r_i(t)$ the positions of the particles at time $t$, and its fluid density $n(\vec x,t) = \sum_i \delta(\vec x_i - \vec r_i(t))$. Consider the connected correlation functions at two space-time points in a thermodynamic equilibrium state, $\bra n(\vec x,t)n(0,0)\ket^{\rm c} = \bra n(\vec x,t)n(0,0)\ket - \bra n(\vec x,t)\ket\bra n(0,0)\ket$. By linear response, such correlation functions obey the hydrodynamic equation. Diffusive effects give rise to the spreading of the correlation around the ballistic trajectories of normal modes, such as the sound mode. Around a ballistic trajectory of velocity $\vec v$, at large time $t\to\infty$, normal diffusion, from hydrodynamic equations such as Fourier's law, imply
\beq\label{cf}
	\bra n(\vec x,t)n(0,0)\ket^{\rm c} \sim \frc{\chi}{g(t)^d} f\Big(\frc{\vec x-\vec vt}{g(t)}\Big),
\eeq
where $d$ is the dimension of space and $\chi$ is the static susceptibility ($\int \dd^d u\,f(\vec u) = 1$), with
\beq
	g(t) = (\lambda t)^{1/z}
\eeq
and dynamical exponent $z=2$, and with the Gaussian form $f(\vec u)= (2\pi)^{-d/2} e^{-|\vec u|^2/2}$. The constant $\lambda>0$ is the diffusion constant. By contrast, one-dimensional classical systems such as chains of anharmonically coupled oscillators display superdiffusion. In these cases, the spreading of density correlation functions around the ballistic trajectory respects \eqref{cf}, but with the dynamical exponent $z=3/2$, characteristic of the Kardar-Parisi-Zhang (KPZ) universality class. In fact, by the powerful theory of nonlinear fluctuating hydrodynamics (NFH) \cite{SpohnNonlinear}, the shape of the spreading $f(z)$ is a universal correlation function from the KPZ universality class. This function numerically looks like a Gaussian, but with faster-decaying tails.

On the other hand, a practical way to access diffusive effects in classical fluids is to consider the growth of the square-displacement of a tagged particle, say $|\Delta \vec r_1(t)|^2$ with $\Delta r_i(t) = \vec r_i(t)-\vec r_i(0)$. The variance with normal diffusive behaviour is
\beq\label{Deltar}
	\bra |\vec \Delta r_1(t)|^2\ket^{\rm c} \sim G(t)^2
\eeq
with
\beq
	G(t) = (\Lambda t)^{1/Z}
\eeq
and again dynamical exponent $Z=2$, where $\Lambda$ is related to the diffusion constant (see below). In two-dimensional fluids such as the hard- or soft-sphere fluids, numerical observations and mode-coupling theory suggest that $G(t)$ receives logarithmic corrections, $G(t) = \sqrt{\Lambda t}(\log t)^{\Gamma}$ where one observes, depending on the model or analysis, $\Gamma=1/2$ or $\Gamma = 1/4$ \cite{shinNormal2018}.

The ``correlation spread" $g(t)$ and ``trajectory spread" $G(t)$ both represent dynamical effects, and are related to each other thanks to the conservation laws. Indeed, consider the local current $\vec j(x,t) = \sum_i \vec v_i(t) \delta(\vec x_i - \vec r_i(t))$, where $\vec v_i(t) = \dd \vec r_i(t)/\dd t$ is the velocity of particle $i$. Then one finds
\beq
	\bra |\vec J(t)|^2\ket^{\rm c} =
	\sum_{ij} \bra \vec \Delta r_i(t) \cdot \vec \Delta r_j(t)\ket^{\rm c}
\eeq
where the space-time integrated current on the system's volume $V$ is $\vec J(t) = \int_0^t \dd s\int_V \dd^d x \,\vec j(\vec x,s)$. In the thermodynamic limit, where translation invariance is recovered and assuming that the particles are not correlated, we get $\sum_{ij} \bra \vec \Delta r_i(t) \cdot \vec \Delta r_j(t)\ket^{\rm c}\sim V\rho \,\bra  |\vec \Delta r_1(t)|^2\ket^{\rm c}$ where $\rho = \bra n\ket$ is the average density, and thus from \eqref{Deltar},
\beq\label{Gtcurrent}
	t \int_0^t \dd s\int \dd^d x\,\bra \vec j(\vec x,t)\cdot\vec j(0,0)\ket^{\rm c} \sim \rho\, G(t)^2.
\eeq
Using the conservation law $\p_t n + \vec \nabla\cdot \vec j=0$, and assuming the form \eqref{cf} where $f(\vec u)$ has finite variance $\Delta f = \int \dd^d u\, |\vec u|^2 f(\vec u)$, one obtains (see Appendix \ref{appcurrentcorr})
\beq\label{Ggcons}
	G(t)^2 \sim \frc{\chi \Delta f}{\rho} t g'g.
\eeq
For the power law behaviours described above, $g(t) = (\lambda t)^{1/z}$ and $G(t) = (\Lambda t)^{1/Z}$, this gives $\Lambda = \big(\frc{\chi \Delta f}{z\rho}\big)^{z/2}\lambda$ and
\beq\label{Zz}
	Z=z.
\eeq
(the diffusive case $z=2$ gives $\Lambda = \frc{d\chi\lambda}{2\rho}$). Logarithmic corrections to diffusive behaviours $g(t) = \sqrt{\lambda t}(\log t)^\gamma$ and $G(t) = \sqrt {\Lambda t} (\log t)^{\Gamma}$ are related as 
\beq\label{Gg}
	\Gamma = \gamma.
\eeq

Two comments are in order concerning this simple analysis.\medskip

\noindent {\bf 1.} In dimensions greater than one, some hydrodynamic modes propagate into shells of co-dimension 1, not described by the form \eqref{cf}. Further, in general (and generically in one dimension), there may be many ballistic modes with different velocities $\vec v_a$. These have associated partial susceptibilities $\chi_a$ (summing to the full susceptibility $\chi = \sum_a \chi_a$), correlation types $g_a(t)$ and shapes $f_a(\vec u)$. In the above analysis, one has to sum over these modes, keeping the dominant correlation type in the large-$t$ asymptotics. A more accurate treatment makes use of normal modes, as explained in the sections that follow. Here, I keep the simple assumption that $\eqref{cf}$ is sufficient in order to illustrate the ideas.

\medskip
\noindent {\bf 2.} The assumption that $f(\vec u)$ has finite variance is immediate in the case of diffusion, thanks to the Gaussian distribution. It also holds for KPZ superdiffusion -- which we may refer to as ``normal superdiffusion". However, there are types of superdiffusion that break it, for instance those related to Levy processes, as for the heat mode in one dimension \cite{SpohnNonlinear,PopkovFibonacci2015}. We will refer to this as ``fat superdiffusion". In these cases, relations \eqref{Zz} and \eqref{Gg} are broken. There is currently no known {\em a priori} criterium for establishing if the variance of $f$ is finite or not, the cases being studied to date relying on numerical observation or the phenomenological theory of NFH. However, from an analysis of the known examples, in $d=1$, it appears as though there is a special class fat superdiffusion that is relevant. The phenomenology for how to modify \eqref{Ggcons}, and the result, are as follows.

First, results from NFH suggest that, in all fat-superdiffusive cases, the normalisation coefficient in the fat tail of $\bra n(x,t)n(0,0)\ket^{\rm c}$ at large $|x|$, grows linearly in time, $\bra n(x,t)n(0,0)\ket^{\rm c}\sim c t |x|^\nu$ at large $|x|$ for some $-3< \nu<-1$. Here the bounds on $\nu$ come from the constraints of a divergent variance but a finite, time-independent susceptibility. Therefore,
\beq\label{fatf}
	f(u) \sim c'|u|^{-(1+z)}\quad (|u|\to\infty).
\eeq
In particular, this means that $\nu = -(1+z)$, and so $z\in(0,2)$, in agreement with superdiffusion. Second, in \eqref{Ggcons}, the current-current correlation function must be integrated only on the region of space relevant to the fat-superdiffusive mode. This region grows linearly in time, and is bounded by the trajectories, of velocities $v_1<v,\; v_2>v$, of the modes that surround the fat-superdiffusive mode, of velocity $v$ (for instance, the heat mode is bounded by the two sound modes in typical momentum-conserving models). From this, one obtains (see Appendix \ref{appcurrentcorr})
\beq
	G(t)^2 \sim \frc {c(a^{2-z}+b^{2-z})}{(2-z)\rho} t^{3-z}
\eeq
where $a = |v_1-v|,\,b=|v_2-v|$. Thus one has, instead of \eqref{Zz}, the relation
\beq\label{Zfat}
	Z = \frc2{3-z} \in(2/3,2) \quad \mbox{(fat superdiffusion, $d=1$).}
\eeq

\subsection{(Super)diffusion bounds from static correlations}

The above relations between correlation and trajectory spreads follow in a relatively simple fashion from the conservation laws. The results of this paper amount to a new relation, which makes use of the Cauchy-Schwartz inequality and fundamental principles of many-body physics. This new relation is powerful enough to show that, under a certain natural condition, {\em there must be diffusion or superdiffusion in any dimension $d\in[0,2]$}. Further, it shows that in $d=1$, we must have $z\leq 3/2$ for finite-variance superdiffusion.

This section is an overview of some of the calculations  in Section \ref{ssectsuper}. Sections \ref{ssectconv} and \ref{ssectlinresp} present similar calculations, but based on weaker assumptions, showing only, respecticely, positive diffusion (thus not necessarily superdiffusion, and leading to \eqref{viscosity}), and infinite diffusion (thus superdiffusion, but without information about the exponents).

%The result covers all dimensions between 0 and 2, inclusing fractional dimensions, which are accessible by considering fluids on fractal graphs.

In order to express the result, it is convenient to introduce the ordering $\succeq$ between functions $g(t), h(t)$ which, at large $t$, grow monotonically. For any two such functions, the limit $\lim_{t\to\infty} g(t)/h(t)$ exists in the extended non-negative real numbers: it is either $\infty$, a finite number, or 0. We say that $g(t)\succeq h(t)$ if $\lim_{t\to\infty} g(t)/h(t)>0$

The main result which we overview here is as follows: define the 3-point self-coupling as
\beq\label{nonzeroself}
	C:= {\rm min}_i \Big\{\int \dd^d x\,\dd^dy \, \bra n(\vec x,0)n(\vec y,0) j^i(\vec 0,0)\ket^{\rm c}\Big\} \neq 0.
\eeq
Then
\beq\label{main}
	G(t)^2g(t)^d \succeq (Ct)^2.
\eeq
That is, correlations in space, as characterised by $C$, impose a relation between correlation spreading, and trajectory spreading, which are dynamical quantities. This is a type of ``fluctuation-dissipation" theorem, involving the third cumulant for the statistical fluctuations of observables.

In particular, assuming the power laws with the relation \eqref{Zz} valid for finite variance, $G(t)\propto t^{1/z}$ and $g(t) \propto t^{1/z}$, this gives
\beq
	z\leq 1+\frc{d}2.
\eeq
For $d=1$, we recover the KPZ exponent as an upper bound, $z\leq 3/2$, and the condition of nonzero  3-point self coupling agrees with that found from NFH; and thus we derive the presence of superdiffusion without using noisy dynamics. For $d=2$ we find normal diffusion as an upper bound, $z\leq 2$. Assuming logarithmic corrections to normal diffusion and the relation \eqref{Gg}, $g(t) \propto \sqrt{t}(\log t)^\gamma$ and $G(t) \propto \sqrt t (\log t)^{\gamma}$, the relation can only be satisfied in $d\geq 2$, and gives
\beq
	\gamma\geq 0.
\eeq
That is, logarithmic corrections may be superdiffusive, but may not be subdiffusive, in agreement with previous studies. In higher dimensions $d>2$, there may be subdiffusion.

For $d=1$ and fat superdiffusion (infinite variance), using \eqref{Zfat} we obtain the bound
\beq\label{ineqfat}
	z \leq \frc{1+\sqrt 5}2.
\eeq
This is a new result. We observe that this value is the limit of the Fibonacci sequence obtained for superdiffusive modes associated to Levy processes in \cite{PopkovFibonacci2015}. Levy processes have infinite variance, so these are indeed fat-superdiffusive modes. These modes have zero 3-point self coupling in the Fibonacci sequence theory (this includes the heat mode in generic anharmonic chains, which has $z=5/3>\frc{1+\sqrt 5}2$), so are not constrained by \eqref{ineqfat}; however the limit tends, formally, to a mode with nonzero 3-point self coupling, and thus it makes sense that the limit be constrained by \eqref{ineqfat}.
%3-z + 1/z > 2 => 1-z+1/z > 0
% z^2-z-1 < 0
% 1/2 \pm 1/2 sqrt{1+4} : (1 pm sqrt{5})/2
% z = 0: <0 ok. so z \leq (1+sqrt 5)/2

\begin{figure}
\bc\includegraphics[width=6cm]{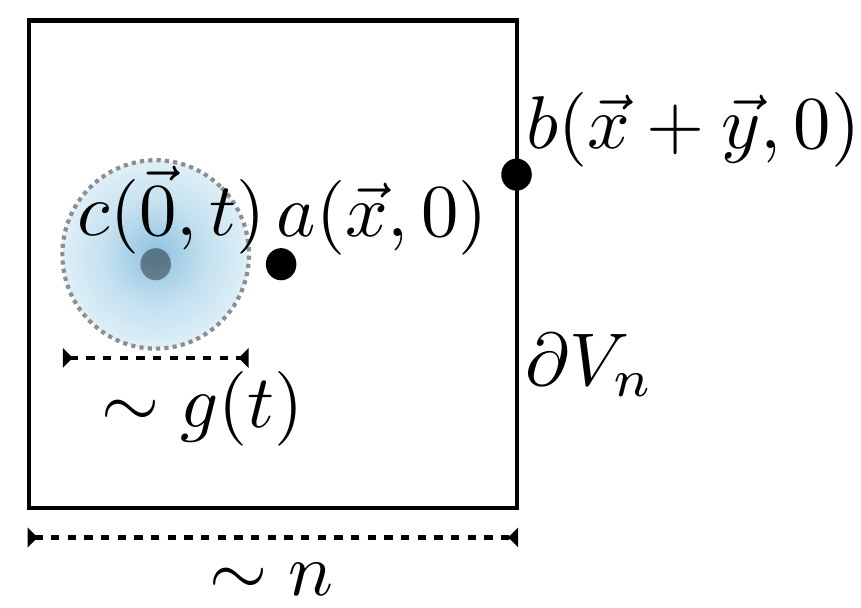}\ec
\caption{Pictorial representation of the condition on three-point correlation function. This is an extension to three-point functions of the correlation spreading form \eqref{cf} for two-point functions, involving the spreading $g(t)$. We see that if $n\gg g(t)$, it is not possible for both $a(\vec x,0)$ and $b(\vec x+\vec y,0)$ to be simultaneously within the blue shaded region, and thus by clustering the connected correlation function vanishes.}
\label{correlation}
\end{figure}

The derivation presented below takes without loss of generality zero expectation values of densities and currents (which can be achieved by appropriate shifts), and for simplicity assumes that the fluid mode's velocity is zero $\vec v=\vec 0$. It also relies on two assumptions: (a) That 4-point functions in space cluster fast enough; this is expected to be easy to prove in various models and states with current methods. (b) That dynamical 3-point functions cluster, in accordance with the correlation spread of \eqref{cf}:
\beq\label{3pointprop}
	\bra a(\vec x,0) b(\vec x+\vec y,0) c(\vec 0,t)\ket^{\rm c} \to 0 \qquad (|\vec x|\gg g(t)\ \mbox{or}\ 
	|\vec x+\vec y|\gg g(t))
\eeq
See Fig.~\ref{correlation}. This (or the required precise form) is much harder to prove, although with the less restrictive choice $z=1$ this is expected to follow from the Lieb-Robinson bound \cite{LiebRobinson} (to my knowledge the proof has not been worked out yet).

In order to derive the main result \eqref{main}, consider the observables
\beq
	Q_n(\vec x) = \int_{V_n} \dd^d y\,n(\vec x+\vec y,0)n(\vec x,0)
\eeq
where $V_n = [-n/2,n/2]^d$ is the $d$-dimensional cube of linear size $n$. These are examples of bilinear charges, see Section \ref{ssectbilinear}. Then, by using clustering of the state $\bra\cdots\ket$, we obtain
\beq\label{4pointfct}
	\int \dd^d x\, \bra Q_n(\vec x) Q_n(\vec 0)\ket^{\rm c} \sim 2n^d \Big[\int \dd^d x\,\bra n(\vec x,0) n(\vec 0,0)\ket^{\rm c}\Big]^2
	=2n^d \chi^2
\eeq
where on the left-hand side, $\bra Q_n(\vec x) Q_n(\vec 0)\ket^{\rm c} = \bra Q_n(\vec x) Q_n(\vec 0)\ket - \bra Q_n(\vec x)\ket \bra Q_n(\vec 0)\ket$. This is obtained by using clustering of the 4-point function, which implies that the leading large-$n$ asymptotics is given by the two Wick contractions that are not subtracted on the left-hand side. On the other hand, we may evaluate
\beq
	\int \dd^dx\,\bra Q_n(\vec x) \p_t \vec j(\vec 0,t)\ket^{\rm c} =
	-\int \dd^dx\,\bra \p_t Q_n(\vec x) \vec j(\vec 0,t)\ket^{\rm c}
\eeq
by using the conservation laws:
\beq
	=\int \dd^d x\int_{\p V_n}\dd^{d-1}y\cdot \Big(
	\bra  \vec
	j(\vec x+\vec y) n(\vec x,0) \vec j(\vec 0,t)\ket^{\rm c}
	+
	\bra 
	n(\vec x+\vec y) \vec j(\vec x,0) \vec j(\vec 0,t)\ket^{\rm c}
	\Big).
\eeq
Using three-point clustering property \eqref{3pointprop}, we conclude that $\int \dd^dx\,\bra Q_n(x) \p_t j(0,t)\ket^{\rm c}=0$ for all $t$ such that $g(t)\ll n$ (see Fig.~\ref{correlation}). Therefore, we may average over a time period of extent $0<t_n\ll g^{-1}(n)$,
\beq
	\int \dd^dx\,\bra Q_n(\vec x) \vec j(\vec 0,0)\ket^{\rm c} = \frc1{t_n} \int_{-t_n/2}^{t_n/2} \dd s\,
	\int \dd^dx\,\bra Q_n(\vec x) \vec j(\vec 0,s)\ket^{\rm c}.
\eeq

Now consider $\int \dd^dx\,\bra a^\dag(\vec x) b(0)\ket^{\rm c}$ as a function of the observables $a,b$. It is a sesquilinear form $(a,b)$, and non-negative, as $(a,a) = \lim_{n\to\infty} n^{-d}\,\bra A_n^\dag A_n\ket^{\rm c}$ where $A_n = \int_{V_n} \dd^dx\, a(x)$. This is the basis for the construction of the ``ballistic hydrodynamic space", where spatial coordinates are integrated over, see Section \ref{secthydro} and especially Section \ref{ssectbaldif}. Therefore, we write (assuming the conserved densities and currents to be real)
\beq
	\int \dd^dx\,\bra Q_n(\vec x) \vec j(\vec 0,0)\ket^{\rm c} = t_n^{-1} (Q_n,\vec j_n)
\eeq
where $j_n = \int_{-t_n/2}^{t_n/2} j(0,s)$. By the Cauchy-Schwartz inequality, this is bounded as
\beq
	|\int \dd^dx\,\bra Q_n(\vec x) j^i(0,0)\ket^{\rm c}|^2
	\leq
	(Q_n,Q_n)\,t_n^{-2} ( j_n^i, j_n^i)
\eeq
and thus
\beq\label{mainineq}
	C^2
	\leq
	(Q_n,Q_n)\,t_n^{-2} ( \vec j_n \cdot, \vec j_n).
\eeq
On the right-hand side, $( \vec j_n \cdot, \vec j_n)$ is (simply related to) the inner product at the basis of the ``diffusive hydrodynamic space", where time has been integrated over, see again Section \ref{ssectbaldif}. The inequality obtained is a special case of the general inequalities found from an analysis of the hydrodynamic spaces, see Section \ref{sectext}; in particular $Q_n$ gives rise to a ``quadratically extensive charge".

By time-translation invariance and \eqref{Gtcurrent}, we find $(\vec j_n\cdot,\vec j_n) \sim t_n \int_0^{t_n} \int \dd^d x\,\bra\vec j(\vec x,t)\cdot \vec j(0,0)\ket^{\rm c}\sim \rho\, G(t_n)^2$ and using \eqref{4pointfct}, we obtain$(Q_n,Q_n)= \int \dd^d x\,\bra Q_n(x) Q_n(0)\ket^{\rm c} \sim 2n^d\chi^2$. Combining with \eqref{mainineq},
\beq
	2\rho\chi^2 n^d G(t_n)^2 t_n^{-2} \succeq C^2.
\eeq
If $G(t)$ grows slower than, or as fast as, $t$, then the tightest condition is obtained by taking $t_n \sim g^{-1}(n)$. Writing $n=g(t)$, and using $\rho,\chi^2>0$, this becomes \eqref{main}.

\section{Hydrodynamic spaces} \label{secthydro}

In this section I describe an abstract and general construction which organises observables of many-body systems into equivalence classes representing their contributions at various scales of hydrodynamics. One-dimensional systems are the focus, although the extension to higher dimensions is natural.

The construction does not require any specific underlying many-body system. For generality, we consider a countable-dimensional linear space $\hi V$ over $\C$, the space of local or quasi-local observables, forming an algebra on which there is an anti-linear involution $\la a \mapsto \la a^\dag$, and over which there is a state $\bra\cdots\ket$, a positive linear functional representing the averages. Out of this, a more general setup is extracted, a Hilbert space $\hi H$ with a positive-definite sesquilinear form $\bra \cdot,\cdot\ket_{\hi H}$. This second setup is obtained from the first by taking the Cauchy completion of $\hi V$ under the topology induced by a natural inner product constructed from $\bra\cdots\ket$, as in the Gelfand-Naimark-Segal construction. Examples of inner products are
\beq\label{ipa}
	\bra\la a,\la b\ket_{\hi H} = \bra \la a^\dag \la b\ket\quad\mbox{or}\quad
	\bra\la a,\la b\ket_{\hi H} = \frc12 \bra \la a^\dag \la b + \la b \la a^\dag\ket \equiv \bra \la a^\dag \cdot \la b \ket,
\eeq
or one may use the Kubo-Mori-Bogoliubov (KMB) inner product, with $\la a\cdot\la b = \int_0^1 \dd s\,\rho^{-s} \la a \rho^{s}\la b$ is $\bra\cdots\ket$ is described by the density matrix $\rho$. Here for simplicity it is assumed that the inner product is not degenerate on $\hi V$.

The viewpoint here is that of the $C^*$ algebra formulation of quantum and classical statistical mechanics. In the context of spin-$1/2$ quantum spin chains in the thermodynamic limit (infinite length), for instance, one has a uniformly hyperfinite $C^*$ algebras \cite{BratelliRobinson12}; strong results exist and many of the properties we require are shown rigorously, in particular many of the rigorous results of \cite{DoyonProjection} can be applied. In this context, the local observables $\hi V$ can be chosen as those supported on finite numbers of sites:
\beq\label{Vspin}
	\hi V = {\rm span}\lt\{\prod_{x\in \Lambda} \sigma_{j_x}(x): \Lambda \subset\Z,\, |\Lambda|<\infty\rt\}
\eeq
where $\sigma_j(x)$ acts nontrivially only on site $x\in\Z$, as the Pauly matrix $\sigma_j$ (for $j=1,2,3$). In the modern understanding, one might consider extending this space to adjoin the quasi-local observables \cite{IlievskietalQuasilocal}, but these details are not crucial below. There is no need for $\hi V$ to take all elements of the $C^*$ algebra (we do not need $\hi V$ to be complete with respect to the $C^*$ norm). A state $\bra\cdots\ket$ could be any Kubo-Martin-Schwinger state based on a local Hamiltonian. Similar considerations hold for classical (non-quantum) models.

Note that the Hilbert space $\hi H$ is not related to any underlying quantum-mechanical Hilbert space of the many-body model; this abstract construction is valid for classical and quantum models, deterministic or stochastic.

\subsection{Hydrodynamic reduction}\label{ssectred}

Suppose that there is some action of a one-parameter group of unitary operators $U_s$ on $\hi H$, with $U_s U_{s'} = U_{s+s'}$. I take $s\in\R$, but the main property for the present discussion is that the group be non-compact; there is no difficulty in adapting to $s\in\Z$. In the continuous notation, I assume that $U_s$ is ``well behaved", that is, that the resulting inner products are continuous and differentiable. By unitarity, $U_s$ preserves the inner product,
\beq\label{inv}
	\bra U_s\la a, U_s\la b\ket_{\hi H} =\bra \la a, \la b\ket_{\hi H}.
\eeq
For instance, $U_s$ could be the group of space- or time-translations, or any other symmetry of the underlying model, under which the state is chosen to be invariant. Again, note that this does not require the underlying model to be quantum or even deterministic: $U_s$ acts, and is invertible, on the abstract Hilbert space $\hi H$, and unitarity is the result of an invariance.

The Hilbert space $\hi H$ and its inner product contain a large amount of information about the microscopic many-body model. At large scales, the number of degrees of freedom is reduced, and a small amount of information is sufficient. We would like to extract these emergent properties. As a guiding principle, emergent properties of many-body systems with respect to the group $U_s$ are essentially those obtained by looking at invariants. There are two natural ways of extracting invariants.

First, one may simply restrict to the subspace
\beq
	\hi Q = \cap_s \ker(U_s-1).
\eeq
This is the subspace of all observables which are invariant under all $U_s$. This is a closed subspace, hence we can make orthogonal projections. Let $\mathbb P$ denote the orthogonal projection onto $\hi Q$. I call this process {\em hydrodynamic projection}, and denote it by
\beq\label{pichydroproj}
	\begin{matrix}\hi H \\ \quad  \downarrow \mathbb P \\ \hi Q
	\end{matrix}\,.
\eeq

The second way goes into more details of the behaviour at large $s$. We instead consider the equivalence classes with respect to $U_s$, providing this with a Hilbert space structure by integrating over $s$. As the group is non-compact, convergence needs to be imposed, and for this the kernel needs to be taken away. More precisely, suppose that every element $\la a$ in some subspace $\whi V \subset \hi H$ projects onto $\hi Q$ at infinity fast enough: for every $\la a,\la b\in\whi V$, there exists $b>0,\,c>0$ and $d>1$ such that
\beq\label{req}
	|\bra U_s \la a,\la b\ket_{\hi H}
	- \bra \mathbb P \la a, \la b\ket_{\hi H}| \leq c(b+|s|)^{-d}.
\eeq
The quantity $c$ is the overall normalisation, $b$ is in order to encode uniform boundedness, and $d$ is the decay exponent. The precise power law form \eqref{req}, with $d>1$, is here chosen as it is one of the weakest decay assumption we may impose, yet being strong enough for our purposes.

Then we can form the {\em hydrodynamic Hilbert space} $\hi H'$. We consider the new sesquilinear form\footnote{Note the equivalent ways of writing:
\beq
	\bra (1 -\mathbb P)U_s \la a,\la b\ket_{\cal H}=
	\bra (1 -\mathbb P)U_s \la a,(1-\mathbb P)\la b\ket_{\cal H} = \bra U_s \la a,\la b\ket_{\hi H}
	- \bra \mathbb P \la a, \mathbb P \la b\ket_{\hi H}.
\eeq}
\beq\label{hydro}
 	\bra\la a,\la b\ket_{\whi H} =
	\int \dd s\, \bra(1-\mathbb P)U_s \la a,\la b\ket_{\hi H}
\eeq
for all $\la a,\la b\in\whi V$. The integral is taken over $\R$, and convergent. This is positive semidefinite, as by the assumption \eqref{req}, we can write it in an explicitly non-negative way,
\beq\label{intin}
	 \bra\la a,\la b\ket_{\whi H} = \lim_{n\to\infty} \frc1{2n} \int_{-n}^n\dd s \int_{-n}^n\dd t\, \bra(1-\mathbb P) U_s\la a, (1-\mathbb P)U_t\la b\ket_{\hi H}
\eeq
(see e.g. \cite[Lem 4.2]{Doyon2017}). We mod out the associated null space $\whi N$, and Cauchy-complete the resulting space of equivalence classes $\whi V/\whi N$, obtaining a new Hilbert space $\whi H$. On this space, $\hi Q\equiv_{\whi H} \{0\}$ as it is projected out, and $U_s\equiv_{\whi H} 1$, as we integrate over $s$ in \eqref{hydro}. 
%Thus we have
%\beq
%	\whi V\subset\hi H\  \mbox{project onto $\hi Q$},\quad
%	\whi H\supset \whi V / \whi N\ \mbox{Hilbert space completion},\quad \bra\cdot,\cdot\ket_{\whi H}\  \mbox{inner product}.
%\eeq
I call this process the {\em hydrodynamic reduction} of $\hi H$ with respect to $U=\{U_s:s\in\R\}$ with ``nucleus" $\whi V$, and pictorially represent it as
\beq
	\hi H \stackrel{U}\rightarrow \whi H
\eeq
(where I keep the information of the subspace $\whi V\subset \hi H$ hidden).

Note that if I were {\em averaging} over $s$ instead of integrating in \eqref{hydro}, then by von Neumann's ergodic theorem \cite{RudinFunctional}, this would create the projection $\mathbb P$. Here I subtract the projection, and then integrate -- looking, so to speak, at the next, nontrivial order.

The condition \eqref{req} is assumed to hold on $\whi V$. It might not be satisfied for all elements in $\hi H$. It will be convenient to define the ``$\whi H$-lower norm"  on all of $\hi H$, which takes values in the extended non-negative real numbers $\h\R = \R_+\cup\{\infty\}$, by
\beq\label{lowernorm}
	||\la a||_{\whi H}^- = \liminf_{n\to\infty} \frc1{\sqrt{2n}}  \Big|\Big|\int_{-n}^n\dd s (1-\mathbb P) U_s\la a\Big|\Big|_{\hi H}\in\h\R \quad (\la a\in\hi H).
\eeq
This agrees with the norm $||\la a||_{\whi H} = \sqrt{\bra \la a,\la a\ket_{\whi H}}$ if \eqref{req} holds for $\la b=\la a$. The upper norm could also be considered, but as I will be considering lower bounds, the lower norm is more meaningful.

\begin{rema}\label{remaambi}
Under certain conditions \cite{DoyonProjection}, it is possible to show that $\whi H$, as a set, is exactly (the completion of) the set of equivalence classes of $\whi V$ under $\la a\equiv \la a + \la q$ for $\la q\in\hi Q\cap \whi V$, and $\la a \equiv U_s\la a$ for $s\in\R$.
\end{rema}

\begin{rema}\label{remachains}
In quantum spin chains, time translation is a strongly continuous one-parameter group on $\hi H$, hence continuity and differentiability are guaranteed \cite{BratelliRobinson12}. Further, the space $\hi Q$ may be a very small space; for instance, in many natural states of quantum spin chains, the only  observables that are invariant under space translations are those proportional to the ``identity observable" ${\bf 1}$ (the identity operator on the chain)\footnote{On finite chains, homogeneous sums over the full space are trivially translation invariant; but here the chain is infinite, so these sums are not expected to give elements of $\hi H$ (and are not part of the underlying $C^*$ algebra).}, so $\hi Q = \C{\bf 1}$.
\end{rema}

\subsection{Ballistic and diffusive spaces}\label{ssectbaldif}

I now argue that the processes of hydrodynamic projections and reductions are at the basis of the most fundamental objects of hydrodynamics: the Drude weights and the Onsager matrix, which  control the Euler and diffusive scales of the hydrodynamic equations. The general viewpoint adopted above allows me to discuss these objects of hydrodynamics in a way that, I hope, makes clear the parallel between the Euler and diffusive scales, shedding light, at least structurally, on the Onsager matrix. Indeed, it suggests that the Onsager matrix is, in a sense, a susceptibility; in particular, the subtraction of the Drude weight is the subtraction of a ``disconnected" component, much like is done by writing the usual susceptibilities as space-integrated cumulants.

In homogeneous many-body systems in one dimension, two natural symmetries that the states and dynamics may have, leading to non-compact unitary groups on $\hi H$, are translations in space and time, which I will denote $\iota_x$ and $\tau_t$ respectively. Again, here I use the continuous notation, but all results apply in the discrete case as well (for instance, $x\in\Z$ in a quantum chain). In integrable systems, there is in addition an infinite number of such groups: all the flows associated to the higher Hamiltonians of the integrable hierarchy. For generality, I write $\iota_x = \tau_x^{(1)}$, $\tau_t=\tau_t^{(2)}$ and $\eta_s = \tau_s^{(3)}$, and assume in total a certain number (which may be infinite) $N\geq 2$ of nontrivial flows, $\tau_s^{(\ell)}$ for $\ell\geq 1$ with $\tau_s^{(\ell)} = 1$ for $\ell>N$, which act on $\hi H$. For instance, $\eta_s=1$ is trivial if the chain is not integrable $(N=2)$. I also assume that all flows commute.
%In this notation, the case of conventional, non-integrable systems, without nontrivial higher flows and without internal symmetries, is $N=2$.

Consider the following sequential construction. We first construct the hydrodynamic projection $\whi H$ of $\hi H$ with respect to $\iota$, with some nucleus $\whi V\subset \hi H$. Typically $\whi V = \hi V$, all local and quasi-local observables; their correlations cluster fast enough in KMS states for instance \cite{Araki}. On $\whi H$, space translations are trivial, $\iota_x \equiv_{\whi H} 1$ for all $x$. However, time translations are nontrivial, and now act on $\whi H$. I still denote by $\tau_t$ the resulting action on $\whi H$. Then, we construct the hydrodynamic projection $\wwhi H$ of $\whi H$ with respect to $\tau$ (with nucleus $\wwhi V\subset \whi H$). We continue the process until the last nontrivial flow. Pictorially,
\beq\label{HHp1}
	\hi H \stackrel{\iota}\rightarrow \whi H \stackrel{\tau}\rightarrow
	\wwhi H \stackrel{\tau^{(3)}}\rightarrow\cdots\stackrel{\tau^{(N)}}\rightarrow \hi H^{(N)} \stackrel{1}\rightarrow \{0\}.
\eeq
These may be called the hydrodynamic spaces of 0th, 1st, 2nd, ... order. For instance,
\beq\label{hydro2}
 	\bra\la a,\la b\ket_{\whi H} =
	\int \dd x\, \bra(1-\mathbb P_\iota)\iota_x \la a,\la b\ket_{\hi H},\qquad
 	\bra\la a,\la b\ket_{\wwhi H} =
	\int \dd t\, \bra(1-\mathbb P_\tau)\tau_t \la a,\la b\ket_{\whi H}.
\eeq
%Of course, in this construction, there is no need to have $\tau_t^{(\ell)}$ defined on the original space $\hi H$; it only needs to be defined on $\hi H^{(\ell-1)}$.

Let me also consider the spaces of invariants and associated projections. In fact, as the flows commute, it is more natueral to project onto the invariants of all higher flows. I then simply define
\beq\label{Qflows}
	\hi Q^{(\ell)} = \cap_{\ell'> \ell} \cap_s \ker(\tau_s^{(\ell')}-1)
\eeq
(with the notation $\hi Q^{(0)} = \hi Q$, $\hi Q^{(1)} = \whi Q$, etc.), and denote in particular $\mathbb P_\iota:\hi H\to \hi Q$, $\mathbb P_\tau:\whi H\to \whi Q$ and $\mathbb P_{\eta}:\wwhi H\to \wwhi Q$. Note that $\hi Q^{(N)} = \hi H^{(N)}$, and so, in this notation, if $N=2$ (conventional, non-integrable systems) then $\mathbb P_\eta = 1$ is trivial.

I claim that the sequence of spaces of invariants \eqref{Qflows} encode the physics of many-body systems at finer and finer hydrodynamic scales. In particular, I call $\hi Q$ the {\em thermodynamic space}, $\whi Q$ the {\em ballistic space}, and $\wwhi Q$ the {\em diffusive space}:
\beq
	\hi Q = \hi H_{\rm th},\quad
	\whi Q = \hi H_{\rm bal},\quad
	\wwhi Q = \hi H_{\rm dif}.
\eeq
Thus, pictorially,
\beq\label{HHp}
	\begin{matrix} \hi H & \stackrel{\iota}\rightarrow & \whi H & \stackrel{\tau}\rightarrow & 
	\wwhi H & \cdots \\
	\quad  \downarrow \mathbb P_\iota & & \quad  \downarrow \mathbb P_{\tau} &&
	\qquad  \downarrow \mathbb P_{\eta} \\
	\hi H_{\rm th} &&  \hi H_{\rm bal} &&\hi H_{\rm dif} & \cdots
	\end{matrix}
\eeq
I will  use the notation (recall that $\eta$ represent a flow with respect to a higher conserved charge, if any, and $\mathbb P_\eta = 1$ if there aren't any)
\beq
	\bra \la a,\la b\ket_{\bal} = \bra \mathbb P_\tau \la a,\la b\ket_{\whi H},\quad
	\bra \la a,\la b\ket_{\dif} = \bra \mathbb P_\eta \la a,\la b\ket_{\wwhi H}.
\eeq
In Section \ref{ssectkubo}, I justify this claim by connecting with the Drude weights and Onsager matrix, and in Section \ref{ssecthydroeq}, by making the connection with the hydrodynamic equations.

As is discussed below, $\mathbb P_\iota$ is just the subtraction of expectation values -- this is a trivial part of the construction, which I discuss mainly to establish the structural symmetry. In general $\hi H_{\rm bal}$ contains the extensive conserved charges, and so this part of the construction is nontrivial. Finally, if there are no higher flows, for instance in chaotic systems, then $\mathbb P_\eta=1$ and $\hi H_{\rm dif} = \wwhi H$. In this case, the diffusive space is formed of all remaining observables after the $\tau$ hydrodynamic reduction (where there is no space-time translation left). We will see below that, by contrast, if $\eta$ is nontrivial, generic elements of $\wwhi H$ are not in $\hi H_{\rm dif}$, but the currents of conserved densities are; this is relevant to integrable systems.

For simplicity I will keep the same notation of lower-case latin letters $\la a,\la b,\ldots$ for elements of $\hi V, \hi H, \whi H,\wwhi H$. I use the notation $\equiv_{\whi H},\,\equiv_{\wwhi H}$ for equivalences in the indicated spaces, and $[\cdot]_{\whi H},\,[\cdot]_{\wwhi H}$ for the corresponding equivalence classes, when it is necessary for clarity.

\begin{rema}\label{remataut}
The passage from $\tau_t$ acting on $\hi H$ to $\tau_t$ acting on $\whi H$ is nontrival. However, under general conditions (see \cite[Thm 6.3]{Doyon2017} and \cite{DoyonProjection}), including the condition that $\whi V$ be in some sense large enough, $\tau_t$ is indeed on $\whi H$ a well-behaved unitary operator for all $t\in\R$, forming a one-parameter group. For instance, all conditions are satisfied in quantum spin chains with finite local spaces, where we may take $\whi V = \hi V$.
\end{rema}

\subsection{Drude weights and Onsager matrix}\label{ssectkubo}

Consider the algebra $\hi V$ of observables. First, as mentioned in Remark \ref{remachains}, in most cases, the only translation-invariant local observables are those proportional to the identity, $\hi H_{\rm th} = \C{\bf 1}$ (which is also invariant under all higher flows), thus this is assumed here for simplicity. Then, the projection onto $\hi H_{\rm th}$ is just the subtraction of the average of the observable:
\beq
	\mathbb P_\iota\la a = {\bf 1}\bra \bf 1,\la a\ket_{\hi H} = {\bf 1}\bra \la a\ket.
\eeq
That is, the thermodynamic space is nothing but the set of averages of local observables, and this indeed describes the thermodynamics. The presence of a higher-dimensional space of $\iota$-invariant is associated with phase co-existence, see for instance \cite{BratelliRobinson12,IsraelConvexity}. We now see that the condition \eqref{req}, with respect to $U = \iota$, is simply that of strong enough clustering at large distances. Then $\bra \la a,\la b\ket_{\whi H}$ is the space-integrated connected two-point function: \beq\label{ipiota}
	\bra \la a,\la b\ket_{\whi H} = \int \dd x\,\bra \la a^\dag(x)\cdot\la b(0)\ket^{\rm c} = \bra A^\dag\cdot\la b\ket^{\rm c} := \bra A^\dag \cdot \la b\ket - \bra A^\dag\ket\bra \la b\ket
\eeq
where $\la a(x) = \iota_x\la a$ and I denote by capital letters the formal total space integrals, $\int \dd x\,\la a(x) = A$. In quantum spin chains,  in any Gibbs state with respect to a local enough Hamiltonian, the clustering requirement \eqref{req} follows for local observables from a standard result of Araki \cite{Araki}, and as mentioned we may take $\whi V = \hi V$ for the nucleus.

Second, the space of invariants $\hi H_{\rm bal}$ is the space of {\em conserved quantities} of the model. Indeed, if observables $\la q$ and $\la j$ in the nucleus $\whi V$ are related as per the continuity relation
\beq\label{conslaw}
	\p_t \la q(x,t) + \p_x \la j(x,t) = 0
\eeq
where $\la a(x,t) = \iota_x \tau_t \la a$, then clearly $\p_t \la q(x,t) \equiv_{\whi H} 0$, as the action of $\iota_x$ on $\whi H$ is trivial. In fact, it is possible to show that if $\la q$ clusters fast enough and $\p_t \la q(x,t)\equiv_{\whi H} 0$, then there exists $\la j$, which is clustering, such that \eqref{conslaw} holds; see \cite{DoyonProjection} for the precise statement. Therefore, if $\la q$ satisfies \eqref{conslaw}, the equivalence class of $\la q$, an element of $\whi H$, lies in $\hi H_\bal$,
\beq
	[\la q]_{\whi H} \in \hi H_\bal.
\eeq
Under the equivalence, we have $\la q \equiv_{\whi H} \la q + \p_x \la a$ for local $\la a$, and therefore it is natural to identify the equivalence class of $\la q$ with the formal, total space integral $Q = \int \dd x\, \la q(x)$ (that is, $\int \dd x\,\iota_x : \hi V \to \whi H$). This is a conserved quantity.

Denoting by $\la q_i$ a basis for $\hi H_{\rm bal}$ (which is countable-dimensional), we have, on $\whi H$,
\beq\label{proj}
	\mathbb P_\tau\la a = \sum_i \la q_i \mathsf C^{ij} \bra \la q_j,\la a\ket_{\whi H}
\eeq
where $\mathsf C^{ij}$ is the inverse of the static covariance matrix $\mathsf C_{ij} = \bra \la q_i,\la q_j\ket_{\whi H}$ (see Appendix \ref{diagonal}). In particular, there are basis elements $\la q_I$ for $\hi H_\bal$ that diagonalise the flux Jacobian, see Appendix \ref{diagonal}; these are the normal modes, here labelled by capital letters $I,J,K,\ldots$.

Of interest is the large-time limit, or time-average, of overlaps where one observable is evolved in time. By von Neumann's ergodic theorem (see the precise statement in \cite{DoyonProjection}), this projects onto $\hi H_{\rm bal}$,
\beq\label{hpdrude}
	\mathsf D_{\la a,\la b} = \lim_{T\to\infty} \frc1{2T}
	\int_{-T}^T \bra \tau_t \la a,\la b\ket_{\whi H}
	= \bra\la a, \la b\ket_{\bal}.
\eeq
If $\la a = \la j_i$ and $\la b=\la j_j$ are chosen to be currents associated to conserved densities as in \eqref{conslaw}, then $\mathsf D_{\la a,\la b}$ are the widely-studied {\em Drude weights} characterising ballistic transport \cite{Spohn-book}. Thus indeed $\hi H_{\rm bal}$ is associated to ballistic transport. In particular, we obtain the hydrodynamic projection formula for the Drude weights \cite{Spohn-book,SciPostPhys.3.6.039},
\beq\label{hpdrude2}
	\mathsf D_{\la a,\la b} = \sum_{ij}
	\bra \la a ,\la q_i\ket_{\bal} \mathsf C^{ij}
	\bra \la q_j,\la b\ket_{\bal}.
\eeq
The inner product \eqref{ipiota} and the associated Hilbert space $\whi H$ have been studied in the context of many-body systems \cite{Doyon2017}. The space of invariants $\hi H_{\rm bal}$ can be shown rigorously to control the correlation functions at Euler scales \cite{DoyonProjection}.

Finally, what is the meaning of $\wwhi H$, and in particular of the space of invariants $\hi H_{\rm dif}$ of $\wwhi H$? Of course, this invariant subspace is not the full space only if there are nontrivial higher flows, $N>2$. Recall that by construction $\la q_i\equiv_{\wwhi H}0$; hence these are not nontrivial elements of $\hi H_{\rm dif}$. But we get intuition on $\hi H_{\rm dif}$ by noting that, if $[\la j_i]_{\whi H}$, from the current defined in \eqref{conslaw}, projects fast enough onto $\hi H_\bal$ at large times, so that it lies in the nucleus $\wwhi V$, then {\em the elements $[\la j_i]_{\wwhi H} \in \wwhi H$ lie in the subspace of invariants under any higher flows},
\beq\label{jiproj}
	[\la j_i]_{\wwhi H}\in \hi H_\dif.
\eeq
This is a result obtained in a different context \cite{DurninTherma2020}, and reproduced in Appendix \ref{appcur}. The proof can be obtained in the $\hi H$-setup, and relies on the assumption that the conserved density $\la q_i\in \whi V$, associated to $\la j_i$, satisfies continuity equations such as \eqref{conslaw} but with other local currents $\la j_i^{(\ell)}$ when evolved under higher flows $\tau^{(\ell)}_s$.  Then, we can re-write the inner product in $\wwhi H$ as
\beq
	\mathfrak L_{\la a,\la b} = \bra \la a,\la b\ket_{\wwhi H} =
	\int \dd t\,\Big[
	\int \dd x\,\bra \la a^\dag(x,t)\cdot \la b(0,0)\ket^{\rm c}
	-
	\mathsf D_{\la a,\la b}
	\Big].\label{Onsa}
\eeq
Taking again $\la a = \la j_i$ and $\la b=\la j_j$, Eq.~\eqref{Onsa} is nothing else but the {\em Green-Kubo formula for the Onsager matrix}, and we see that
\beq\label{Lijcurrent}
	\mathfrak L_{ij} = \mathfrak L_{\la j_i,\la j_j} = \bra \la j_i,\la j_j\ket_\dif.
\eeq
As the Onsager matrix is simply related to the diffusion matrix  \cite{Spohn-book} (see also \cite{dNBD2})\footnote{In \cite[Sec 2]{dNBD2}, the derivation is correct in the quantum case if products of local observables are replaced by the KMB product $\cdot$. This guarantees that the result of the formal $\beta_j(x)$-differentiation is real in \cite[Eq B.3]{dNBD2}, and that $\mathcal{PT}$-symmetry implies \cite[Eq 2.33]{dNBD2}. This then shows that \eqref{Onsa} and \eqref{Lijcurrent} is indeed a good definition of the Onsager matrix elements in the quantum case.}, this shows that $\hi H_\dif$ is the Hilbert space encoding diffusive processes in hydrodynamics.

Much like in \eqref{hpdrude2}, we can expand in a basis $\{\la w_\alpha\}$ for $\hi H_{\rm dif}$,
\beq\label{diffdecomp}
	\bra  \la a,\la b\ket_{\dif} = \sum_{\alpha,\beta}
	\bra \la a,\la w_\alpha\ket_{\dif} \mathfrak C^{\alpha\beta}
	\bra \la w_\beta,\la b\ket_{\dif},
\eeq
where $\mathfrak C^{\alpha\beta}$ is the inverse of the {\em diffusive covariance matrix} $\mathfrak C_{\alpha\beta}$,
\beq
	\mathfrak C_{\alpha\beta} = \bra \la w_\alpha,\la w_\beta\ket_{\wwhi H}.
\eeq
If we know the elements $\la w_\alpha$ and how to evaluate the components of $\la a$ and $\la b$ on these, we have an {\em exact projection formula} for the Onsager matrix. In order for this  to be useful, we need to characterise the basis elements $\la w_\alpha$. Characterising the elements $\la w_\alpha$ is where all the physics of diffusion lies. In order to  address this, I discuss  the relation between the hydrodynamic spaces and {\em extensive charges} in Section \ref{sectext}.

Crucially, however, the clustering requirement \eqref{req} for $\whi H$, instead of $\hi H$, {\em does not} follow on local observables from standard results. The limit defining the Onsager matrix may not exist. Nevertheless, at least for the diagonal elements, one can always {\em define} the lower Onsager matrix elements using the lower norm \eqref{lowernorm}, as
\beq
 	\mathfrak L_{\la a,\la a}^- = \big(||\la a||_{\wwhi H}^-\big)^2,\quad
	\mathfrak L_{ii}^- = \big(||\la j_i||_{\wwhi H}^-\big)^2.
\eeq
These exist in the extended reals for all $\la a\in\whi H$ (including all $\la j_i\in \whi H$). It is the analysis of such lower norms, and their relations to extensive charges, that lead to nontrivial statements about diffusion and superdiffusion.

\begin{rema} The phenomenology from nonlinear fluctuating hydrodynamics (NFH) \cite{SpohnNonlinear} suggests that in typical non-integrable systems, clustering on $\whi H$ is not fast enough for many observables, including some (or all) conserved currents $\la j_i$, and that one has $\mathfrak L_{ii}^- = \infty$. In integrable systems, the Onsager matrix elements appear to be finite for a large class of currents \cite{dNBD,dNBD2,GHKV2018}, but not for all \cite{IlievskiSuper2018,LZP19,GopalaKinetic2019,GopalaAnomalous2019,deNardisAnomalous2019,BulchandaniKardar2019}. As part of the results of this work, sufficient conditions for lower Onsager matrix elements to be infinite are obtained, these conditions agreeing with the results of NFH.
\end{rema}

\begin{rema}\label{remaqi} The conserved densities $\la q_i$ can be considered as lying in various spaces: as local observables in $\hi V$ and in fact of the nucleus $\whi V$, or equivalence classes, elements of $\whi H$, spanning the subspace $\hi H_\bal$. The latter is the more fundamental definition; however the former is more convenient for explicit calculations, and for considerations of the conservation laws \eqref{conslaw}. Naturally, an element $\la q_i$ of $\whi H$ can be defined from an element of $\whi V$, but in general the converse may have obstructions ($\la q_i$ may be a Cauchy sequence which converges in $\whi H$ but not in $\hi H$, or not in more general expectation values $\bra\cdots\ket$) and ambiguities (shifts $\la q_i(x)\mapsto \la q_i(x)+a{\bf 1}$ and $\la q_i(x)\mapsto \la q_i(x)+\p_x \la a(x)$ modify the element of $\hi V$ or $\whi V$, but not its equivalence class in $\whi H$).

Below, I will assume that $\hi V\subset \whi V$ (the local observables are clustering fast enough in space; see Remark \ref{remataut}), that the above ambiguities are the only ones (see Remark \ref{remaambi}), and that it is possible to take a basis $\la q_i\in \whi H$ formed of equivalence classes of local observables (elements of $\hi V$), and I will choose for every such $\la q_i$ a unique representative, still denoted $\la q_i \in \hi V$, which satisfies conservation laws \eqref{conslaw} for local current observables with respect to time evolution $\tau_t$ and every higher flow $\tau_s^{(\ell)}$ (if any).

In non-integrable cases, $\hi H_\bal$ is expected to be a finite-dimensional space generated by  local observables. In integrable quantum chains, $\hi V$ may have to include observables which are quasi-local, supported on infinitely-many sites but with appropriate decay properties \cite{IlievskietalQuasilocal}.
\end{rema}

\begin{rema}\label{remaflows}
It is usually the case that the conserved charges $Q_i = \int \dd x\,\la q_i(x)$ generate flows on the observables in $\hi H$, with infinitesimal generator given, on a dense subspace, by the commutator (resp.~Poisson bracket) in quantum (resp.~classical) systems, $\ri [Q_i,\cdot]$. Further, it is also often the case that many, although not necessarily all, of these flows commute with each other.
\end{rema}

\subsection{Maximal entropy states and hydrodynamic equation}\label{ssecthydroeq}

The state $\bra\cdots\ket$ has up to now been assumed to be invariant under the flows $\tau_s^{(\ell)}$, including space and time translations, and to have appropriate clustering properties, Eq.~\eqref{req} and the various clustering assumptions in Section \ref{sectwave}. A natural set of such states are those which have maximised entropy with respect to all available conserved charges. Their statistical distributions are formally of the form $\re^{-\sum_i \beta^i Q_i}$, where again $Q_i = \int \dd x\,\la q_i(x)$ are the (formal) total conserved charges. A better definition is that the thermodynamic potentials $\beta^i$ generate a flow on the set of states, defined by the equation
\beq
	-\frc{\p}{\p\beta^i} \bra \cdots\ket = \bra Q_i \cdots\ket - \bra Q_i\ket \bra \cdots\ket.
\eeq
This can be taken as the starting point for a rigorous construction of the states, see \cite{Doyon2017}. These maximal entropy states include Gibbs and generalised Gibbs states. An immediate consequence is that {\em variations in $\hi H_{\rm th}$ are related to overlaps in $\hi H_{\rm bal}$}:
\beq\label{derbeta}
	-\frc{\p \bra \la a\ket}{\p\beta^i} = \bra \la q_i,\la a\ket_{\whi H} = \bra \la q_i,\la a\ket_{\hi H_\bal}.
\eeq
Note in particular that the thermodynamic potentials $\beta^i$ are here defined in a gauge-invariant fashion (they do not depend on the specific choice of $\la q_i$ as an element of $\hi H$).

Within this formulation, one can write the hydrodynamic equations in an elegant form that makes clear the role of $\hi H_{\bal}$ and $\hi H_\dif$. Recall that the hydrodynamic equation is an equation for how the state $\bra\cdots\ket_{x,t}$ depends on $x,t$, in the manifold of maximal entropy states. Using the state coordinates $\beta^i(x,t)$, up to diffusive scales, a form of the hydrodynamic equation is
\beq
	\sum_j\mathsf C_{ij} \p_t \beta^j + \sum_j(\mathsf A\mathsf C)_{ij} \p_x \beta^j = \frc12 \p_x \big(\mathfrak L_{ij} \p_x \beta^j\big)
\eeq
where $\mathsf C_{ij} = \bra \la q_i,\la q_j\ket_{\rm bal}$ is the static correlation matrix, and $\mathsf A_{i}^{~j} = \p\bra \la j_i\ket /\p\bra \la q_j\ket$ is the flux Jacobian, with $(\mathsf A\mathsf C)_{ij} = \bra \la q_i ,\la j_j\ket_{\rm bal}$. This can equivalently be written as
\beq\label{hydroeq}
	\bra \la u + \la v,\la q_i\ket_\bal = \frc12 \p_x \bra \la v,\la j_i\ket_{\dif},\qquad
	\la u = \sum_i \la q_i \p_t\beta^i,\ \la v = \sum_i \la j_i \p_x\beta^i,
\eeq
where ballistic and diffusive terms are described by the ballistic and diffusive Hilbert spaces. In particular, the Euler equation is an equation on $\hi H_\bal$,
\beq
	\la u + \la v = 0 \quad (\mbox{on $\hi H_\bal$, at the Euler scale}).
\eeq
At the diffusive scale, the total hydrodynamic entropy production can be written as
\beq\label{entropy}
	\p_t S = \frc12\int \dd x\,||\la v||_{\dif}^2,
\eeq
which is obviously non-negative. The diffusive Hilbert space also gives a measure of the distance of a space-time configuration to the Euler equation, for instance
\beq
	\int_{x_1}^{x_2}\dd x\,\big| \bra \la u + \la v,\la q_i\ket_\bal\big|
	\leq
	(N_{[x_1,x_2]}+1)
	\max_{[x_1,x_2]}\big(\,||\la v||_{\dif} ||\la j_i||_{\dif}\,\big)
\eeq
where $N_{[x_1,x_2]}$ is the number of extrema of $\bra \la v,\la j_i\ket_{\dif}$ in the open interval $(x_1,x_2)$.

The hydrodynamic equation \eqref{hydroeq} is written in an explicitly {\em gauge-invariant form}: the inner products are gauge invariant, and the state coordinates $\beta^i$ also are. Thus, even up to diffusive scale, the hydrodynamic equation is universal: it does not depend on the choice of the conserved densities $\la q_i$ as elements of $\hi H$. In order to extract physical information from \eqref{hydroeq}, however, one must provide a map from $\beta^i$ to averages of local observables in space-time. This map is not universal; in general, one chooses a set of densities $\la q_i$ in $\hi H$ and the map is completely fixed by requiring $\bra \la q_i(x,t)\ket = \bra \la q_i\ket_{x,t}$.

\subsection{Fractional-order spaces and superdiffusion}\label{ssectfractorder}

It may happen that for some elements $\la a\in\hi H$ the requirement \eqref{req} is not satisfied: the vanishing is not fast enough for the integral \eqref{hydro} to exist. Consider the discussion of Subsection \ref{ssectbaldif}. It may be that $||\la a||_{\whi H}^-=\infty$. It is then natural to define Hilbert spaces at fractional orders, lying in a sense between $\hi H$ and $\whi H$. A proposal is, for $u\in(0,1)$,
\beq\label{Hu}
	\bra \la a,\la b\ket_{\hi H^{(u)}} = \lim_{n\to\infty} n^{u-1} \int_{-n}^n \dd x\,\bra (1-\mathbb P_\iota)\iota_x\la a,\la b\ket_{\hi H}.
\eeq
More precisley one may consider elements for which the following asymptotic expansion holds
\beq\label{asympfrac}
	\bra (1-\mathbb P_\iota)\iota_x\la a,\la b\ket_{\hi H} = \omega_{\la a,\la b}^\pm |x|^{-u}(1+o(1))\quad (s\to\pm\infty)
\eeq
for some $\omega_{\la a,\la b}^\pm\in\C$. Then,
\beq
	\bra \la a,\la b\ket_{\hi H^{(u)}} = \frc{\omega_{\la a,\la b}^++\omega_{\la a,\la b}^-}{1-u}.
\eeq
Non-negativity of the inner product is guaranteed by a similar argument as in \eqref{intin}:
\beqa
	\lefteqn{\lim_{n\to\infty} (2n)^{u-2}\int_{-n}^n \dd x\,\int_{-n}^n\dd y
	\,\bra(1-\mathbb P_\iota) \iota_x\la a, (1-\mathbb P_\iota)\iota_y\la b\ket_{\hi H}} &&\n &=&
	\lim_{n\to\infty} (2n)^{u-2}\int_{-2n}^{2n} \dd x\,|2n-x|
	\,\bra(1-\mathbb P_\iota) \iota_x\la a, \la b\ket_{\hi H} 
	= \frc{\omega_{\la a,\la b}^++\omega_{\la a,\la b}^-}{(1-u)(2-u)} = \frc{\bra \la a,\la b\ket_{\hi H^{(u)}}}{2-u}.
	\label{fracquad}
\eeqa
The left-hand side is non-negative whenever $\la a=\la b$. A possible definition of the $u$th-order space with respect to $\hi H$  is therefore as the completion  $\hi H^{(u)}$ of the set of equivalence classes of a subspace $\hi V^{(u)}\subset\hi H$ (the nucleus) in which \eqref{asympfrac} holds pairwise. Likewise, the $u$th-order lower norm is, for any $\la a\in \hi H$,
\beq\label{lowernormHu}
	||\la a||_{\hi H^{(u)}}^- =
	\sqrt{2-u}\liminf_{n\to\infty} (2n)^{\frc{u-2}2}\Big|\Big|\int_{-n}^n \dd x
	\,(1-\mathbb P_\iota) \iota_x\la a\Big|\Big|_{\hi H}\quad(u\in(0,1)).
\eeq
Note that there is at most one value of $u_*\in(0,1)$ such that $0<||\la a||_{\hi H^{(u_*)}}^-<\infty$, with ordering on the extended reals,
\beq\label{ordering}
	||\la a||_{\hi H^{(u)}}^- =0 \ (u<u_*), \quad
	||\la a||_{\hi H^{(u)}}^- = \infty\ (u>u_*).
\eeq

The construction can be made based on $\whi H$ instead of $\hi H$. I will understand $\hi H^{(u)}$ for $u\in (j,j+1)$, $j=1,2,3,\ldots$, as $(u-j)$th-order spaces with respect to $\hi H^{(j)}$. For instance, for $u\in(1,2)$, there is a nucleus $\hi V^{(u)}\in\whi H$ where the asymptotic condition
\beq\label{wasympfrac}
	\bra (1-\mathbb P_\tau)\tau_t\la a,\la b\ket_{\whi H} = \omega_{\la a,\la b}^\pm |t|^{-(u-1)}(1+o(1))\quad (t\to\pm\infty),
\eeq
holds; the inner product is
\beq\label{wHu}
	\bra \la a,\la b\ket_{\hi H^{(u)}} = \lim_{n\to\infty} n^{u-2} \int_{-n}^n \dd t\,\bra (1-\mathbb P_\tau)\tau_t\la a,\la b\ket_{\whi H},
\eeq
and for every $\la a,\la b\in \hi V^{(u)}$; the space $\hi H^{(u)}$ is obtained by completing the space of equivalence classes, and the lower norm is
\beq\label{loweru12}
	||\la a||_{\hi H^{(u)}}^- = \sqrt{3-u}
	\liminf_{n\to\infty} (2n)^{\frc{u-3}2}\Big|\Big|\int_{-n}^n \dd t
	\,(1-\mathbb P_\tau) \tau_t\la a\Big|\Big|_{\whi H}\quad(u\in(1,2)).
\eeq

For my purposes, the construction of $\hi H^{(u)}$ for $u\in(1,2)$ is the most relevant. By the calculation of Appendix \ref{appcur}, one shows that if a current observable $j_i$ has equivalence classes that lies in $\hi H^{(u)}$, then it lies in the subspace of invariants $\hi Q^{(u)}\subset \hi H^{(u)}$ under higher flows. This space of invariants is, in some sense, ``in-between" the ballistic and diffusive spaces.

In Appendix \ref{appsscaling}, I argue that $\hi H^{(u)}$ (or more precisely, $\hi Q^{(u)}$) represents {\em superdiffusion}. More precisely, from considering the phenomenology of superdiffusive correlation functions in space-time, I obtain a precise definition of what characterises a superdiffusive mode. Consider a hydrodynamic normal mode $\la q_I$. If it is superdiffusive, one expects, at large space-time separations,
\beq\label{qqsup0}
	\bra \la q_I(x,t)\la q_I(0,0)\ket^{\rm c} \sim
	\frc1{|\lambda t|^\alpha} f\Big(\frc{x-v_I^{\rm eff}t}{|\lambda t|^\alpha}\Big)
\eeq
where $f(z)>0$ integrates to 1, $\lambda>0$, $v^{\rm eff}_I$ is the eigenvalue associated to $\la q_I$, and $\alpha \in (1/2,1)$ (the case $\alpha=1/2$ would be normal diffusion). In fact, for a given $\alpha$, one should distinguish two types of superdiffusive behaviours, ``normal" and ``fat", according as the variance of the distribution $f(z)$ behaves: finite or infinite. In the latter case, an analysis of behaviours studied in the literature suggests a mechanism by which a precise power law emerges as a function of $\alpha$. Thus one has either
\beq\label{normalassump0}
	\int \dd z\,z^2f(z)<\infty\quad\mbox{(normal superdiffusion),}
\eeq
or, by the argument presented in Appendix \ref{appsscaling},
\beq\label{fatassump0}
	f(z) \sim c |z|^{-\big(1+\frc1\alpha\big)}\quad\mbox{(fat superdiffusion)}
\eeq
for some $c>0$, at large $|z|$. In Appendix \ref{appsscaling} I then explain how, using conservation laws, Eq.~\eqref{qqsup0}, and these considerations about the variance of $f(z)$, the following conditions on the lower norm \eqref{loweru12} of the associated current $\la j_I$ are obtained:
\beq\label{exponent} ||j_I||_{\hi H^{(u)}}^- = 
	\lt\{\ba{ll}
	0 & \mbox{for $u<3-2\alpha$}\\
	\infty & \mbox{for $u>3-2\alpha$}
	\ea\rt. \quad \mbox{(normal superdiffusion)}
\eeq
and
\beq\label{exponentfat}
	||j_I||_{\hi H^{(u)}}^- = \lt\{\ba{ll}
	0 & \mbox{for $u<1/\alpha$}\\
	\infty & \mbox{for $u>1/\alpha$}
	\ea\rt. \quad \mbox{(fat superdiffusion).}
\eeq
These are taken here as precise definitions of normal and fat superdiffusive modes. That is, a current observable $\la j_I\in \whi H$ has normal (fat) superdiffusion exponent $\alpha$ if Eq.~\eqref{exponent} (Eq.~\eqref{exponentfat}) holds. These conditions will form the basis for the results on superdiffusion in the following sections.

\section{Extensive charges and lower bounds on diffusion and superdiffusion} \label{sectext}

Extensive charges are sequences of observables which represent, in a more concrete fashion, the equivalence classes discussed in the previous section. They serve here two purposes: first to make the abstract hydrodynamic Hilbert spaces more concrete, and second to provide a way of bounding (from below) physical quantities of interest, in particular the Onsager matrix elements and superdiffusion exponents.

\subsection{Linearly extensive charges}\label{ssectext}

Let me consider the general setup of Subsection \ref{ssectred}, the reduction $\hi H\stackrel{U}\rightarrow \whi H$ with nucleus $\whi V$. The concept of extensive charges emerged from Prosen's work on pseudolocal charges \cite{ProsenPseudo1,ProsenPseudo2}, and was developed in integrable spin chains \cite{ProsenQuasilocal1,ProsenQuasilocal2}. I gave a precise construction in \cite{Doyon2017} in the context of quantum spin chains. I follow these lines here, and define {\em linearly extensive charges} in the present more general context.

Consider a sequence $Q = (Q_n\in\hi H:n\in\N)$. Assume without loss of generality that $Q_n = (1-\mathbb P)Q_n$ for all $n$. Suppose there is a subspace $\whi D(Q)\subset \hi H$, the domain of $Q$, with $\whi V\subset \whi D(Q)$, such that the sequence satisfies the following conditions:
\bi
\item[1.] There exists $\gamma>0$ such that $||Q_n||_{\hi H}^2 \leq \gamma n$ for all $n\in\N$ large enough.
\item[2.] The limit $ Q'(\la a): = \lim_{n\to\infty} \bra Q_n,\la a\ket_{\hi H}$ exists for all $\la a\in\whi D(Q)$.
\item[3.] There exists $k>0$ such that $\lim_{n\to\infty} \sup_{s,t\in[-kn,kn]} |\bra Q_n,U_s \la a-U_t\la a\ket_{\hi H}| = 0$ for all $\la a\in\whi D(Q)$.
\ei
A linearly extensive charge for $\hi H$ is an equivalence class of such sequences, under the equivalence $Q' \equiv \t Q'$ if and only if  $Q'(\la a) = \t Q'(\la a)\,\forall\,\la a\in\whi V$. That is, the set of such limit-actions of such sequences is the set of extensive charges.

The first point above states the linear extensivity of the charge. The third specifies that the charge is extensive with respect to $U_s$, and that it is homogeneous. The second specifies that it acts well on $\whi D(Q)$. I emphasise that there is here {\em no condition on these charges being ``conserved"}: in the general setup there is no such notion, as only the one-parameter group $U$ is considered.

One can show (see below) that $Q'(\cdot)$ is bounded with respect to the lower norm on $\whi H$. As a consequence, on $\whi V$ it is continuous with respect to the norm on $\whi H$, and by the Riesz representation theorem, there exists $\la b\in\whi H$ such that
\beq\label{linhil}
	Q'(\la a) = \bra\la b,\la a\ket_{\whi H} \quad \forall\;\la a\in\whi V.
\eeq
Since $\whi V$ is dense in $\whi H$, the map can be extended, by continuity, to all of $\whi H$.

The proof of boundedness is simple. I reproduce that of \cite{Doyon2017}. Let $\la a\in\whi D(Q)$. By Point 3 above and the assumption that $Q_n = (1-\mathbb P)Q_n$, there exists $\delta_n$ with $\lim_{n\to\infty}\delta_n=0$, such that
\beq
	\bra Q_n,\la a\ket_{\hi H} = \frc1{2kn} \int_{-kn}^{kn}\dd s\, \bra Q_n,(1-\mathbb P)U_s \la a\ket_{\hi H} + \delta_n.
\eeq
The norm of the first term on the right-hand side is bounded by using the Cauchy-Schwartz inequality:
\beq\label{leq}
	\leq \sqrt{\frc{|| Q_n||^2_{\hi H}}{2kn}}
	\sqrt{\frc1 {2kn}\int_{-kn}^{kn}\dd s \int_{-kn}^{kn}\dd t \,\bra (1-\mathbb P)U_s\la a, (1-\mathbb P)U_t\la a\ket_{\hi H}}
\eeq
and by using \eqref{lowernorm}, we obtain
\beq\label{bounhp}
	|Q'(\la a)| \leq \sqrt{\frc{\gamma}{2k}}
	||\la a||_{\whi H}^-.
\eeq
If $\la a\in\whi V$ then we can use $||\la a||_{\whi H}^- = ||\la a||_{\whi H}$ the norm on $\whi H$. With \eqref{linhil} we find
\beq\label{linhilbd}
	||\la b||_{\whi H}\leq \sqrt{\frc{\gamma}{2k}}.
\eeq

Simple examples of such extensive charges are the (equivalence classes of the) sequences of the form
\beq\label{simpleext}
	Q_n = \int_{-n/2}^{n/2}\dd s\, U_s \la b
\eeq
for $\la b\in\whi V$, where one can take any $0<k<1/2$. In this simple case, the limit action is $Q'(\la a) = \bra\la b,\la a\ket_{\whi H}$, and the norm is exactly $||\la b||_{\whi H} = \sqrt{\gamma}$.

In \cite{Doyon2017}, this construction was used with $U=\iota$ space translations. It was also shown that there is in fact a {\em bijection} between the set of linearly extensive charges, and the Hilbert space $\whi H$. The proof can be applied to the general construction presented here as well. Therefore
\beq
	\mbox{linearly extensive charges for $\hi H$}
	\ \cong\ 
	\whi H.\label{chargeinclH}
\eeq
This extends the second equation of \eqref{ipiota} to arbitrary linearly extensive charges $Q'$ instead of the particular $A=\int \dd x\,\la a(x)$ constructed from elements $\la a\in\whi V$. This establishes that the correct completion of the space of such formal integrals of local densities is the Hilbert space $\whi H$. I will not need the isomorphism \eqref{chargeinclH} in what follows.

\subsection{Quadratically extensive charges}\label{ssectquadext}

Now concentrate on the specific setup of Subsections \ref{ssectbaldif} and \ref{ssectkubo}. There are (at least) two one-parameter groups, space translations $\iota_x$ and time translations $\tau_t$. As the construction in Subsection \ref{ssectext} is completely general, we can apply it for $U=\iota$ with $\mathbb P = \mathbb P_\iota$, and this is how we will understand Points 1,2,3. But also we can go further and construct the linearly extensive charges for $\whi H$. We consider a sequence $Q = (Q_n\in\whi H:n\in\N)$, and assume without loss of generality that $Q_n = (1-\mathbb P_\tau)Q_n$ for all $n$. We suppose there is a subspace $\wwhi D(Q)\subset \whi H$, the domain of $Q$, with $\wwhi V\subset \wwhi D(Q)$, such that the sequence satisfies the following conditions:
\bi
\item[$1'$.] There exists $\gamma>0$ such that $||Q_n||_{\whi H}^2 \leq \gamma n$ for all $n\in\N$ large enough.
\item[$2'$.] The limit $Q''(\la a): = \lim_{n\to\infty} \bra Q_n,\la a\ket_{\whi H}$ exists for all $\la a\in\wwhi D(Q)$.
\item[$3'$.] There exists $k>0$ such that $\lim_{n\to\infty} \sup_{s,t\in[-kn,kn]} |\bra Q_n,\tau_s \la a-\tau_t\la a\ket_{\whi H}| = 0$ for all $\la a\in\wwhi D(Q)$.
\ei
By the general result of Subsection \ref{ssectext}, every such charge $Q''$ is bounded on its domain by the lower norm for $\whi H$, and gives rise, when acting on $\wwhi V$, to an element of $\wwhi H$. In fact, the space of these is isomorphic to $\wwhi H$ (which I again state without proof as this will not be used). I will refer to these as the {\em quadratically extensive conserved charges for $\hi H$}:
\beqa
	&&\mbox{quadratically extensive conserved charges for $\hi H$} \n &=&
	\mbox{linearly extensive charges for $\whi H$}\n
	&\cong& 
	\wwhi H.\label{chargeincl}
\eeqa
For every such charge, there exists $\la b\in \wwhi H$ such that
\beq\label{linhil2}
	Q''(\la a) = \bra \la b,\la a\ket_{\wwhi H}
\eeq
for all $\la a\in\wwhi V$. Note that these are ``conserved" by construction, as in $\wwhi H$, the group $\tau_t$ acts trivially. Naturally, sequences of the form $Q_n = \int_{-n/2}^{n/2} \dd t\,\tau_t \la b$ for $\la b\in\wwhi V$ give rise to linearly extensive charges for $\whi H$. Most interestingly, we obtain a bound on the $\wwhi H$-lower norm -- or Onsager matrix element -- of elements of $\wwhi D(Q)$ on which $Q''$ is nonzero:
\begin{theorem} \label{theoquad1} Let $Q''$ be a quadratically extensive charge for $\hi H$, and $\la a\in\wwhi D(Q)$. Then
\beq\label{bounhpp}
	\mathfrak L_{\la a,\la a}^- \geq \frc{2k}{\gamma} |Q''(\la a)|^2.
\eeq
If $\la j_i\in\wwhi D(Q)$ and $Q''(\la j_i)\neq 0$, then the associated lower Onsager matrix element $\mathcal L_{ii}^-$ is strictly positive (and may be infinite).
\end{theorem}
\proof This follows immediately from the above discussion and \eqref{bounhp}. \eproof

Why should linearly extensive charges for $\whi H$ be seen as quadratically extensive in $\hi H$? Intuitively, extensivity for $\whi H$ requires, from the viewpoint of $\hi H$, an integral on an interval of length $n$, whose position is integrated over the full space. As $n$ goes to infinity, we have a double integral over space. This will be made clearer in Section \ref{sectwave}, where explicit examples are discussed. In particular, there, it is seen that, with $Q_n$ supported on regions of length $n$, and if a Lieb-Robinson bound is available,
\beq\label{kquadbd}
	0<k < \frc1{2v_{\rm LR}}\quad\mbox{($Q_n$ on length $n$ in space, Lieb-Robinson bound).}
\eeq

Another reason for the nomenclature lies in a construction by Prosen \cite{Prosenquad}, which I show in Appendix \ref{sectprosen} leads to elements of $\wwhi H$. This construction appears to be slightly more restrictive than the  general definition of quadratically extensive conserved charges that I propose in \eqref{chargeincl}, because it requires an apparently stronger clustering property, Eq.~\eqref{uniclus}. It is nevertheless useful in practice.

\subsection{Fractionally extensive charges}\label{fracext}

There is a natural way of modifying the construction of linearly extensive charges of Subsection \ref{ssectext} in order to connect them to the fractional-order spaces of Subsection \ref{ssectfractorder}. I define $u$-extensive charges for $\hi H$ by assuming Point $1^{(u)}=1$, and modifying Points 2 (for the notation) and 3 (more fundamentally) as follows, for $u\in(0,1)$:
\bi
	\item[$2^{(u)}$.] The limit $ Q^{(u)}(\la a):= \lim_{n\to\infty} \bra Q_n,\la a\ket_{\hi H} $ exists for all $\la a\in\hi D^{(u)}(Q)$.
	\item[$3^{(u)}$.] There exists $k>0$ such that $\lim_{n\to\infty} \sup_{x,y\in[-kn^{1/u},kn^{1/u}]} |\bra Q_n,\iota_x \la a-\iota_y\la a\ket_{\hi H}| = 0$ for all $\la a\in D^{(u)}(Q)$.
\ei
We can bound the resulting charge by using
\beq\label{Qnubounding}
	\bra Q_n,\la a\ket_{\hi H} = \frc1{2kn^{1/u}} \int_{-kn^{1/u}}^{kn^{1/u}}\dd x\, \bra Q_n,(1-\mathbb P_\iota)\iota_x \la a\ket_{\hi H} + \delta_n
\eeq
and by bounding the first term as
\beq
	\leq \sqrt{\frc{|| Q_n||^2_{\hi H}}{4k^2n}}
	\sqrt{\frc1 {n^{2/u-1}}\int_{-kn^{1/u}}^{kn^{1/u}}\dd x \int_{-kn^{1/u}}^{kn^{1/u}}\dd y \,\bra (1-\mathbb P_\iota)\tau_x\la a, (1-\mathbb P_\iota)\tau_y\la a)_{\hi H}}.
\eeq
Replacing $n$ by $kn^{1/u}$ in \eqref{lowernormHu}, we obtain
\beq
	|Q^{(u)}(\la a)|\leq
	\frc{(2k)^{-u/2}\sqrt\gamma }{\sqrt{2-u}}||\la a||_{\hi H^{(u)}}^-.
\eeq
Thus, there exists $\la b\in \hi H^{(u)}$ such that, for all $\la a\in\hi V^{(u)}$,
\beq
	Q_{\hi H^{(u)}}(\la a) = \bra \la b,\la a\ket_{\hi H^{(u)}},\quad
	||\la b||_{\hi H^{(u)}} \leq \frc{(2k)^{-u/2}\sqrt\gamma }{\sqrt{2-u}}.
\eeq

Again one can make a similar construction one level up: I define, for $u\in(1,2)$, the $u$-extensive charges for $\hi H$ as the $(u-1)$-extensive charges for $\whi H$. That is, Point $1^{(u)} = 1'$ holds, and, for $u\in(1,2)$,
\bi
	\item[$2^{(u)}$.] The limit $ Q^{(u)}(\la a):= \lim_{n\to\infty} \bra Q_n,\la a\ket_{\whi H} $ exists for all $\la a\in\hi D^{(u)}(Q)$.
	\item[$3^{(u)}$.] There exists $k>0$ such that $\lim_{n\to\infty} \sup_{s,t\in[-kn^{1/(u-1)},kn^{1/(u-1)}]} |\bra Q_n,\tau_s \la a-\tau_t\la a\ket_{\whi H}| = 0$ for all $\la a\in\hi D^{(u)}(Q)$.
\ei
We obtain
\beq\label{Qhbdfrac}
	|Q^{(u)}(\la a)|\leq
	\frc{(2k)^{-(u-1)/2}\sqrt\gamma }{\sqrt{3-u}}||\la a||_{\hi H^{(u)}}^-
\eeq
and the following:
\begin{theorem} \label{theoboundhu} Let $Q^{(u)}$ be a $u$-extensive charge for $\hi H$ with $u\in (1,2)$, and $\la a\in\hi D^{(u)}(Q)$. Then
\beq\label{bounhu}
	||\la a||_{\hi H^{(u)}}^- \geq 
	\frc{(2k)^{(u-1)/2}\sqrt{3-u}}{\sqrt\gamma }\,
	|Q^{(u)}(\la a)|.
\eeq
If $\la j_I\in\hi D^{(u)}(Q)$ and $Q^{(u)}(\la j_I)\neq 0$, and if $\la j_I$ has normal (resp.~fat) superdiffusion exponent $\alpha$, Eq.~\eqref{exponent} (resp.~Eq.~\eqref{exponentfat}), then this exponent is bounded as
\beq
	\alpha \geq \frc{3-u}2\quad \Big(\mbox{resp.~}\alpha\geq \frc1u\Big).
\eeq
\end{theorem}

\section{Bilinear charges}\label{sectwave}

The construction of Section \ref{sectext}, and in particular the bounds obtained, are still relatively abstract. Concrete bounds are obtained by choosing appropriate extensive charges. In this section, I construct such charges and analyse their consequences. The strength of the bound obtained depends on the assumption we can make on clustering properties of multi-point correlation functions. I will make in turn various assumptions increasing in strength, obtaining stronger results: basic clustering properties which follow from Lieb-Robinson bounds, stronger properties based on hydrodynamic velocities obtained by a linear response argument, and a still stronger clustering assumption from the superdiffusive phenomenology. The latter will give lower bounds on the superdiffusive exponents, reproducing that of the KPZ universality class and other exponents found in NFH. Under the clustering assumptions stated, and within the appropriate $C^*$-algebraic framework, the results are rigorous; however the clustering assumptions based on hydrodynamic velocities seem to be rather hard to show rigorously.

For simplicity, here and in the rest of the paper, I assume that $q_i^\dag = q_i$, and that the state $\bra\cdots\ket$ is invariant under the flows generated by all conserved charges $Q_i$; these assumptions are however not necessary for similar results to hold.

\subsection{Bilinear charges}\label{ssectbilinear}

We are looking for a natural space of quadratically or fractionally extensive conserved charges for $\hi H$, as per the general definitions  Points $1',\,2',\,3'$, or Points $1^{(u)},\,2^{(u)},\,3^{(u)}$ with $u\in(1,2)$. Quadratically extensive charges for $\hi H$ are linearly extensive for $\whi H$, and with linear extensivity in time, the sequences of time integrals $\int_{-n}^n \dd t\,\tau_t \la a$ for $\la a\in\hi V$ (as in \eqref{simpleext}), are examples. However these do not immediately lead to new results. Instead, we use the general construction and consider charges which are linearly extensive for $\whi H$ because of a linear growth in {\em space}. Certainly, because of Point $3'$, they still must be invariant under time translation, hence these must be conserved charges.

A natural guess are the bilinear expressions in the total conserved charges, formally $Q_i Q_j$. This closely resembles the quadratic expansion of the current made in \cite{MedenjakDiffusion2019} following the philosophy of NFH, and I will establish the full connection in Section \ref{sectadd}. Defining precise objects in $\whi H$, let me then consider the sequence in $n$, parametrised by $i,j$, defined as
\beq\label{Qijn}
	Q_{ij;n} = (1-\mathbb P_\tau) \Big[\int_{-n}^n\dd x\, \la q_i(x) \la q_j\Big]_{\whi H}
\eeq
where $\la q_i(x) = \iota_x\la q_i$ is the space-translate, and $\mathbb P_\tau$ is the projector onto the space of conserved densities, Eq.~\eqref{proj}. I show below that these indeed result in quadratically and fractionally extensive charges. They will be referred to as ``bilinear charges", as they are formed of bilinear expressions in linearly extensive charges for $\hi H$.

In this section I assume, without loss of generality, that $\la q_i$'s have zero expectation values $\bra \la q_i\ket=0$.

\begin{rema}
In definition \eqref{Qijn}, the $\la q_i$'s are seen as elements of $\hi V$, the set of local observables, on which there is an algebra. The result of the product $\la q_i(x)\la q_j = \la q_i(x)\la q_j(0)$ is then seen as an element of $\whi H$ (that is, its equivalence class is taken), and then the projection $1-\mathbb P_\tau$ is applied on its integral. It will be simple to see that the resulting charge $Q_{ij}$ from \eqref{Qijn} does not depend on the choice of element $\la q_i$ of the $\whi H$-equivalence class. Therefore, $Q_{ij}$ is indeed a bilinear in the charges $Q_i$, $Q_j$. The constant-shift ambiguity is immediately irrelevant: since $\mathbb P_\tau \la q_i = \la q_i$, we can replace $\la q_{i}$ by $\la q_{i} - {\bf 1}\bra \la q_{i}\ket$ in \eqref{Qijn}. For the gauge transformation, this follows from the clustering assumption \eqref{clustering} below. See Remark \ref{remaqi} for the assumptions made on the $\la q_i$'s.
\end{rema}

\begin{rema}
Below, the projection formula \eqref{proj} is used, and assumed to converge, on local observables $\la a \in\hi V$ inside correlation functions. Further, the clustering condition \eqref{clustering} is assumed to hold uniformly enough on a basis of conserved densities $\la q_i$. For simplicity, the flux Jacobian is also assumed to have discrete spectrum, and sums over normal modes are assumed to converge. These assumptions would be difficult to show rigorously, of course, only if $\hi H_\bal$ is infinite-dimensional.
\end{rema}

\subsection{Diffusion from Lieb-Robinson-type correlations}\label{ssectconv}

In this subsection, I obtain bounds on diagonal Onsager matrix elements using the charges \eqref{Qijn}, by assuming a clustering condition on correlation functions. The idea behind the clustering assumption is the Lieb-Robinson bound, but the assumption is weak enough in order for the theorem to be as widely applicable as possible.

I assume the existence of $v>0$ such that, for all $\la a_i\in\hi V$, connected correlation functions $\bra \la a_1(x_1,t_1)\la a_2(x_2,t_2) \cdots\ket^{\rm c}$ decay fast enough whenever there is some $i$ such that $|x_i-x_j|- v|t_i-t_j|\gg 0$ for all $j$. For the proofs below, it is sufficient to restrict ourselves to connected correlation functions with up to four observables. More precisely, given $\la a_i\in\hi V$ for $i=1,\ldots,r$, there exists a bounded function $f(z)>0$ ($z\in\R$), and real numbers $b>0,\,c>0$ and $d> r-1$, such that
\beq\label{clustering}\begin{aligned}
	|\bra \la a_1(x_1,t_1)\cdots \la a_r(x_r,t_r)\ket^{\rm c}|
	\leq f\Big(\max_i \big\{\min_j\{z_{ij}\}\big\}\Big)\\ 
	z_{ij}= |x_i-x_j|-v|t_i-t_j|,\quad f(z)= c(b+z)^{-d}\ \forall\ z>0.\end{aligned}
\eeq
Here and below, I use again the notation $\la a(x) = \iota_x\la a$ and $\la a(x,t) = \iota_x\tau_t \la a$.

This assumption naturally encodes the uniform boundedness of connected correlation functions via the introduction of the shift $b>0$, and their decay whenever any one of the observable is far enough in space. The decay occurs as soon as the maximal ``extent" of the time-evolved observable, as determined by the velocity $v$, does not overlap with that of any of the other observables. If a Lieb-Robinson bound exists, with $v_{\rm LR}$ the Lieb-Robinson velocity, then it is expected that we can take any $v=v_{\rm LR}$ and $d$ arbitrary large; in fact, $f(z)$ may be chosen to decay exponentially at large $z$. This is rigorously proven in the case $r=2$ using the Lieb-Robinson bound for states which cluster in space exponentially \cite{LiebRobinson,BHC06}, and general results exist for algebraically clustering states \cite{Doyon2017}; the proof techniques are generalisable to $r>2$. Note that \eqref{req}, for $U_s$ being space translations, is exactly the special, non-dynamical case $r=2$ and $t_1=t_2$ of \eqref{clustering}, hence we may take
\beq
	\whi V = \hi V .
\eeq
To be precise, below, we only need the non-dynamical case of \eqref{clustering} (with equal times) for $r=2,4$, and the dynamical case with $t_1= t_2\neq t_3$ for $r=3$. In the latter case, the bound is represented by Fig.~\ref{correlation}, with the linear (ballistic) scaling $g(t)\propto t$, characteristic of the Lieb-Robinson bound.

It is simple to show that \eqref{clustering} implies that the equal-time, space-integrated $r$-point correlation functions are finite. For instance, the following integral on $\R^2$ exists and is finite, the condition $d>2$ being sufficient:
\beqa
	\bra A_1 A_2 \la a_3\ket^{\rm c} &=& \int \dd x_1\dd x_2\,
	\bra \la a_1(x_1)\la a_2(x_2) \la a_3(x_3)\ket^{\rm c} \n
	&=& \int_{z_1>z_2>0} \dd z_1\dd z_2\,\sum_{\sigma\in S_3}
	\bra :\hspace{-0.1cm}\la a_{\sigma_1}(z_1)\la a_{\sigma_2}(z_2) \la a_{\sigma_3}(0)\hspace{-0.1cm}:\ket^{\rm c}
\eeqa
where the ``normal ordering" $:\hspace{-0.1cm}\cdots\hspace{-0.1cm}:$ places the observables in the order $\la a_1(\cdot) \la a_2(\cdot) \la a_3(\cdot)$.  Using \eqref{clustering},
\beqa
	\bra A_1 A_2 \la a_3\ket^{\rm c} &\leq & 6c\int_{z_1>z_2>0} \dd z_1\dd z_2\,\big[b+\max\{z_2,|z_1-z_2|\}\big]^{-d}\n
	&= & 12c\int_{z>w>0} \dd z\dd w\,(b+z)^{-d}\n
	&\leq & 12c\int_{z>0} \dd z\,(b+z)^{1-d}\n
	&=& \frc{12cb^{2-d}}{d-2} < \infty\quad \mbox{for $d>2$}.
\eeqa

Using Theorem \ref{theoquad1}, I will show the following:
\begin{theorem} \label{theoquad2} Let $\la a\in \hi V$. Under the assumption \eqref{clustering}, we have
\beq\label{bounhppquad}
	\mathfrak L_{\la a,\la a}^- \geq \frc{\Big|
	\bra Q_i Q_j  (1-\mathbb P_\tau)\la a\ket^{\rm c} \Big|^2}{2v\,\Big(|| \la q_i||_{\whi H}^2 || \la q_j||_{\whi H}^2
	+\bra \la q_i,\la q_j\ket_{\whi H}^2\Big)}
\eeq
for any $i,j$. In particular, if
\beq\label{QiQjjk}
	\bra Q_i Q_j  \la j_k\ket^{\rm c} - \sum_{lm} \bra Q_i Q_j  \la q_l\ket^{\rm c}\mathsf C^{lm}
	\bra Q_m  \la j_k\ket^{\rm c}\neq 0
\eeq
for some $i,j$, then the lower Onsager matrix element $\mathfrak L_{kk}^-$ is strictly positive (and may be infinite).
\end{theorem}
All quantities in Theorem \ref{theoquad2} can be evaluated purely from the thermodynamics of the system. This implies nonzero diffusion whenever \eqref{QiQjjk} holds.

Consider for instance a conventional Galilean gas in a Gibbs state at rest. There are three conserved charges: the total number of particles $Q_0$, the total momentum $Q_1$ and the total energy $Q_2$, and the weight of a configuration takes the form $\exp\big[-\beta (Q_2-\mu Q_0)\big]$. Diffusion, characterised by the Onsager matrix element $\mathfrak L_{11}$, may be bounded by taking $k = 1$ ($\la j_k$ being the momentum current, that is, the pressure) and $i=0$, $j=0$ (the quadratic charge corresponding to the square of the total number of particles). Generically, the inequality \eqref{QiQjjk} holds as basic symmetries do not forbid it, hence there is a nontrivial lower bound. The bound \eqref{bounhppquad} can in fact be written in terms of conventional thermodynamic quantities. For simplicity, let us assume that all average quantities (densities of particles and energy, and pressure) vary very little with the inverse temperature $\beta$. Then we can concentrate solely on $Q_0$ and $Q_1$, keeping $\beta$ fixed, and $\mathsf C$ is a 2 by 2 matrix. By parity symmetry, the only element that contributes to the sum in \eqref{QiQjjk} is $\mathsf C^{00}$. Let $P = \bra \la j_1\ket$ be the pressure,  $\rho = \bra \la q_0\ket$ the particle density, and $\chi = \beta^{-1}\p\rho/\p\mu$ the susceptibility.  Note that $\dd P/\dd\rho = v_{\rm s}^2$ is the square of the sound speed. Then
\beq\label{L11bound}
	\mathfrak L_{1, 1}^- \geq \frc{\big(\chi v_{\rm s} \p_\rho v_{\rm s}\big)^2}{v}.
\eeq
Recall that $v$ is the velocity controlling the decay of correlation functions, which can be taken as the Lieb-Robinson velocity. The Onsager matrix element $\mathfrak L_{11}$ is simply related to the bulk viscosity $\zeta$ of the gas, as
\beq
	\zeta = \frc{\rho \mathfrak L_{1,1}}{2\chi v_{\rm s}^2}.
\eeq
Hence this gives the explicit positive lower bound on the viscosity \eqref{viscosity}, solely derived from clustering properties.

By nonlinear fluctuating hydrodynamics, diffusion is in fact expected to be infinite in many non-integrable systems, as there is superdiffusion; hence the above inequality is not expected to be saturated. Bounds on superdiffusion are discussed in Subsections \ref{ssectlinresp} and \ref{ssectsuper}.

\proof I show that the sequence $Q_{ij} = (Q_{ij;n}:n\in\N)$ satisfies Points $1'$, $2'$ and $3'$ (Subsection \ref{ssectquadext}) of linearly extensive charges for $\whi H$, giving the charge $Q_{ij}''$ with $\hi V\subset \wwhi D(Q_{ij})$.

Point $1'$ is the requirement that $||Q_{ij;n}||_{\whi H}^2\leq\gamma n$. We have, explicitly subtracting the projection onto conserved densities,
\beq\label{Qsq}
	||Q_{ij;n}||_{\whi H}^2 = \int_{-n}^n \dd x \int _{-n}^n\dd y\,
	\Big[\bra \la q_i(x)\la q_j ,\la q_i(y)\la q_j\ket_{\whi H}
	- \sum_K \bra \la q_i(x)\la q_j ,\la q_K\ket_{\whi H}
	\bra \la q_K, \la q_i(y)\la q_j\ket_{\whi H}\Big]
\eeq
where I used the diagonal basis $\la q_K$ described in Appendix \ref{diagonal}. The second term in the square brackets on the right-hand side is finite under integration as $n\to\infty$ by three-point function clustering \eqref{clustering}, and $\int \dd x\, \bra \la q_i(x)\la q_j ,\la q_K\ket_{\whi H} = \bra Q_iQ_j\cdot\la q_K\ket^{\rm c}$. Recall that we assume $\bra \la q_i\ket=0$, thus the connected three-point functions appear. The first term in \eqref{Qsq} is
\beq
	\int_{-n}^n \dd x \int_{-n}^n \dd y \int \dd z\,\Big[
	\bra \la q_i(z+x)\la q_j(z) \cdot \la q_i(0) \la q_j(y)\ket - 
	\bra \la q_i(z+x)\la q_j(z)\ket\bra \la q_i(0) \la q_j(y)\ket
	\Big].
\eeq
As the same argument applies to both terms of the symmetric product $\la q_i(z+x)\la q_j(z) \cdot \la q_i(0) \la q_j(y) = \frc12 (\la q_i(z+x)\la q_j(z) \la q_i(0) \la q_j(y) +  \la q_i(0) \la q_j(y) \la q_i(z+x)\la q_j(z))$, let me just consider the first term. This can be written as
\beqa\label{linearpp}
	\lefteqn{\int_{-n}^n \dd x \int_{-n}^n \dd y \int \dd z\,\Big[
	\bra \la q_i(z+x)\la q_j(z) \la q_i(0) \la q_j(y)\ket - 
	\bra \la q_i(z+x)\la q_j(z)\ket\bra \la q_i(0) \la q_j(y)\ket
	\Big]} && \n
	&& =
	\int_{-n}^n \dd x \int_{-n}^n \dd y \int \dd z\,\Big[
	\bra \la q_i(z+x)\la q_j(z) \la q_i(0) \la q_j(y)\ket^{\rm c} +\n && \qquad +\ 
	\bra \la q_i(z+x)\la q_i(0)\ket^{\rm c}\bra \la q_j(z) \la q_j(y)\ket^{\rm c} +
	\bra \la q_i(z+x)\la q_j(y)\ket^{\rm c}\bra \la q_j(z) \la q_i(0) \ket^{\rm c}
	\Big].
\eeqa

In \eqref{linearpp}, the first term in the square brackets on the right-hand side gives under integration a finite value as $n\to\infty$ by clustering of the four-point function.

For the second term, a change of variable gives
\begin{multline}
	\int_{-n}^n \dd x \int_{-n}^n \dd y \int \dd z\,\bra \la q_i(z+x)\la q_i(0)\ket^{\rm c}\bra \la q_j(z) \la q_j(y)\ket^{\rm c}
	=\\
	\int \dd x \dd y\,(2n-|x-y|)\Theta(2n-|x-y|)\bra \la q_i(x)\la q_i(0)\ket^{\rm c}\bra \la q_j(y) \la q_j(0)\ket^{\rm c}\label{boundingproduct}
\end{multline}
where $\Theta(\cdots)$ is the step function. By convergence of the integral of $\bra \la q_i(x)\la q_i(0)\ket^{\rm c}\bra \la q_j(y) \la q_j(0)\ket^{\rm c}$ over $(x,y)\in\R^2$, we have
\beq
	\lim_{n\to\infty}\int \dd x \dd y \,\Theta(2n-|x-y|)\bra \la q_i(x)\la q_i(0)\ket^{\rm c}\bra \la q_j(y) \la q_j(0)\ket^{\rm c}
	= ||\la q_i||^2_{\whi H} ||\la q_j||^2_{\whi H}.
\eeq
Therefore, on the right-hand side of \eqref{boundingproduct}, the term in the parenthesis proportional to $2n$ gives the contribution
\beq\label{part2n}
	2n||\la q_i||^2_{\whi H} ||\la q_j||^2_{\whi H}.
\eeq
On the other hand, we bound, according to \eqref{clustering} (the case $r=2$, with $d>1$)
\begin{multline}
	\Big|\int \dd x \dd y\,|x-y|\Theta(2n-|x-y|)\bra \la q_i(x)\la q_i(0)\ket^{\rm c}\bra \la q_j(y) \la q_j(0)\ket^{\rm c}\Big|\leq\\
	c^2\int \dd x \dd y\,|x-y|\Theta(2n-|x-y|)(b+|x|)^{-d} (b+|y|)^{-d}.\label{intermsas}
\end{multline}
We may estimate, say for $z>0$,
\beqa
	\lefteqn{\int \dd x\,(b+|x|)^{-d} (b+|z+x|)^{-d}} && \n &=& 
	\Bigg[\int_{-\infty}^{-z/2} + \int_{-z/2}^\infty\Bigg] \dd x\,(b+|x|)^{-d} (b+|z+x|)^{-d} \n
	&\leq&
	\int_{-\infty}^{-z/2} \dd x\,(b+z/2)^{-d} (b+|z+x|)^{-d} 
	+ \int_{-z/2}^\infty\dd x\, (b+|x|)^{-d} (b+z/2)^{-d} \n
	&=& \frc{4b^{1-d}(b+z/2)^{-d}-2(b+z/2)^{1-2d}}{d-1} \leq
	\frc{4b^{1-d}(b+z/2)^{-d}}{d-1} .
\eeqa
Therefore, the right-hand side of \eqref{intermsas} is bounded as
\beq
	\leq c^2 \int_{-2n}^{2n} \dd z\,|z|\frc{4b^{1-d}(b+|z|/2)^{-d}}{d-1}.
\eeq
This converges as $n\to\infty$ if $d>2$, and diverges proportionally to $n^{2-d}$ is $d<2$. Since $d>1$, the divergence as $n^{2-d}$ is sub-dominant compared to $n$, hence in all cases the part \eqref{part2n} dominates and we have
\beq
	\lim_{n\to\infty} (2n)^{-1} \int_{-n}^n \dd x \int_{-n}^n \dd y \int \dd z\,\bra \la q_i(z+x)\la q_i(0)\ket^{\rm c}\bra \la q_j(z) \la q_j(y)\ket^{\rm c} = ||\la q_i||^2_{\whi H} ||\la q_j||^2_{\whi H}.
\eeq

The third term in the square brackets on the right-hand side in \eqref{linearpp} has the same structure as the second term by a simple change of the $z$ variable. Hence the same derivation can be applied. In the quantum case, we need to be careful with the ordering of the observables in order to identify the integrated two-point correlation functions with inner products in $\whi H$.  By our assumption that the state is invariant under the actions of $Q_i$, we have $\bra Q_i \la q_j(0)\ket^{\rm c} = \bra \la q_j(0) Q_i \ket^{\rm c}$, so that the ordering does not matter. This could also have been obtained by the assumption that all densities $\la q_i$'s are conserved under all flows generated by $Q_\ell$'s, with ${\rm i}[Q_\ell, \la q_i(x)] + \p_x \la j_i^{(\ell)}(x)=0$.

Overall, we obtain
\beq\label{scalingqij}
	\lim_{n\to\infty}\frc{||Q_{ij;n}||_{\whi H}^2}{2n} =|| \la q_i||_{\whi H}^2 || \la q_j||_{\whi H}^2
	+\bra \la q_i,\la q_j\ket_{\whi H}^2.
\eeq
This is linear extensivity for $\whi H$, where we may take any 
\beq\label{gammaquad}
	\gamma > 2\Big(|| \la q_i||_{\whi H}^2 || \la q_j||_{\whi H}^2
	+\bra \la q_i,\la q_j\ket_{\whi H}^2\Big).
\eeq

As for Point $2'$, we must show that $\lim_{n\to\infty} \bra Q_{ij;n},\la a\ket_{\whi H}$ exists for all $\la a\in\hi V$. Using 3-point clustering, this is immediate,
\beqa
	Q''_{ij}(\la a) = \lim_{n\to\infty} \bra Q_{ij;n},\la a\ket_{\whi H}
	&=&\int \dd x\, \bra\la q_i(x) \la q_j,(1-\mathbb P_\tau)\la a\ket_{\whi H} \n
	&=& \bra Q_i Q_j \cdot (1-\mathbb P_\tau)\la a\ket^{\rm c}.\n
	&=& \bra Q_i Q_j (1-\mathbb P_\tau)\la a\ket^{\rm c}.\label{Qij}
\eeqa
where in the last line, I use the invariance of the state under the actions of $Q_i$ and $Q_j$. Note that this also implies that $Q_{ij}''$ is invariant under all higher flows $\tau_s^{(\ell)}$ with $\ell>2$ (if any), as follows from the associated conservation laws and three-point clustering:
\beq\label{invhigher}
	Q_{ij}''\big(\tau_s^{(\ell)}\la a\big) = Q_{ij}''(\la a).
\eeq

Finally, for Point $3'$, we use the fact that $Q_{ij;n}$ is itself a ``conserved quantity", in the following sense: From the conservation laws \eqref{conslaw}, and the fact that $\p_t \tau_t \mathbb P_\tau\la a = \p_t  \mathbb P_\tau\la a=0$ for any $\la a$, we find
\beq\label{dtinvr}
	\p_t\bra\tau_tQ_{ij;n}, \la a\ket_{\whi H}\big|_{t=0} 
	= 
	\bra \la j_i(-n)\la q_j -  \la j_i(n)\la q_j
	- \la q_i(-n) \la j_j
	+\la q_i(n) \la j_j,\la a\ket_{\whi H}.
\eeq
Therefore, by time-translation invariance of the inner product, and integrating,
\beq\label{tinvr}
	\bra Q_{ij;n}, \tau_{s_1}\la a - \tau_{s_2}\la a)\ket_{\whi H}
	=
	\int_{s_1}^{s_2} \dd t\,\bra\la j_i(-n)\la q_j -  \la j_i(n)\la q_j
	- \la q_i(-n) \la j_j
	+\la q_i(n) \la j_j,\tau_t \la a\,\ket_{\whi H}.
\eeq
I now show that if both $s_1,s_2$ lie in the open interval $(-n/(2v),n/(2v))$, then this vanishes as $n\to\infty$, giving Point $3'$ with any $0<k< 1/(2v)$. Indeed, intuitively, in this case the observable $\tau_t \la a$ in \eqref{tinvr} ``covers" a region of length strictly less than $n$, and in a connected three-point function with, for instance, $\la j_i(x+ n)\la q_j(x)$ (both at time 0), for every $x$, at least one of $\la j_i(x+ n)$ or $\la q_j(x)$ lies outside of this interval\footnote{To be precise, the boundary $1/(2v)$ can also be included, as under integration its nonvanishing contribution has zero measure; but this does not affect the final result.}.

More precisely, we can use the condition \eqref{clustering} for 3-point functions in order to show that each term in the inner product on the right-hand side of \eqref{tinvr} gives a vanishing contribution as $n\to\infty$. Consider a term of the form $\int_{s_1}^{s_2} \dd t\,\bra \la b(n) \la c , \tau_t\la a\ket_{\whi H}$ with $\bra \la b\ket = \bra \la c\ket =0$. This is bounded as
\beqa
	\Big|\int_{s_1}^{s_2} \dd t\,\bra \la b(n) \la c , \tau_t\la a\ket_{\whi H}\Big|&\leq&
	\int_{s_1}^{s_2} \dd t\,\int \dd x\,
	|\bra \la a(x,t)\cdot\la b(n,0)  \la c(0,0) \ket^{\rm c}|\n
	&=&
	n^2 \int_{-k}^{k} \dd t\int \dd x\,
	|\bra \la a(nx,nt)\cdot\la b(n,0) \la c(0,0) \ket^{\rm c}|\label{proofquad2}
\eeqa
where it is sufficient to consider $s_1 = -nk$, $s_2=nk$ with $k>0$. For now I am not imposing any additional condition on $k$. Using \eqref{clustering} for the 3-point function on the last line, three values are to be compared: $z_{12} = n$, $z_{13} = n\big(|1-x|-v|t|\big)$, $z_{23} = n\big(|x|-v|t|\big)$, with $|t|<k$. I analyse the double integral in \eqref{proofquad2} by looking at two distinct regions:
\beq\label{R1R2}
	R_1 = \{(x,t):|x-1/2|>3/2+kv, |t|<k\}, \quad R_2 = \{(x,t):|x-1/2|\leq 3/2+kv,|t|\leq k'\}
\eeq
for some $k'<1/(2v)$. Doing it this way is useful for the proofs of other theorems below. For these to cover the full integration region, the condition $k\leq k'$ is then imposed.

On the one hand, in $R_1$, the quantity $z=\min\{z_{13},z_{23}\}$ is the maximal value, and is positive. Thus we can use the bound $f(z) = c(b+z)^{-d}$, and we bound the integral over this region on the right-hand side of \eqref{proofquad2} as
\begin{multline}
	cn^2 \int_{-k}^{k} \dd t\int_{|x-1/2|>3/2+kv} \dd x\,
	\big[b+n\,(\min\{|1-x|,\,|x|\}-v|t|)\big]^{-d}\\
	=
	\frc{c}{(d-1)(d-2)v}\big[(b+n(1+2kv) )^{2-d}
	- (b+n )^{2-d}]
	\label{derideri}
\end{multline}
where the $x$ integral is convergent as $d>1$. Since in fact $d>2$, this result vanishes in the limit $n\to\infty$.

On the other hand, in $R_2$, the lower bound of the minimum of $|1-x|-v|t|$ and $|x|-v|t|$ is positive: the former is positively lower bounded on $x\leq 1/2$, the latter on $x\geq 1/2$. As $z_{12}$ is positive, we find that $z=\max_i \big\{\min_j\{z_{ij}\}\big\}$ is positive, and we can again use the bound $f(z) = c(b+z)^{-d}$. The minimal value of $z$ for $|x-1/2|\leq 3/2+kv$ and $|t|\leq k'$ is obtained at $x=1/2$ and $t=\pm k'$ and given by $n(1/2-k'v)$. Taking this minimum value, we bound the integral over this region on the right-hand side of \eqref{proofquad2} as
\beq
	cn^2 \int_{-k'}^{k'} \dd t\int_{|x-1/2|\leq 3/2+kv} \dd x\,
	\big[b+n(1/2-k'v)\big]^{-d}
	=
	(3/2+kv)2kc n^2
	\big[b+n(1/2-k'v)\big]^{-d}.
\eeq
Since $d>2$, this result vanishes in the limit $n\to\infty$.

With $k\leq k'$, this cover the full integration region in \eqref{proofquad2}, and we have shown that it vanishes as $n\to\infty$. This therefore imposes $k\leq 1/(2v)$.

Therefore, the resulting map \eqref{Qij} is a quadratically extensive charge for $\hi H$, with $\hi V\subset\wwhi D(Q)$. Using \eqref{gammaquad} and Theorem \ref{theoquad1}, and optimising the lower bound, we have shown the theorem.
\eproof

\subsection{Diffusion from hydrodynamic correlations}\label{ssectlinresp}

The Lieb-Robinson-type bound \eqref{clustering} is rigorously established at least for $r=2$ in quantum chains, but it is not state dependent, and not optimal. In general, one expects clustering \eqref{clustering} to hold for velocities $v$ smaller than the Lieb-Robinson velocity $v_{\rm LR}$, up to an optimal value that is state-dependent; this value can be further optimised for certain observables. A natural guess is that $v$ in \eqref{clustering} and \eqref{bounhppquad} can be set to the largest hydrodynamic velocity (the largest propagation velocity of normal modes, see Appendix \ref{diagonal}). In fact, one can derive stronger clustering statements by applying simple hydrodynamic-type linear response arguments, of the type used in \cite{doyoncorrelations}. In this subsection, I propose a general clustering statement based on the results of such linear response arguments, from which I derive stronger bounds on diagonal Onsager matrix elements.

It is sufficient to concentrate on three-point functions. Hydrodynamic principles give results for Euler-scale correlation functions $\bra\cdots\ket^{\rm eul}$, where space and time coordinates have been sent to infinity in fixed proportions; for discussions of the Euler scaling limit, see  \cite{Spohn-book,doyoncorrelations}. In Appendix \ref{appproj}, I obtain two results concerning Euler-scale three-point functions of the type $\bra\,(1-\mathbb P_\tau)\la a(x,t) \la b(y,0) \la c (z,0)\,\ket^{\rm eul}$: (1) in such a three-point function, local observables $\la b$ and $\la c$ project onto conserved densities, Eq.~\eqref{first}; and (2) conserved densities propagate ballistically, Eq.~\eqref{second}. These statements are obtained using linear response arguments in the classical context, but are expected to hold in the quantum context as well.

The latter statement, expressed in terms of normal modes, takes the form
\beq\label{second2}
	\bra\,(1-\mathbb P_\tau)\la a(x,t) \cdot \la q_I(y,0) \la q_J (z,0)\,\ket^{\rm eul} =
	\bra Q_I Q_J(1-\mathbb P_\tau)\la a\ket^{\rm c} \,\delta(x-y-v^{\rm eff}_It)\delta (x-z-v^{\rm eff}_Jt).
\eeq
As the currents of normal modes project diagonally, $\mathbb P_\tau \la j_I = v^{\rm eff}_I \la q_I$, combining with the former statement, a similar formula holds with currents instead of densities, for instance
\beq\label{second2prime}
	\bra\,(1-\mathbb P_\tau)\la a(x,t) \cdot \la j_I(y,0) \la q_J (z,0)\,\ket^{\rm eul} =v_I^{\rm eff}
	\bra Q_I Q_J(1-\mathbb P_\tau)\la a\ket^{\rm c} \,\delta(x-y-v^{\rm eff}_It)\delta (x-z-v^{\rm eff}_Jt).
\eeq

In \eqref{second2} and \eqref{second2prime}, the delta-functions arise from the Euler scaling; microscopic three-point functions are expected to have support around these ballistic trajectories with power-law decay. The clustering assumption I propose makes this more precise. As I do not know of rigorous results in this direction for any specific class of models, I will make the assumption as general as possible, in accordance with the expected phenomenology and the above formulae.

Thus, in addition to \eqref{clustering}, I assume that, for all $\ep>0$, and for $\la b,\la c$ being densities or currents $\la q,\la j$, connected three-point functions $\bra (1-\mathbb P_\tau)\la a(x,t)\cdot \la b_{I_1}(y_1,0) \la c_{I_2}(y_2,0)\ket^{\rm c}$ decay fast enough whenever $|x-y_1 - v^{\rm eff}_{I_1}t|- \ep |t|\gg 0$, or $|x-y_2 - v^{\rm eff}_{I_2}t|- \ep |t|\gg 0$; that is, whenever $(x,t)$ lies away from either of the ballistic trajectories emanating from $(y_1,0)$ and $(y_2,0)$. More precisely, given $\la a\in\hi V$,  normal-mode indices $I_1,I_2$, and $\ep>0$, there exist a real number $d>2$, and a function $f(z,t)>0$ ($z>0$, $t\in\R$) that is $(-d)$-homogeneous, $f(nz,nt) = n^{-d}f(z,t)\;\forall\;n>0$, monotonically decreasing in $z$, $f(z,t)<f(z',t)$ if $z>z'$, and uniformly bounded on $\{(z,t):t\in T\}$ for any $z\neq 0$ and compact subset $T\not\ni \{0\}$, such that
\beq\label{clusteringhydro}
\begin{aligned}
	|\bra (1-\mathbb P_\tau)\la a(x,t)\cdot\la b_{I_1}(y_1,0) \la c_{I_2}(y_2,0)\ket^{\rm c}|
	\leq f(\max_i\{z_i\},t)  \\
	\mbox{if }\max_i\{z_i\}>0,\quad z_i = |x-y_i - v^{\rm eff}_{I_i}t|- \ep |t|.
	 \end{aligned}
\eeq
Again, the somewhat complicated specifications of the function $f(z,t)$ are for generality of the theorem. For instance, the functions $f(z,t) = z^{-d+\nu}|t|^{-\nu}$ for $\nu< d$ would do, as well as any linear combination thereof with positive coefficients. Pictorially, this condition can be represented by Fig.~\ref{correlation}, with a sub-linear long-time scaling $g(t)\ll t$ and conditions on some of the observables being normal modes (as above).

This gives rise to a strengthening of Theorem \ref{theoquad2}, in particular proving that, if certain conditions are met, there must be superdiffusion:
\begin{theorem} \label{theoquad3} Assume that \eqref{clustering} holds. Let $\la a\in \hi V$, and let $I,J$ be normal-mode indices, and assume that \eqref{clusteringhydro} hold for this $\la a$ and $I_1=I$, $I_2=J$. If $v_I^{\rm eff} \neq v_J^{\rm eff}$, then
\beq\label{bounhppquad2}
	\mathfrak L_{\la a,\la a}^- \geq \frc{\Big|
	\bra Q_I Q_J  (1-\mathbb P_\tau)\la a\ket^{\rm c} \Big|^2}{|v_I^{\rm eff}-v_J^{\rm eff}|}.
\eeq
If $v_I^{\rm eff} = v_J^{\rm eff}$ (including the case $I=J$) and
\beq\label{QIQJa}
	\bra Q_I Q_J  (1-\mathbb P_\tau)\la a\ket^{\rm c} \neq 0,
\eeq
then $\mathfrak L_{\la a,\la a}^-=\infty$. In particular, if \eqref{QIQJa} holds for a current observable $\la a= \la j_i$ and some $I=J$, then the associated lower Onsager matrix element is infinite, $\mathfrak L_{ii}^-=\infty$ (hence there is superdiffusion).
\end{theorem}
\proof
The proof of Points $1'$, $2'$ and $3'$ that $Q=(Q_{IJ;n}:n\in\N)$ gives a linearly extensive charge in Subsection \ref{ssectconv} still holds true. However, with \eqref{clusteringhydro}, we can strengthen the analysis of \eqref{tinvr} for Point $3'$, and thus optimise the range of values of $k$. We look for $k>0$ such that $\lim_{n\to\infty} \sup_{s,t\in[-kn,kn]} |\bra Q_{IJ;n},\tau_s \la a-\tau_t\la a\ket_{\hi H}| = 0$. Because $Q_{IJ;n} = (1-\mathbb P_\tau) Q_{IJ;n}$, we can make the replacement $\la a \to (1-\mathbb P_\tau)\la a$ in \eqref{dtinvr} and \eqref{tinvr}. Again, we bound a typical term as
\beq
	\Big|\int_{s_1}^{s_2} \dd t\,\bra \la b_{I_1}(n) \la c_{I_2} , (1-\mathbb P_\tau)\tau_t\la a\ket_{\whi H}\Big|\leq
	n^2 \int_{-k}^{k} \dd t\int \dd x\,
	|\bra (1-\mathbb P_\tau)\la a(nx,nt)\cdot \la b_{I_1}(n,0) \la c_{I_2}(0,0) \ket^{\rm c}|
	\label{proofquad3}
\eeq
with $s_1 = -nk$, $s_2=nk$ for some $k>0$ to be determined. The paragraph just after \eqref{proofquad2} shows that, using the bound \eqref{clustering} for the 3-point function on the right-hand side, the contributions in the regions $R_1$ and $R_2$ defined by \eqref{R1R2} vanish as $n\to\infty$ for any $k'<1/(2v)$, where $v$ is the velocity from the assumption \eqref{clustering}. Therefore, it is sufficient to concentrate on the rest,
\beq
	R_3 =\{(x,t):|x-1/2|\leq 3/2+kv, k'<|t|<k\}.
\eeq

In order to use the bound in \eqref{clusteringhydro}, we must guarantee that, in this region, $\max_i\{z_i\}$ has a positive lower bound, where $z_1 = n\big( |x-1+v_1t|-\ep |t|\big)$ and $z_2 = n\big( |x+v_2t|-\ep |t|\big)$ (with $v_i = v_{I_i}^{\rm eff}$). Let $y=x-1/2+(v_2+v_1)t/2$ and $y_t = 1/2+(v_2-v_1)t/2$. Then $z_1/n = |y-y_t|-\ep|t|$ and $z_2/n = |y+y_t|-\ep|t|$. Note that $y_0>0$. For every $y_t>0$, the maximum $z=\max_i \{z_{i}\}$ is $z_2$ ($z_1$) if $y>0$ ($y<0$). Over $y$, the minimal value of $z$ is at $y=0$, and is $n(y_t-\ep|t|)$. Thus if $k$ is chosen such that $y_t > \ep|t|$ for all $|t|\leq k$, this minimal value is positively bounded on the time integration region. Then, with $m = \inf_{|t|< k}(y_t-\ep |t|)$ and by monotonicity and $(-d)$-homogeneity of $f(z,t)$, we have
\beqa
	\lefteqn{n^2 \int_{R_3} \dd t\dd x\,
	|\bra (1-\mathbb P_\tau)\la a(nx,nt)\cdot\la b_{I_1}(n,0) \la c_{I_2}(0,0) \ket^{\rm c}|}
	&& \n &\leq&
	n^2 \int_{k'<|t|<k} \dd t\int_{|x-1/2|\leq 3/2 + vk} \dd x\,
	f(nm,nt) \n
	&=&
	n^{2-d} \int_{k'<|t|<k} \dd t\int_{|x-1/2|\leq 3/2 + vk} \dd x\, f(m,t).
\eeqa
As the time integration region is finite and excludes a neighbourhood of $t=0$, the function $f(m,t)$ is uniformly bounded on this region. Since $d>2$, the result vanishes as $n\to\infty$. Considering terms where $I_1$ and $I_2$ are exchanged, the conditions $y_t > \ep|t|$ are all satisfied if $|t| < 1/(|v_1-v_2| + 2\ep)$. As a consequence one can choose any $0<k<1/(|v_{I}^{\rm eff}-v_{J}^{\rm eff}|+2\ep)$ for $\lim_{n\to\infty} \sup_{s,t\in[-kn,kn]} |\bra Q_{IJ;n},\tau_s \la a-\tau_t\la a\ket_{\hi H}| = 0$. Thus, using $||\la q_I||_{\whi H}^2 = ||\la q_J||_{\whi H}^2 = 1$ and $\bra\la q_I,\la q_J\ket_{\whi H}=0$ and taking $\ep>0$ arbitrarily small, Theorem \ref{theoquad1} implies the theorem.
\eproof

\begin{rema}\label{remasuperpheno1}
The assumption \eqref{clusteringhydro} should be interpreted in view of the discussion of superdiffusion in Subsection \ref{ssectfractorder} and Appendix \ref{appsscaling}. In the normal superdiffusion case, exponential decay, or decay with high enough power law, away from the ballistic trajectory can be assumed, and thus the assumption is expected to hold with $d$ arbitrarily large. In the fat superdiffusion case, by \eqref{fatassump} one may expect weaker power-law clustering in the region around the ballistic trajectory. For  two-point functions of conserved densities one has $O\big(|t|\,|x-y_i-v^{\rm eff}_{I_i}t|^{-1-1/\alpha}\big)$, from \eqref{qqsup} with \eqref{fatassump}. For two-point functions involving currents, powers of $|t|$ decrease and of $|x-y_i-v^{\rm eff}_{I_i}t|$ increase in equal amount. Higher-point functions are subject to different power laws, and Euler scaling predicts one less power for every observable \cite{doyoncorrelations}. Thus, one might expect $O\big(|t|^{-\nu}\,|x-y_i- v_it|^{\nu-1/\alpha-1}\big)$ for values $\nu<2$; which is indeed in agreement with \eqref{clusteringhydro} for $\alpha\in(1/2,1)$, where one finds $d=1/\alpha+1>2$. A more in-depth analysis of three-point function clustering would be useful.
\end{rema}

\subsection{Superdiffusion}\label{ssectsuper}

I now propose a natural clustering property, inferred from the superdiffusion phenomenology, under which the conditions and related superdiffusion exponents found from NFH naturally emerge. This will show that these characteristics of superdiffusion, usually explained using NFH, actually arise without the need for the hydrodynamic equation, mode-coupling theory or the addition of noise.

Note that Theorem \ref{theoquad3} implies that there is superdiffusion for the normal mode $K$ if the coupling $\bra Q_I^2(1-\mathbb P_\tau)\la j_K\ket^{\rm c}\neq0$ for some $I$.  This condition is exactly that found from NFH \cite{SpohnNonlinear} for superdiffusive behaviours of normal modes: these couplings are the ``$G$-couplings" introduced in this context. Here, it is seen to hold as well in the quantum context.

In NFH, a variety of superdiffusion exponents and scaling functions are found, depending on which $G$-coupling(s) is (are) nonzero. In particular, it is found, in the classical context, that if the fully diagonal coupling is nonzero, $\bra Q_I^2(1-\mathbb P_\tau)\la j_I\ket^{\rm c}\neq0$, then superdiffusion is within the Kardar-Parisi-Zhang (KPZ) universality class, with exponent $\alpha_I=2/3$. This is the case of sound modes in generic anharmonic chains and other momentum conserving one-dimensional systems  \cite{SpohnNonlinear}. Further, using mode-coupling theory, it is found that if the fully diagonal coupling is zero, $\bra Q_K^2 (1-\mathbb P_\tau)\la j_K\ket^{\rm c}=0$, but a partially diagonal coupling is nonzero, $\bra Q_I^2 (1-\mathbb P_\tau)\la j_K\ket^{\rm c}\neq0$ for some KPZ-superdiffusive mode $I$ and some $K\neq I$, then superdiffusion for mode $K$ is controlled by the L\'evy  $(\alpha_K^{-1})$-stable distribution, with exponent $\alpha_K=3/5$. This is the case of the heat mode in generic anharmonic chains  \cite{SpohnNonlinear}. More generally, it is argued that if the nonzero couplings are a single fully diagonal one, $\bra Q_I^2 (1-\mathbb P_\tau)\la j_I\ket^{\rm c}\neq0$, at the root of a sequence of nonzero partially diagonal ones, $\bra Q_K^2 (1-\mathbb P_\tau)\la j_{K+1}\ket^{\rm c}\neq0$, $K=I,I+1,\ldots,K_{\rm max}-1$, then the superdiffusion scaling functions are Levy distribution with exponents obtained from the Fibonacci sequence, see \cite{PopkovFibonacci2015} for details.

Here I obtain a set of relations between conditions of nonzero $G$-couplings and bounds superdiffusion exponents that, in particular, exactly reproduce the above structure, in both the classical and quantum contexts.

Recall the result \eqref{second2}, which follows from a linear response analysis. With $I=J$, we have co-propagation of the normal modes, and this is at the basis of the divergence of the Onsager matrix element in Theorem \ref{theoquad3} (as $v_I^{\rm eff} = v_J^{\rm eff}$). However, as mentioned, the actual propagation of normal modes is expected to lead, at least from the viewpoint of correlation functions, to a power-law extension around the ballistic trajectory. That is, a strong enough clustering property should be applicable in the full region away from this power-law expanding domain. Considering this full region, one can get a more precise characterisation of superdiffusion. 

According to \eqref{qqsup}, one would expect this domain to expand with the superdiffusion exponent $\alpha_I$ for mode $I$. Thus, I assume the following. Let $\la a\in\hi V$, let $\la b,\la c$ be densities or currents $\la q,\la j$, let $I$ be a normal-mode index, with superdiffusive exponent $\alpha = \alpha_I$ in Eq.~\eqref{qqsup}, and let
\beq\label{ualphabd}
	u>\alpha_I+1 \in (1,2).
\eeq
Then there exist real numbers $p\geq 0$ and $d>2+(2-u)p$, and a function $f(z,t)>0$ ($z>0$, $t\in\R$) that is $(-d)$-homogeneous, $f(nz,nt) = n^{-d}f(z,t)\;\forall\;n>0$, monotonically decreasing in $z$, $f(z,t)<f(z',t)$ if $z>z'$, and such that $|t|^{1+p}f(z,t)$ is uniformly bounded on $\{(z,t):t\in T\}$ for any $z\neq 0$ and compact subset $T$, such that:
\beq\label{clusteringu}
\begin{aligned}
	|\bra (1-\mathbb P_\tau)\la a(x,t)\cdot \la b_{I}(y_1,0) \la c_{I}(y_2,0)\ket^{\rm c}|
	\leq f(\max_i\{z_i\},t)  \\
	 \mbox{if } \max_i\{z_i\}>0 \quad (z_i = |x-y_i - v^{\rm eff}_{I}t|-  |\kappa t|^{u-1}).
	 \end{aligned}
\eeq
The main difference with \eqref{clusteringhydro} is that the cone, which was controlled by $\ep>0$, around the ballistic trajectory beyond which there is clustering, is now replaced by the power-law expanding region controlled by the strength $\kappa$ and the power $u-1$ (here I use my convention in Subsection \ref{ssectfractorder} with $u\in(1,2)$). Another difference lies in the function $f(z,t)$ controlling the decay beyond this power-law region. Near $t=0$ it is constrained to diverge at most as a power law, and the bound on $d$ is, generically, greater than 2. Again, the functions $f(z,t) = z^{-d+\nu}|t|^{-\nu}$ for $\nu< d$ would do, as well as linear combinations with positive coefficients. In these cases, one takes $p=\max\{0,\nu-1\}$, and a calculation shows that the values $d>u/(u-1)$ always satisfy the required bound. See Remarks \eqref{remasuperpheno1} and \eqref{remasuperpheno2} for a connection with the superdiffusive phenomenology. Pictorially, this condition can be represented by Fig.~\ref{correlation}, with a spreading $g(t)$ as explained in Section \ref{sectoverview} (and the correspondence $1/z = \alpha_I$).

With this, I show that the sequence $Q_{II;n}$, Eq.~\eqref{Qijn}, is a $u$-extensive sequence on the observable $\la a$. The corresponding $u$-extensive charge acts again as $Q^{(u)}(\la a) = \bra Q_I^2 (1-\mathbb P_\tau)\la a\ket^{\rm c}$. Theorem \ref{theoboundhu} then shows that if this is nonzero for some current $\la a = \la j_K$ with normal (resp.~fat) superdiffusion exponent $\alpha_K$, then $\alpha_K \geq (3-u)/2$ (resp.~$\alpha_K\geq1/u$). The nontrivial results in the following theorem then arise by connecting the conditions on $u$ from, on the one hand, the superdiffusion exponent $\alpha_I$ controlling the extent of the domain beyond which the $I^{\rm th}$ mode clusters, Eq.~\eqref{ualphabd}, and, on the other hand, Theorem \ref{theoboundhu}, which relies on the superdiffusion class of the current $\la j_K$ -- the exponent necessary for convergence of space-time integrated connected current-current correlation functions. Note that, although there is no full rigour here, there is no arbitrariness in the conditions imposed: Eq.~\eqref{ualphabd} expresses the superdiffusive behaviour with exponent $\alpha_I$ around the ballistic trajectories as applied to three-point functions; and Theorem \ref{theoboundhu} follows from the the superdiffusive behaviour of two-point functions. In particular, taking $I=K$, this leads to specific numerical lower bounds for $\alpha_I$. 
\begin{theorem} Assume the clustering properties \eqref{clustering}. Assume that the additional clustering condition \eqref{clusteringu} with \eqref{ualphabd} holds for some mode $I$, and, in Point A (resp.~B) below, for $\la a = \la j_I$ (resp.~$\la a = \la j_K$). Then
\bi
\item[A.] If $\bra Q_I^2 (1-\mathbb P_\tau)\la j_I\ket^{\rm c}\neq 0$, and if $\la j_I$ has normal (resp.~fat) superdiffusion exponent $\alpha_I$, Eq.~\eqref{exponent} (resp.~Eq.~\eqref{exponentfat}), then
\beq
	\alpha_I \geq \frc23\quad\big(\mbox{resp.~} \alpha_I\geq \frc2{\sqrt 5+1}\big).
\eeq
\item[B.] If $\bra Q_I^2 (1-\mathbb P_\tau)\la j_K\ket^{\rm c}\neq 0$, and if $\la j_K$ has normal (resp.~fat) superdiffusion exponent $\alpha_K$, Eq.~\eqref{exponent} (resp.~Eq.~\eqref{exponentfat}), then
\beq
	\alpha_K \geq 1-\frc{\alpha_I}2\quad\big(\mbox{resp.~} \alpha_K\geq \frc1{\alpha_I+1}\big).
\eeq
\ei
\end{theorem}
\proof I show that the clustering condition \eqref{clusteringu} with \eqref{ualphabd} implies that $Q_{II;n}$ is a $u$-extensive sequence for the observable $\la a$, for all $u>\alpha_I+1$. For simplicity I assume $v_I^{\rm eff}=0$; the case $v_I^{\rm eff}\neq0$ is easily dealt with by similar arguments, using an additional shift $x\to x+v^{\rm eff}_It$ in the integration variable below. The result will then follow from Theorem \ref{theoboundhu}.

Points $1'$ and $2^{(u)}$ of Subsection \ref{fracext} with $u\in(1,2)$, are immediate from the proof of Theorem \ref{theoquad2}. For Point $3^{(u)}$, we need to show from \eqref{tinvr} that there exists $k>0$ such that
\[
	\lim_{n\to\infty} \sup_{s,t\in[-kn^{1/(u-1)},kn^{1/(u-1)}]} |\bra Q_{II;n},\tau_s \la a-\tau_t\la a\ket_{\whi H}| = 0.
\]
Again, we bound a typical term in \eqref{tinvr} (under the replacement $\la a\to (1-\mathbb P_\tau)\la a$) as
\beq\label{trpqo}
	\Big|\int_{s_1}^{s_2} \dd t\,\bra \la b_{I}(n) \la c_{I} , (1-\mathbb P_\tau)\tau_t\la a\ket_{\whi H}\Big|\leq
	m^2 \int_{-k}^{k} \dd t\int \dd x\,
	|\bra (1-\mathbb P_\tau)\la a(mx,mt)\cdot\la b_{I}(n,0) \la c_{I}(0,0) \ket^{\rm c}|
\eeq
where $m=n^{1/(u-1)}$ ($\gg n$ as $n\to\infty$) and $s_1 = -mk$, $s_2=mk$ with $k>0$ to be determined. As in the paragraph just after \eqref{proofquad2}, we may use the bound \eqref{clustering} for the 3-point function on the right-hand side. The regions \eqref{R1R2} are now expressed as
\beqa
	R_1 &=& \{(x,t):|x-n/(2m)|>3n/(2m)+kv, |t|<k\}\n
	R_2 &=& \{(x,t):|x-n/(2m)|\leq 3n/(2m)+kv,|t|\leq nk'/m\}
\eeqa
for some $k'<1/(2v)$, where $v$ is the velocity of the bound \eqref{clustering}. In these regions, the contribution of the integral on the right-hand side of \eqref{trpqo} vanishes as $n\to\infty$. Thus we may restrict to
\beq
	R_4 = \{(x,t):|x|\leq 1+vk, nk'/m<|t|<k\}.
\eeq
We now use \eqref{clusteringu}. With $m^{u-1}=n$, we have $z_1 = |mx-n|-n|\kappa t|^{u-1}$ and $z_2 = m|x|-n|\kappa t|^{u-1}$. The minimal value over $x$ of $\max_i\{z_i\}$ is obtained at $x=n/(2m)$ and is $n(1/2-|\kappa t|^{u-1})$. This is positively bounded on $t\in[-k,k]$ if $k < \kappa^{-1} 2^{1/(1-u)}$. Then, with $s = \inf_{|t|< k}((1/2-|\kappa t|^{u-1})$ and by monotonicity and $(-d)$-homogeneity of $f(z,t)$, we have
\beqa
	\lefteqn{m^2 \int_{R_4} \dd t\dd x\,
	|\bra (1-\mathbb P_\tau)\la a(mx,mt)\cdot\la b_{I}(n,0) \la c_{I}(0,0) \ket^{\rm c}|} &&\n
	&\leq& m^2 \int_{nk'/m<|t|<k}\dd t\int_{|x|\leq 1+vk}\dd x\,f(ms,mt) \n
	&=& m^{2-d}\, 2(1+vk) \int_{m^{u-2}k'<|t|<k}\dd t\,f(s,t). \label{fghht}
\eeqa
Under the conditions on the function $f(z,t)$ expressed above \eqref{clusteringu}, the time integral is finite for $p=0$, or else diverge as $m^{p(2-u)}$. The requirement that $d>2+(2-u)p$ then guarantees that the limit $m\to\infty$ of the right-hand side of \eqref{fghht} vanishes.

The theorem is proved using Theorem \ref{theoboundhu} and minimising $u$.
\eproof

The sound modes discussed above have nonzero fully diagonal coupling. As the KPZ scaling function has finite variance, the superdiffusion is normal. The lower bound in Point A, $\alpha_I\geq 2/3$, is then in agreement with the KPZ exponent $2/3$, which saturates it. The heat mode discussed above has zero fully diagonal coupling, but nonzero partially diagonal couplings to the sound modes. As the Levy distribution has infinite variance, superdiffusion is fat in this case. The lower bound in Point B, taking $\alpha_I = 2/3$, is then $\alpha_K\geq 3/5$. This is again in agreement with the superdiffusion exponent found in this case, the bound being saturated. More generally, taking the sequence of nonzero couplings as discussed above, using the fact that the superdiffusion is fat, and assuming the bound in Point B to be saturated, we have $\alpha_{K+1} = 1/(\alpha_K+1)$, and this reproduces the exponents based on the Fibonacci sequence  \cite{PopkovFibonacci2015}.

Interestingly, we obtain strong statements concerning the relation between exponents and tails of the distribution. For instance, we deduce that if the partially diagonal coupling of mode $K$ with mode $I$ is nonzero, $\bra Q_I^2(1-\mathbb P_\tau)\la j_K\ket^{\rm c}\neq0$, if mode $I$ is of KPZ type, and if the exponent of mode $K$ is smaller than $2/3$, as it is indeed observed to be ($3/5<2/3$), then the scaling function of mode $K$ must have divergent variance. This is indeed the case, as it is the Levy distribution. This is a nontrivial consequence on the scaling function, inferred solely from the associated exponent.

Note that one should expect the bounds to be saturated. Indeed, for instance, $Q_{\hi H^{(u)}}(\la j_I)$ is finite, and one would expect to be able to bound it as in \eqref{Qhbdfrac} by a finite upper bound on the optimal $u$; thus $||\la j_I||_{\hi H^{(u)}}^-$ should be finite on this optimal $u$.

It is also possible to argue for the KPZ lower bound in the normal superdiffusion case using factionally extensive charges of the Prosen type, see Appendix \ref{appsuper}.

\begin{rema}\label{remasuperpheno2}
Continuing on Remark \ref{remasuperpheno1}, recall that the phenomenological expectation of the behaviour of the three-point function in \eqref{clusteringu} is as $O\big(|t|^{-\nu}\,|x-y_i- v_it|^{\nu-1/\alpha_I-1}\big)$, for values $\nu<2$. This is indeed in agreement with the specifications of the clustering requirement \eqref{clusteringu} for $\alpha_I\in(1/2,1)$, where one finds $d=1/\alpha_I+1>u/(u-1)$ under the inequality \eqref{ualphabd}.
\end{rema}

\begin{rema} If \eqref{clusteringu} held for $u=\alpha_I+1$, then we could use this value in Theorem \ref{theoboundhu}, and apply it in order to obtain a bound on the strength of the superdiffusion spreading $\lambda$ in Eq.~\eqref{qqsup}. However, I do not expect \eqref{clusteringu} to hold for $u=\alpha_I+1$, for any $\kappa$. Indeed, exactly on the curve of the superdiffusion growth, there are corrections and clustering is not uniformly vanishing. One might also hope to weaken Point $3^{(u)}$ in order to include finite corrections, which might be supported in small regions and therefore would give vanishing contributions to $\delta_n$ in \eqref{Qnubounding} (here replacing $u$ by $u-1$). However, although the nonzero correlations are only ``near" the curve of superdiffusion growth, I expect the extent of the spatial region on which they lie to grow with the same power, as $t^{u-1}$. Therefore, the correction term $\delta_n$ receives nonvanishing contributions. Hence, the techniques proposed here seem to be too rough to give bounds on the superdiffusion strength.
\end{rema}

\section{Additional remarks}\label{sectadd}

\subsection{Two-body wave scattering}\label{ssecttwobody}

Relation \eqref{second2}, which is derived in Appendix \ref{appproj} and used in Subsection \ref{ssectlinresp} in order to justify the clustering assumption used there, allows us to go further and {\em determine the exact element of $\wwhi H$ which represent the charge $Q_{IJ}''$}. That is, as per \eqref{linhil2}, there must exist $\la w_{IJ}\in\wwhi H$ such that
\beq\label{Qw}
	Q_{IJ}''(\la a) = \bra \la w_{IJ},\la a\ket_{\wwhi H}.
\eeq
I show in Appendix \ref{apprep}, using Euler-scale arguments, that
\beq\label{wIJ}
	\la w_{IJ} = 
	\lim_{X\to\infty} \frc{|v^{\rm eff}_I-v^{\rm eff}_J|}{2X} \int_{-X}^X \dd x\,\la q_I(x) \la q_J.
\eeq
The limit in $X$ is to be taken within $\wwhi H$, that is, one evaluates this limit {\em after} taking the limit over time in the second equation in \eqref{hydro2}. The expression within the limit in \eqref{wIJ} is a Cauchy sequence within this Hilbert space, and it is the limit that gives an element of $\wwhi H$. From this, the overlaps between $\la w_{IJ}$ and $\la w_{KL}$ are evaluated by combining \eqref{Qw} and \eqref{wIJ}:
\beqa
	\lefteqn{\bra \la w_{IJ},\la w_{KL}\ket_{\wwhi H} } && \n
	&=& Q_{IJ}''(\la w_{KL}) \n
	&=& \frc{|v^{\rm eff}_K-v^{\rm eff}_L|}{2X}\lim_{X\to\infty} \int_{-X}^X \dd x\,Q_{IJ}''(\la q_K(x) \la q_L) \n
	&=& \frc{|v^{\rm eff}_K-v^{\rm eff}_L|}{2X}\lim_{X\to\infty} \int_{-X}^X \dd x \,\Big[% \n && \;
	\bra Q_I Q_J \la q_K(x)\la q_L\ket^{\rm c} + \bra Q_I \la q_K(x)\ket^{\rm c}\bra Q_J\la q_L\ket^{\rm c}
	+ \bra Q_I \la q_L(x)\ket^{\rm c}\bra Q_J\la q_K(x)\ket^{\rm c} \n && \hspace{4cm}
	-\sum_M \bra Q_I Q_J \la q_M\ket^{\rm c}\bra Q_M \la q_K(x)\la q_L\ket^{\rm c}
	\Big].
\eeqa
The first three terms inside the bracket are the expression of the three-point connected function in \eqref{Qij} in terms of a four-point connected function. The last term inside the bracket integrates to $\sum_M \bra Q_I Q_J \la q_M\ket^{\rm c}\bra Q_M Q_K\la q_L\ket^{\rm c}$ which is finite, whence this contribution vanishes after dividing by $X$. The first term integrates to $\bra Q_I Q_J Q_K \la q_L\ket^{\rm c}$ which is also finite. The second and third terms give contributions that diverge with $X$ (because of space-translation invariance), respectively $2X \bra Q_I\la q_K\ket^{\rm c} \bra Q_J\la q_L\ket^{\rm c}$ and $2X \bra Q_I\la q_L\ket^{\rm c} \bra Q_J\la q_K\ket^{\rm c}$. Using diagonality $\bra Q_I\la q_K\ket^{\rm c} = \delta_{IK}$, we find
\beq\label{normw}
	\bra \la w_{IJ},\la w_{KL}\ket_{\wwhi H} = |v^{\rm eff}_K-v^{\rm eff}_L| \big(\delta_{IK}\delta_{JL} + \delta_{IL}\delta_{JK}\big).
\eeq
Note that this implies $||\la w_{IJ}||_{\wwhi H} = \sqrt{|v_I^{\rm eff}- v_J^{\rm eff}|}$. In particular, for $I=J$, or whenever $v_I^{\rm eff} = v_J^{\rm eff}$, we find
\beq\label{zerowII}
	||\la w_{IJ}||_{\wwhi H} = 0 \quad (v_I^{\rm eff} = v_J^{\rm eff}).
\eeq
This implies that in these cases, $\la w_{IJ}$ are {\em null} in $\wwhi H$. This is at the root of divergent Onsager matrix elements in Theorem \ref{theoquad2}.

The set of elements $\la w_{IJ}\subset \wwhi H$ spans (a dense subset of) what may be called the {\em wave scattering subspace},
\beq\label{Hscat}
	\hi H_\scat = \overline{{\rm span}\{\la w_{IJ}\}}\subset \hi H_\dif.
\eeq
This is indeed a subspace of $\hi H_\dif$, as a consequence of \eqref{invhigher}. By symmetry, a basis is obtained by considering $I> J$. With this, we may project onto $\hi H_\scat$, and we obtain a lower bound for the matrix $\mathfrak L_{\la a,\la b}$ defined in \eqref{Onsa}, and also for the Onsager matrix $\mathfrak L_{ij}$ constructed out of the conserved currents. The bound is the matrix
\beq\label{Lscat}
	\mathfrak L^{\rm scat}_{\la a,\la b}
		= \sum_{I>J\atop v^{\rm eff}_I\neq v^{\rm eff}_J} \frc{\bra \la a,\la w_{IJ}\ket_{\dif} \bra \la w_{IJ},\la b \ket_{\dif}}{|v^{\rm eff}_I-v^{\rm eff}_J|},
\eeq
that is
\beq
	\mathfrak L \geq \mathfrak L^\scat.
\eeq
In particular,
\beq
	\mathfrak L_{\la a,\la a}  \geq \sum_{I>J\atop v^{\rm eff}_I\neq v^{\rm eff}_J} \frc{\big| \bra Q_I Q_J \cdot (1-\mathbb P_\tau)\la a \ket^{\rm c}\big|^2}{|v^{\rm eff}_I-v^{\rm eff}_J|}.
\eeq
In fact, Formula \eqref{wIJ} suggests the definition of a bilinear map
\beq\label{phiscat}\begin{aligned}
	\phi_{\rm scat} :&& \overline{\hi H_{\rm bal}\wedge \hi H_{\rm bal}} &\to \hi H_{\rm dif} \\ &&
	(\la q_I,\la q_J)&\mapsto \phi_{\rm scat}(\la q_I,\la q_J) = \la w_{IJ}.
	\end{aligned}
\eeq
(By \eqref{zerowII}, the diagonal of $\hi H_\bal\otimes \hi H_\bal$ is indeed omitted in the domain of $\phi_{\rm scat}$.) This maps onto the subspace $\hi H_\scat\subset \hi H_\dif$,
\beq
	{\rm Ran}(\phi_{\rm scat}) = \hi H_\scat.
\eeq

%Equation \eqref{phiscatmain} means that the result of the action of the element $\phi_{\rm scat}(\la q_I,\la q_J)$ via the inner product \eqref{hydro2}  for $\wwhi H$, where the total time-integral of the element $\phi_{\rm scat}(\la q_I,\la q_J)$ is involved, is given by \eqref{covder}, where, instead, connected correlation functions of total space-integrated densities are involved. This transposition from space to time integrals is a nontrivial step.

Recall that $\la q_I$ as an element of $\hi H$ is defined up to constant shifts $\la q_I(x)\mapsto \la q_I(x)+a{\bf 1}$ and gauge transformations $\la q_I(x)\mapsto \la q_I(x)+\p_x \la a(x)$. Thus in order for the map \eqref{phiscat} to be well defined, we must verify that the result is invariant under these transformation. This is indeed the case: since $\la q_I\equiv0$ in $\wwhi H$, then the constant shift leaves the expression invariant; and the gauge transformation produces, in the integral, boundary terms which are finite in $\wwhi H$, hence their contribution vanishes in the limit $X\to\infty$.

I call $\phi_{\rm scat}$ the {\em wave scattering map}. An interpretation of it is as follows. The elements $\phi_\scat(\la q_I,\la q_J)$ represent hydrodynamic {\em two-body scattering states for ballistic waves}, where two interacting, coherent, ballistic waves propagate (the fact that the product of conserved densities are coherently and ballistically transported is expressed mathematically in Formula \eqref{second} in Appendix \ref{appproj}). The average over space $x$ in \eqref{wIJ} represents the average over all distances between the waves. The factor $|v_I^{\rm eff}-v_J^{\rm eff}|$ relates to a ``density of state": the normalisation factor by which we divide the integral in \eqref{wIJ} is not the total distance $2X$ between the observables, but the total time required for them to reach a distance $2X$ under ballistic propagation. The range $\hi H_\scat$ of the map is the {\em wave scattering subspace} of $\hi H_\dif$. The overlaps $\bra\phi_\scat(q_I,q_J),\la a\ket_\dif = \bra \la w_{IJ},\la a\ket_\dif$ represent the ``probability amplitudes" for the wave scattering state to be created by the perturbation $\la a$. Physically, it can be interpreted in terms of the {\em spreading of the observable} $\la a$ as it evolves: it is the strength of ballistically propagating coherent pulses of modes $I$ and $J$ at the fronts. In general, the spreading is formed of a linear combination of many such fronts. One may interpret the part of the hydrodynamic entropy production \eqref{entropy} due to wave scattering $\hi H_\scat$ (coming from the part of $||\la v||_{\dif}$ obtained by projection onto $\hi H_\scat$) as that coming from a redistribution of large-scale convective structures towards structures at the lower diffusive scale. Applied to the Onsager matrix, with $\la a = \la j_i$ and $\la b = \la j_j$, the lower bound \eqref{Lscat} represents the contribution to diffusion from the scattering of ballistic waves. The projection $\bra \phi_{\rm scat}(\la q_I,\la q_J),\la j_i\ket_{\dif}$ is the amplitude for the current observables to absorb or produce two coherent ballistic waves, and the denominator represents the amplitude of overlap between the waves. 

Finally, I observe that for the current observables $\la j_i$, the expression for $\mathfrak L_{ij}^{\rm scat} = \mathfrak L_{\la j_i,\la j_j}^{\rm scat}$ from \eqref{Lscat} agrees with the formula obtained in \cite{MedenjakDiffusion2019} from a proposal akin to nonlinear fluctuating hydrodynamics. It is worth explaining the relation between the formalism of this work and that developed here. In particular, I believe my derivation clarifies three important points, which may be useful for future studies:
\bi
\item First, the heuristic expansion of the current observable in powers of density observables in \cite{MedenjakDiffusion2019} has the more precise meaning of an expansion within the Hilbert space $\hi H_\scat$. Hence, the proposal there is valid as an exact formula for observables that lie within $\hi H_\scat$ only. It is important to note that $\la j_i$ generically does not lie within this space, and thus there are corrections (as anticipated in \cite{MedenjakDiffusion2019}). From the theory developed here, one can further establish that corrections are present for any observable $\la a\in \wwhi H$ that is not invariant under the higher flows, as these do not lie in $\hi H_\dif$.
\item
Second, the heuristic arguments given in \cite{MedenjakDiffusion2019} for the ballistic propagation of two normal modes, which is crucial in order to relate space-integrals to time-integrals, is shown here to arise from the linear-response formula proven in Appendix \ref{appproj}. This states how products of conserved densities in three-point functions can be evolved according to their ballistic transport. Crucially, this shows that one needs to project out the conserved space in order for this ballistic transport formula to hold, clarifying why the projection has to be performed in \cite{MedenjakDiffusion2019}.
\item
Third, the formula \eqref{Lscat} for $\mathfrak L^{\scat}_{\la a,\la b}$ is seen in the present work as arising from a projection on a Hilbert space. This shows that this formula gives in general a lower bound. This is important, for instance, in order to fully justify, as in Theorem \ref{theoquad3}, the argument made in \cite{MedenjakDiffusion2019} that superdiffusion must occur -- diffusion is infinite -- if (in my notation) $\bra \la w_{II},\la a\ket_{\wwhi H}\neq0$.
\ei

\subsection{Geometry of the manifold of states}\label{ssectgeom}

Recall Equation \eqref{derbeta} relating $\beta^i$-derivatives of averages $\bra\la a\ket$ to inner products in $\hi H_\bal$. It has a clear geometrical meaning, obtained by seeing an observable $\la a$ as a {\em function on the manifold of states}. In symbols, if $\omega\in \mathcal M$ are the maximal entropy states, we may see, for any given $\la a\in\hi V$, the average $\bra \la a\ket = \la a(\omega)$ as a function of the state $\omega$. Then, \eqref{derbeta}  is an equation for the derivative of this function with respect to $\beta^i$. Thus we should identify {\em $\hi H_\bal$ with the tangent space at $\omega$}, and $\la q_i$ with a basis of tangent vectors. In particular, on the right-hand side of \eqref{derbeta}, $\la a\in\hi H_\bal$ is the abstract vector field,  which can be expanded in the basis $\{\la q_i\}$, that represents the derivative of the function $\la a(\omega)$. The static covariance matrix $\mathsf C_{ij}$ is the metric of this Riemannian manifold. Thus, according to the results of Section \ref{sectext}, the tangent space is the space of linearly extensive conserved charges for $\hi H$. This is expanded upon in \cite{Doyon2017}, and gives some clarification to the sense in which linearly extensive conserved charges, as used in the literature on GGEs, are indeed those that generate the maximal entropy states.

Two natural questions arise: if variations in $\hi H_{\rm th}$ give overlaps in $\hi H_\bal$, then do variations in $\hi H_\bal$ give overlaps in $\hi H_\dif$? And, having a Riemannian manifold, could we construct covariant derivatives? These questions are related. The covariant derivatives can be constructed explicitly using the Levi-Civita connection for the metric $\mathsf C_{ij}$. Since $\mathsf C_{ij} = -\frc{\p^2 f}{\p\beta^i\p\beta^j}$ is the Hessian of the specific free energy $f$, the connection in fact simplifies to $\Gamma^i_{jk} = \frc12 \sum_l \mathsf C^{il}\frc{\p}{\p\beta^j} \mathsf C_{lk}$. The vector elements of the abstract vector field $\la a$, given by $\bra \la q^j,\la a\ket_{\bal} = \sum_k\mathsf C^{jk}\bra \la q_k,\la a\ket_{\bal}$, are scalars under parallel transport, and thus by definition of the covariant derivative $\nabla_i$, we have
\beq
	\nabla_i \bra \la q^j,\la a\ket_{\bal} = \frc{\p}{\p\beta^i} \bra \la q^j,\la a\ket_{\bal}.
\eeq
By metric covariance, $\nabla_i \bra \la q^j,\la a\ket_{\bal}$ is a covariant tensor, hence $\nabla_i \bra \la q_j,\la a\ket_{\bal} = \sum_k\mathsf C_{jk}\nabla_i\bra \la q^k,\la a\ket_{\bal}$. Thus,
\beq\label{covder}
	\nabla_i \bra \la q_j,\la a\ket_{\bal}
	= \frc{\p}{\p\beta^i} \bra \la q_j,\la a\ket_{\bal}
	-
	\sum_{k,l}\Big[\frc{\p}{\p\beta^i} \mathsf C_{jk}\Big]\mathsf C^{kl}
	\bra \la q_l,\la a\ket_{\bal} = -\bra Q_iQ_j (1-\mathbb P_\tau)\la a\ket^{\rm c}
\eeq
where we use the explicit form \eqref{proj} for the projection, as well as $\bra \la q_j,\la a\ket_{\bal} = \bra Q_j\la a\ket^{\rm c}$ and
\beq\label{dbetaQ}
	-\frc{\p}{\p\beta^i} \bra Q_j\la a\ket^{\rm c} = \bra Q_i Q_j\la a\ket^{\rm c}.
\eeq

We recognise the right-hand side of \eqref{covder} as (up to a minus sign) {\em the action of the bilinear charge $Q_{ij}''$} constructed in Subsection \ref{ssectconv}, see Eq.~\eqref{Qij}. That is, the covariant derivative is, on the vector field $\la a$, a quadratically extensive conserved charge for $\hi H$. Thus, by the results of Subsection \ref{ssecttwobody}, the space of covariant derivatives (the tangent space to the tangent space) is a subspace of $\hi H_\dif$; it is the wave scattering space $\hi H_\scat$ spanned by the bilinear charges, Eq.~\eqref{Hscat}.

The remark that covariant derivatives must lie in $\hi H_\dif$ also has a geometric underpinning. Since the metric is covariant, we have $\nabla_i \bra \la q_j,\la q_k\ket_{\bal}=0$. Therefore, seeing the left-hand side of \eqref{covder} as a linear functional on $\la a$, it is null on $\hi H_\bal$. This suggests the hydrodynamic reduction $\whi H\to \wwhi H$, which trivialises $\hi H_\bal$. Further, it is clear from \eqref{covder} that it is invariant under time evolution $\tau_t$. Hence, it acts well on the space of equivalence classes on which $\wwhi H$ is built. The result of Subsection \ref{ssectconv} is of course stronger, showing that, as a linear functional on $\la a$, the covariant derivative is in fact bounded with respect to $\wwhi H$.

The interpretation as a covariant derivative gives a geometric  meaning for the bilinear map \eqref{phiscat}, with
\beq\label{scatov}
	-\nabla_i\bra \la q_j,\la a\ket_{\bal} = \bra \phi_{\rm scat}(\la q_i,\la q_j),\la a\ket_{\dif}.
\eeq

\subsection{Integrability}

This section is more speculative and concerns the relation between many-body integrability and the structure of the Hilbert spaces discussed above.

The classical notion of integrability is based on the Liouville condition, that the number of commuting conserver quantities must agree with the number of conservation laws. This leads to the well known solution by quadrature of the system. The extension of this concept to thermodynamic systems, which possess infinitely-many degrees, such as the thermodynamic limit of quantum or classical gases, or field theories, is a challenging problem. Is the presence of infinitely-many conserved quantities sufficient for integrability? The problem is even starker in quantum chains: in finite volumes, there always are as many conserved quantities in involution as the dimension of the quantal Hilbert space: the projectors on the Hamiltonian eigenspaces. How should we define, without recourse to any particular technical solution procedures, the notion of integrability in infinite chains?

A basis for the answer to these questions has been suggested for a long time: one must consider conserved charges with appropriate locality properties. The Hilbert spaces constructed in this paper, and in particular the results \cite{DoyonProjection} on Euler-scale hydrodynamic projections, formalise the notion of locality in a mathematically precise fashion. Thus it is natural to ask if we can define integrability using these. Here I propose that this is the case.

First, one can verify, as is shown in \cite{MedenjakDiffusion2019}, that the expression for $\mathfrak L_{ij}^{\rm scat}$ from \eqref{Lscat} reproduces the exact result of \cite{dNBD,dNBD2} for the Onsager matrix $\mathfrak L_{ij}$ in integrable models. Hence, in integrable models, the {\em wave scattering bound is saturated}
\beq\label{Lijscatint}
	\mathfrak L_{ij} = \mathfrak L_{ij}^{\rm scat}\qquad \mbox{(integrable systems).}
\eeq
This observation is in agreement with the interpretation of the calculation in \cite{dNBD,dNBD2}, which used a spectral decomposition of the current-current two-point function in terms of matrix elements of particle and holes excitations above a finite-density Bethe background. In this calculation, it is indeed seen that only the 2-particle-hole-pair terms in the spectral decomposition contribute to the Onsager matrix, the result being interpreted as a two-body scattering process. The partial expansion formula \eqref{Lscat} is a generalisation to arbitrary many-body models of this calculation and of its underlying physical interpretation.

One might want to define integrable systems by the requirement that their diffusion be solely due to wave scattering states. I propose that this observation in fact points to a more fundamental definition of integrability in terms of its space of conserved charges. Recall the setup of Subsection \ref{ssectbaldif}, in particular the higher flows and their invariants \eqref{Qflows}. Recall that the wave scattering space -- the space of covariant derivatives -- is a subspace of the diffusive space,
\beq
	\hi H_\scat \subset \hi H_\dif,
\eeq
itself invariant under all higher flows (if any), $\hi H_\dif = \wwhi Q$. Clearly, if there is only a finite number of conserved flows, then $\hi H_\dif$ is relatively big, while $\hi H_\scat$ is relatively small. The extreme case is if $\iota$ and $\tau$ are the only flows; then $\hi H_\dif = \wwhi H$ while $\hi H_\scat$ is composed of a single element $\la w_{12}$. As the number of conserved quantities increase, the space $\hi H_\dif$ decreases and $\hi H_\scat$ increases: the conserved quantities restrict diffusion, but increase the wave scattering contributions within it. Thus, {\em with enough conserved quantities, the spaces may become equal}. Therefore I propose the following definition: a many-body system is integrable if and only if there are enough conserved flows in involution such that
\beq\label{integdef}
	\hi H_\scat = \hi H_\dif\qquad \mbox{(definition of integrability).}
\eeq
That is, in integrable systems, once the complete set of conserved densities $\{\la q_i\}$ has been found at the ballistic level, giving $\hi H_\bal\subset \whi H$ and generating conserved flows in involution, then at the diffusive level, in $\hi H_\dif\subset \wwhi H$, there are no new conserved degrees of freedom. The geometric meaning is that not only the tangent space of the state manifold generates the space of flow invariants in the first-order hydrodynamic space $\whi H$, but also the covariant derivatives generate the space of flow-invariants in the second-order hydrodynamic space $\wwhi H$.

This definition immediately explains why the wave scattering subspace saturates diffusion, Eq.~\eqref{Lijscatint}. Indeed, as mentioned current observables are in general invariant under higher flows, hence elements of the diffusive subspace $\hi H_\dif$, Eq.~\eqref{jiproj}. As a consequence, the definition \eqref{integdef} implies that in integrable systems
\beq
	\la j_i \in \hi H_\scat \mbox{ as an element of $\hi H_\dif$}\qquad\mbox{(integrable systems)}
\eeq
and this immediately implies \eqref{Lijscatint}.

This proposal appears to be supported by the form factor calculations of \cite{dNBD2}. Performing a similar calculation for an arbitrary observable instead of currents, one obtains an infinite series, not only the two particle-hole terms (I refer to \cite{dNBD2} for the discussion of particle-hole states). However, requiring that the observable be in $\hi H_\dif$, hence invariant under all higher flows (as an element of $\wwhi H$, not as a local observable, which would be too strong), all processes in this series must conserved momenta. As only two integrals are made (over space and time), it must be that only the two particle-hole terms remain, and therefore the inner product involving this observable in $\hi H_\dif$ can be projected onto $\hi H_\scat$.

Finally, it is also possible to describe {\em free systems}. I note three proposal or sets of results which I now put together. First, it is proposed in \cite{SpohnInteracting} that the difference between free and interacting integrable models is that in the former, diffusion vanishes, while in the latter it does not. Second, it is proposed in \cite{DoMy19} that free models are exactly those for which the natural flux Jacobian\footnote{The natural flux Jacobian is that obtained in any system of coordinates, on the manifold of maximal entropy states, that is linear in the average conserved densities $\bra\la q_i\ket$. For instance, in general, the system formed by the normal modes at every point on the manifold is nonlinear. See \cite{DoMy19}.} $\mathsf A_i^{~j} = \p\bra \la j_i\ket/\p\bra\la q_j\ket$, is independent of the state. Third, it is shown \cite{FagottiLocally} in a large class of spin chain models with ``free-particle" description that there is a choice of gauge (a choice of explicit conserved densities and currents in the spin chain) such that the current operators is itself a conserved density. I now show that these statements, along with the proposal \eqref{integdef}, agree with the following definition of free models:
\beq\label{freedef}
	\la j_i =0 \mbox{ as an element of $\hi H_\dif$}\qquad\mbox{(definition of free systems)}
\eeq
Indeed, if, in an appropriate gauge, current observables are conserved densities, then $\la j_i \in \hi H_\bal$ as elements of $\whi H$. As a consequence, since $\hi H_\bal \to \{0\}$ in the hydrodynamic reduction to the diffusive space, \eqref{freedef} holds. Another way of seeing this is that if the natural flux Jacobian is state-independent, then the diagonal basis in Appendix \ref{diagonal} is state-independent, and therefore $\bra \la w_{IJ},\la j_j\ket_{\wwhi H}=0$, from Eq.~\eqref{Qij}. Hence, $\mathbb P_\scat j_i=0$, and combined with my proposal for integrability \eqref{integdef} that $\la j_i \in \hi H_\scat$, this gives \eqref{freedef}. Clearly, with \eqref{freedef}, the Onsager matrix vanishes, in agreement with the proposal \cite{SpohnInteracting}.

\section{Conclusion}\label{sectcon}

I have described a procedure for constructing certain Hilbert spaces which give information about the hydrodynamic properties of a many-body system. This is a general procedure, which generalises the standard setup of ballistic hydrodynamic projections. In particular, I have defined the second order hydrodynamic space $\wwhi H$, in which the diffusive Hilbert space lies $\hi H_\dif\subset \wwhi H$. The diffusive space is the space of invariants under higher Hamiltonian flows, if there are any such flows (such as in integrable systems); otherwise it is all of $\wwhi H$. I have explained how the Onsager matrix, describing diffusion, is written as an inner product in $\hi H_\dif$. I have also defined fractional-order spaces $\hi H^{(u)}$, and those lying between the ballistic and diffusive spaces, $u\in(1,2)$, are associated with superdiffusion.

I have shown that quadratically extensive conserved charges give elements of $\wwhi H$.  The construction gives a slightly different, and stronger, version of the lower bound for diffusion found by Prosen \cite{Prosenquad}. I have explained how one can write a hydrodynamic projection formula for the Onsager matrix which extends such lower bounds to a formally exact expression, much like standard hydrodynamic projections for the Drude weights \cite{Spohn-book,SciPostPhys.3.6.039} extend the Mazur bound \cite{mazur69,CasZoPre95,ZoNaPre97,ProsenQuasilocal1,ProsenQuasilocal2} based on linearly extensive charges. 

I have constructed particular elements of $\hi H_\dif$ associated to bilinear expressions in linearly extensive conserved charges. Their contribution to diffusion describes the effects of the scattering of ballistic waves; their overlaps with local observables can be interpreted as representing the ballistic fronts of observable spreading. These elements span what I call the wave scattering space $\hi H_\scat\subset \hi H_\dif$. Geometrically, they describe the covariant derivatives on the manifold of maximal entropy states.

The action of bilinear charges are completely calculable by the knowledge of the averages of local observables. They give rise to explicit lower bounds for the Onsager matrix, whose strength depends on the clustering properties of three-point functions. Using solely clustering of the Lieb-Robinson type, I show that nonzero diffusion can be established from the evaluation of an integrated three-point function, available from the thermodynamic averages of currents. Using an Euler-scale linear response principle, I have obtained expressions for three-point functions involving conserved densities, which suggest sharper clustering properties based on hydrodynamic propagation velocities. These sharper clustering properties imply a stronger diffusion lower bounds. Further, from the three-point function result, the full projection onto $\hi H_\scat$ is evaluated, giving a matrix lower bound. This matrix lower bound agrees with that conjectured recently using different, less rigorous methods \cite{MedenjakDiffusion2019}, and is saturated in integrable models \cite{dNBD,dNBD2,GHKV2018}. 

Assuming that clustering is controlled by hydrodynamic velocities, I have shown that if the diagonal $G$-coupling of nonlinear fluctuating hydrodynamics (NFH) is nonzero, then there must be superdiffusion (the associated Onsager matrix element is infinity). Assuming clustering properties inferred from the general superdiffusion phenomenology further implies that the superdiffusive exponent is bounded from below by $2/3$ if the scaling function has finite variance, the exponent of the KPZ universality class. I likewise bound the heat-mode exponent by $3/5$ and recover the Fibonacci sequence predicted in the context of NFH. These are obtained without the use of the hydrodynamic equation, and without adding noise or using mode-coupling theory, and give a strong confirmations of results from NFH.

I have speculated that an accurate definition of integrability can be obtained by the coincidence of $\hi H_\scat$ with $\hi H_\dif$: with enough conserved quantities in involution, diffusion is limited enough, yet there is enough wave scattering, so that diffusion is purely from scattering of ballistic waves. This parallels the usual statement that integrability is the presence of as many conserved quantities in involution as degrees of freedom. I noted that generically, observables, in integrable and non-integrable systems, are {\em not} supported on the wave scattering subspace, this being a particularity of current observables in integrable systems. Generic observables therefore have more involved spreading structure. In free models, current observables are null in the diffusive space, hence there is no diffusion.

Adding the complementary space $\hi H_1$ within the diffusive space, $\hi H_\dif = \hi H_\scat \oplus \hi H_1$, and $\hi H_2$ within the second order hydrodynamic space, $\wwhi H = \hi H_\dif  \oplus \hi H_2$, we have
\beq
	\wwhi H = \hi H_\scat \oplus \hi H_1 \oplus \hi H_2.
\eeq
The spaces $\hi H_1$ and $\hi H_2$ will be studied in forthcoming works. In particular, $\hi H_2$ is identified with the physics of thermalisation \cite{DurninTherma2020}. I also point out that one might get insight by considering the various hydrodynamic structures that may form besides ballistically propagating waves. For instance, in general, shocks may develop, which will contribute to diffusion. This points to the idea of constructing asymptotic hydrodynamic states based on such structures, giving rise to hydrodynamic spectral decompositions of multi-point functions paralleling the form factor expansion in integrable models \cite{SmirnovBook,dNBD2,CortesPanfil2019}.

A particular family of fluid equations, of which the hydrodynamics of integrable systems is part, is that of linearly degenerate fluids, see \cite{El-2003,El-Kamchatnov-2005,El2011,bulint17}. Linearly degenerate fluids do not develop shocks, and one might therefore conjecture that they admit purely wave scattering diffusion. It would be interesting to study wave scattering processes in linearly degenerate hydrodynamics, and understand how they connect with integrable systems in more depth.

The sequential construction \eqref{HHp} points to the possibility of studying finer hydrodynamic scales when higher flows are available (such as in integrable systems). Effectively, in the hydrodynamic equation probing the finer scales, at the $n^{\rm th}$ stage of the sequence, one expands the dispersion relation of the $n^{\rm th}$ flow to the $n^{\rm th}$ power of momentum, as the lower powers are projected out by the lower stages. There is no need to choose the sequence of flows in any particular order (for instance, any flow of the hierarchy can be used for generating time evolution), but in certain situations, it may be natural to order the flows by the power of momentum they correspond to (for instance, in Galilean systems the momentum is $p$, the energy $p^2/2$, and other linearly extensive conserved charges of integrable systems carry higher powers of $p$). Naturally, extending \eqref{chargeincl}, these higher spaces should be spanned by charges of higher-order extensivity. Importantly, the construction suggests that there are as many meaningful hydrodynamic scales as there are commuting flows; in particular, in conventional, non-integrable systems ($N=2$), this suggests that there are no hydrodynamic scales beyond the diffusive scale: the hydrodynamic assumption that conserved densities on time slices fully describe the local averages is broken. In integrable systems ($N=\infty$), there are infinitely many hydrodynamic scales, supposedly forming an asymptotic expansion of local averages.

Fractional-order spaces $\hi H^{(u)}$ may also be nontrivial for orders $u\in( 0,1)$, that is, lying between the thermodynamic and ballistic spaces. In one dimension, thermal states for local quantum-chain hamiltonians are exponentially clustering, hence such fractional spaces are trivial. However, at zero temperature, there may be quantum phase transitions, where clustering is power-law and spaces of fractional orders between 0 and 1 should become nontrivial. Hence, at orders $u\in(0, 1)$, nontrivial fractional order spaces are associated with quantum criticality, and at orders $u\in(1, 2)$, they are associated with superdiffusion, which can be argued to correspond to dynamical phase transitions \cite{DoMy19}.

Finally, it is clear that, as for the linearly extensive charges \cite{Doyon2017}, most aspects of the present theory can be generalised to higher-dimensional models, where it would be interesting to connect with proposed general results, e.g.~\cite{Kov03}.

\medskip

{\bf Acknowledgments.} This work benefited from discussions with O. A. Castro-Alvaredo, J. De Nardis, W. De Roeck, M. Medenjak, T. Prosen, T. Sasamoto, H. Spohn and T. Yoshimura. I am particularly grateful to T. Yoshimura for showing me the main result of \cite{MedenjakDiffusion2019} before publication while in Tokyo. I acknowledge funding from the Royal Society under a Leverhulme Trust Senior Research Fellowship, ``Emergent hydrodynamics in integrable systems: non-equilibrium theory", ref.~SRF\textbackslash R1\textbackslash 180103, and for the last additions to the paper (in particular Section \ref{sectoverview}), from the EPSRC under the grant ``Emergence of hydrodynamics in many-body systems: new rigorous avenues from functional analysis", ref.~EP/W000458/1. I am grateful to the Tokyo Institute of Technology for support and hospitality during an invited professorship (October 2019), where this research was started. This research was also supported in part by the International Centre for Theoretical Sciences (ICTS) during a visit for the program Thermalization, Many body localization and Hydrodynamics (code: ICTS/hydrodynamics2019/11).

%relation with entanglement spreading?

%Structure of flux observables on $\wwhi H$ tells about nature of model. Zero: free, no diffusion, no thermalisation. Supported on $\wwhi H_{\rm convective}$: interacting integrable, diffusion, no thermalisation (KAM theorem). Supported beyond: interacting non-integrable, diffusion, thermalisation.

%{\color{red}Euler / Boltzmann subspace. Entropy production of hydro: former from redistribution of structures towards lower scales accessible by higher-order hydro; latter from microscopic beyond hydro. Latter leads to thermalisation in homogeneous quenches and is only contribution there.}

\appendix

\section{Current correlation asymptotics in non-correlated particle systems}\label{appcurrentcorr}

Without loss of generality we assume $\vec v=0$ in \eqref{cf}. We derive Eq.~\eqref{Ggcons} as follows:
\beqa
	\int_0^t \dd s\,\int \dd^d x\,\bra j^i(\vec x,s)
	j^i(0,0)\ket^{\rm c} &=&\frc12 \int_0^t \dd s\,\int \dd^d x\,\p_i\p_k |\vec x|^2\,\bra j^i(\vec x,s) j^k(0,0)\ket^{\rm c} \n
	&=& \frc12 \int_0^t \dd s\,\int \dd^d x\,|\vec x|^2\,\p_i\p_k \bra j^i(\vec x,s) j^k(0,0)\ket^{\rm c} \n
	&=& - \frc12 \int_0^t \dd s\,\int \dd^d x\,|\vec x|^2\,\bra \p_i j^i(\vec x,s) \p_k j^k(0,0)\ket^{\rm c} \n
	&=& - \frc12 \int_0^t \dd s\,\int \dd^d x\,|\vec x|^2\,\bra \p_s n(\vec x,s) \p_{s'} n(0,s')|_{s'=0}\ket^{\rm c} \n
	&=& \frc12 \int_0^t \dd s\,\int \dd^d x\,|\vec x|^2\,\p_s^2 \bra  n(\vec x,s) n(0,0)\ket^{\rm c} \n
	&=& \frc12 \p_s \int \dd^d x\,|\vec x|^2\, \bra  n(\vec x,s)  n(0,0)\ket^{\rm c}\Big|_{s=0}^t \n
	&\sim& \frc\chi2 \p_t \big[g(t)^2 \Delta f\big]\n
	&=& \chi \Delta f\, g'(t) g(t)
\eeqa
where
\beq
	\Delta f = \int \dd^d x\,|\vec x|^2 f(\vec x)
\eeq
is assumed finite.

In the fat-superdiffusion case with $d=1$, where $\Delta f= \infty$, we use $\bra n(x,t)n(0,0)\ket^{\rm c}\sim c t |x|^{-1-z}$ and integrate over a spatial region growing proportionality to $t$  (see the discussion around Eq.~\eqref{fatf}), to get
\beqa
	\int_0^t \dd s\,\int_{-at}^{bt} \dd x\,\bra j( x,s)
	j(0,0)\ket^{\rm c} 
	&=& \frc12 \int_{-at}^{bt} \dd x\,|x|^2\, \p_s \bra  n( x,s)  n(0,0)\ket^{\rm c}\Big|_{s=0}^t \n
	&\sim& \frc {c(a^{2-z}+b^{2-z})}{2-z} t^{2-z}.
\eeqa
The velocities $a,b>0$ are here phenomenological; in a more accurate treatment, they are obtained from the bounding modes surrounding the fat-superdiffusive mode. The coefficient $c$ is not simply related to the susceptibility; it simply controls the asymptotics of the density two-point function.

\section{Diagonal basis of conserved quantities}\label{diagonal}

In a thermodynamic state, assumed invariant under all flows generated by the $Q_i$'s, one defines the following {\em hydrodynamic matrices}:
\beqa
	\mathsf C_{ij} &=& \bra \la q_i,\la q_j\ket_{\whi H} = \bra Q_i \la q_j\ket^{\rm c}\n
	\mathsf B_{ij} &=& \bra \la q_i,\la j_j\ket_{\whi H} = \bra Q_i  \la j_j\ket^{\rm c}\n
	\mathsf A_{i}^{~j} &=& \frc{\p \bra \la j_i\ket}{\p\bra \la q_j\ket}
	= \sum_k \mathsf B_{ik} \mathsf C^{kj}
\eeqa
where $\mathsf C^{ij}$ is the inverse of the matrix $\mathsf C_{ij}$. By general principles \cite{SpohnNonlinear,PhysRevX.6.041065,dNBD2,KarevskiCharge2019}, $\mathsf B_{ij} = \mathsf B_{ji}$, implying
\beq
	\mathsf A \mathsf C = \mathsf C \mathsf A^{\rm T}.
\eeq
The matrix $\mathsf C_{ij}$ is like a ``metric", used to raise and lower indices.

It is known that one can diagonalise the flux Jacobian $\mathsf A$,
\beq\label{AR}
	\mathsf A = \mathsf R^{-1} v^{\rm eff} \mathsf R
\eeq
where $v^{\rm eff} = {\rm diag}(v^{\rm eff}_I)$ is the diagonal matrix of normal-mode velocities of the Euler hydrodynamics of the model. The columns of the matrix $\mathsf R^{-1}$ are the eigenvectors of $\mathsf A$. These eigenvectors can be chosen to form an orthonormal basis with respect to the form given by the matrix $\mathsf C^{-1}$, and therefore we can choose $\mathsf R$ to satisfy
\beq
	\mathsf R\mathsf C\mathsf R^{\rm T} = {\bf 1}\qquad\Big(
	\sum_{jk}\mathsf R_{i}^{~j} \mathsf C_{jk} \mathsf R_{l}^{~k} = 
	\delta_{il}\Big).
\eeq

It is convenient to introduce the normal-mode basis, labelled by capital-letter indices $I,J,K,\ldots$, by transforming vectors as
\beq
	\la q_I = \sum_i \mathsf R_I^{~i}\la q_i.
\eeq
In this basis,
\beq
	\mathsf C_{IJ} = \bra \la q_I,\la q_J\ket_{\hi H'} =
	\sum_{ij}\mathsf R_I^{~i} \mathsf C_{ij} \mathsf R_{J}^{~j}
	= \big[\mathsf R \mathsf C \mathsf R^{\rm T}\big]_{IJ}
	=\delta_{IJ}.
\eeq
Hence in this basis, the metric is the identity. In particular, we have
\beq
	\mathbb P_\tau\la a = \sum_I \la q_I \bra \la q_I,\la a\ket_{\whi H}
\eeq
and
\beq
	\mathsf D_{\la a,\la b} = \sum_I \bra \la a, q_I\ket_{\hi H'} \bra q_I,\la b\ket_{\hi H'}
	= \sum_I \bra \la a \cdot Q_I\ket^{\rm c}\bra Q_I\cdot \la b\ket^{\rm c}.
\eeq
The diagonal basis is state-dependent, except in free systems.

\section{Currents as invariants}\label{appcur}

This is based on Subsection \ref{ssectkubo}. It is sufficient to consider the infinitesimal generators of the flows, $\delta_\ell = \lim_{s\to 0}(\tau_s^{(\ell)}-1)/s$. By \eqref{Qflows}, all conserved densities $\la q_i$'s are invariant under all flows in $\whi H$. Then, in $\whi H$, we have $\mathbb P_\tau (\delta_\ell\la a) = 0$ because $\bra \delta_\ell\la a, \la q_i\ket_{\whi H} = -\bra \la a, \delta_\ell \la q_i\ket_{\whi H}=0$. Therefore, in terms of local observables instead of elements of $\whi H$, every flow generates its currents, so that $\delta_\ell\la q_i + \p_x \la j_i^{(\ell)}=0$. Hence, we have (here $\la j_i = \la j_i^{(2)}$ is the ``usual" current)
\beqa
	\bra \delta_\ell\la j_i,\la a\ket_{\wwhi H} &=& \lim_{T\to\infty}
	\int_{-T}^T \dd t\,\int \dd x\,\bra \delta_\ell\la j_i(x,t)\cdot \la a(0,0)\ket^{\rm c}\n
	&=& -\lim_{T\to\infty}
	\int_{-T}^T \dd t\,\int \dd x\,x\bra \delta_\ell(\p_x\la j_i(x,t))\cdot\la a(0,0)\ket^{\rm c}\n
	&=& \lim_{T\to\infty}
	\int_{-T}^T \dd t\,\int \dd x\,x\bra \delta_\ell(\p_t\la q_i(x,t))\cdot\la a(0,0)\ket^{\rm c}\n
	&=& \lim_{T\to\infty}
	\int \dd x\,x\bra \delta_\ell(\la q_i(x,T) - \la q_i(x,-T))\cdot\la a(0,0)\ket^{\rm c}
\eeqa
and using the current for the $\ell$th flow,
\beqa
	&=& -\lim_{T\to\infty}
	\int \dd x\,x \p_x \bra (\la j_i^{(\ell)}(x,T) - \la j_i^{(\ell)}(x,-T))\cdot\la a(0,0)\ket^{\rm c}\n
	&=& \lim_{T\to\infty}
	\int \dd x\,\bra (\la j_i^{(\ell)}(x,T) - \la j_i^{(\ell)}(x,-T))\cdot\la a(0,0)\ket^{\rm c}\n
	&=& \lim_{T\to\infty}
	\bra \la j_i^{(\ell)}(T) - \la j_i^{(\ell)}(-T),\la a\ket_{\whi H}\n
	&=& 
	\bra \mathbb P_\tau (j_i^{(\ell)} - \la j_i^{(\ell)}),\la a\ket_{\whi H}\n
	&=& 0.
\eeqa
In the penultimate step, I used a weak assumption of projection at infinity for $\whi H$, simply that the large time limit is the projection. Therefore, $[Q_\ell,\la j_i]=0$ in $\wwhi H$, and thus $\la j_i\in \wwhi Q = \hi H_{\rm dif}$.

\section{Superdiffusive scaling of current correlation functions}\label{appsscaling}

For the general argument leading to these precise conditions, consider the $\hi V$-setup of local observables and their expectation values, and recall the phenomenology of superdiffusivity. This purports that the two-point functions of a superdiffusive normal mode $q_I$ (see Appendix \ref{diagonal}) take a scaling form at large $x,t$
\beq\label{qqsup}
	\bra \la q_I(x,t)\la q_I(0,0)\ket^{\rm c} \sim
	\frc1{|\lambda t|^\alpha} f\Big(\frc{x-v_I^{\rm eff}t}{|\lambda t|^\alpha}\Big)
\eeq
where $f(z)$ integrates to 1, $\lambda>0$ and $\alpha \in (1/2,1)$ (the case $\alpha=1/2$ would be normal diffusion). For the purpose of this discussion, it is sufficient to assume that $v_I^{\rm eff}=0$ and that $f(z)$ be symmetric. This does not affect the scaling analysis; for instance, $v_I^{\rm eff}=0$, for a given $I$, in many cases can be achieved by appropriate Gallilean or relativistic boosts. In \eqref{qqsup} and below, we must take $\la q_I\in \hi V$, and use \eqref{proj} on $\hi V$. See Remark \ref{remaqi} and Subsection \ref{ssectbilinear}.

The precise shape of $f(z)$, which can be obtained from NFH, is not important for the present consideration. It will be crucial, however, to distinguish two classes: ``normal superdiffusion", where $f(z)$ {\em decays fast enough at large $|z|$} in order to have finite variance
\beq\label{normalassump}
	\int \dd z\,z^2f(z)<\infty\quad\mbox{(normal superdiffusion),}
\eeq
and ``fat superdiffusion", where $f(z)$ has infinite variance. In the latter case, a more accurate characterisation is necessary. Results from NFH \cite{SpohnNonlinear} suggest that {\em the normalisation coefficient of the fat tail grows proportionally to $t$}, $\bra \la q_I(x,t)\la q_I(0,0)\ket^{\rm c} \propto |t| |x|^\nu$ for some $\nu$, for $|x|\gg |\lambda t|^\alpha$ and at large $|t|$. A possible phenomenological explanation is that at very large distances, the fat tail of superdiffusion should be controlled by ballistic processes, and that with these, the correlation grows, at fixed position, linearly in time. Thus, in the fat superdiffusion case, one may expect
\beq\label{fatassump}
	f(z) \sim c |z|^{-\big(1+\frc1\alpha\big)}\quad\mbox{(fat superdiffusion)}
\eeq
for some $c>0$, at large $|z|$.

By applying time derivatives and space integrals, one extracts the appropriately decaying current-current correlation function. Then \eqref{qqsup} implies that
\beq\label{jjcorsup}
	\bra (1-\mathbb P_\tau)\la j_I(x,t)\,(1-\mathbb P_\tau)\la j_I(0,0)\ket^{\rm c} \sim
	\frc{\alpha\lambda^2}{|\lambda t|^{2-\alpha}} \big[(1-2\alpha)h(z) + \alpha \p_z (zh(z))\big],\quad z = \frc{x}{|\lambda t|^\alpha}
\eeq
where $h(z) = \int_{-\infty}^z \dd y\,yf(y)$. This is shown as follows. For normal modes, as the flux Jacobian has been diagonalised, $(1-\mathbb P_\tau)\la j_I = \la j_I - v^{\rm eff}_I \la q_I$, and as mentioned we assume $v^{\rm eff}_I=0$, which simplifies the calculation. Taking time derivatives,
\beq
	\p_t^2 \bra \la q_I(x,t)\la q_I(0,0)\ket^{\rm c} \sim \frc{\alpha}{(\lambda t)^{\alpha} t^2}\big[\alpha\p_z^2(z^2f(z))+(1-\alpha)\p_z (zf(z))\big]
\eeq
where $z = x/(\lambda t)^\alpha$. By symmetry,
\beq
	\int \dd z\,zf(z) = 0
\eeq
hence we can set $z f(z) = \p_z h(z)$ with $h(z)\to0$ as $|z|\to\infty$. We find
\beqa
	\p_t^2 \bra \la q_I(x,t)\la q_I(0,0)\ket^{\rm c}
	&\sim& \p_x^2\Big(
	\frc{\alpha (\lambda t)^{\alpha} }{t^2}\big[\alpha z^2f(z)+(1-\alpha)h(z)\big]
	\Big)\n
	&=& \p_x^2\Big(
	\frc{\alpha (\lambda t)^{\alpha} }{t^2}\big[\alpha \p_z (zh(z))+(1-2\alpha)h(z)\big]
	\Big)
\eeqa
from which we deduce \eqref{jjcorsup} (using space-time translation invariance and the conservation law \eqref{conslaw} for $\la q_I$ and $\la j_I$).

We must now distinguish the two cases. In the normal superdiffusion case, integrating over $x$, we have
\beq\label{jjsup}
	\bra (1-\mathbb P_\tau)\tau_t\la j_I,\la j_I\ket_{\whi H} =|t|^{2\alpha-2}\,\lambda^{2\alpha}\alpha(2\alpha-1) \int \dd z\,z^2 f(z)\,(1+o(t))
\eeq
at large $|t|$. As the variance is finite, this is the correct asymptotics.

In contrast, in the fat superdiffusion case, notwithstanding the overall normalisation coefficient, the space integral of the right-hand side of \eqref{jjcorsup} does not exist, as it decays as $|x|^{1-1/\alpha}$. However, the space integral of the current-current connected correlation function {\em still exists}, as this is a two-point function of local observables. The resolution is that the asymptotic power law in \eqref{jjcorsup} is valid, for any given $t$, only in a (large) region of space around $x=0$. This region grows at most proportionally to $t$ if there is a Lieb-Robinson bound. In fact, one would expect the region to be determined by the available hydrodynamic velocities. For instance, for the heat mode in anharmonic chains, which is found to have fat superdiffusion by NFH, the region lies between the two sound modes, see the discussion in \cite{SpohnNonlinear}. Therefore, integrating over this region, one obtains
\beq\label{jjsupfat}
	\bra (1-\mathbb P_\tau)\tau_t\la j_I,\la j_I\ket_{\whi H} \propto |t|^{1-\frc1\alpha}(1+o(t))
\eeq
where the nonzero proportionality constant depends on the growth rate of the region, and is not important in what follows. This discussion suggests that fat superdiffusion can only occur for modes whose ballistic trajectories lie between that of other modes; these provide the linearly increasing region necessary to contain the power-law decay (however I will not investigate this aspect further).

Note that using \eqref{fatassump} in the exact asymptotic \eqref{jjcorsup}, one in fact finds a vanishing leading-asymptotic coefficient, $(1-2\alpha)h(z) + \alpha \p_z (zh(z))\sim 0\times z^{1-1/\alpha}$. However, the precise coefficient of the asymptotic as evaluated from \eqref{fatassump} is not expected to be meaningful, as it is affected by other modes at large distances $|x|\gg |\lambda t|^\alpha$. Thus only the power law of the asymptotic should be meaningful, and \eqref{jjsupfat} is expected to hold.

The discussion from Eq.~\eqref{qqsup} to Eq.~\eqref{jjsupfat} is based on the phenomenology of superdiffusion. The form \eqref{qqsup}  says that the normal modes propagate ballistically, with a superdiffusive extension controlled by the exponent $\alpha$ around the ballistic trajectory of velocity $v^{\rm eff}_I$. Both results \eqref{jjsup} and \eqref{jjsupfat} lead to the divergence of the Onsager matrix element $\mathcal L_{II}$, because the resulting power law is not integrable. Thus, $\la j_I$ is not an element of $\wwhi H$, but instead
\beq\label{jIspace}\begin{aligned}
	&\la j_I \in \hi H^{(3-2\alpha)}&& \mbox{(normal superdiffusion)}\\
	&\la j_I \in \hi H^{(1/\alpha)}&& \mbox{(fat superdiffusion)}
	\end{aligned}
\eeq
according to \eqref{wasympfrac}. This implies in particular the slightly weaker statements \eqref{exponent} and \eqref{exponentfat}.
%Below, when obtaining results concerning superdiffusion, I will further make precise clustering assumptions based on the  phenomenology of \eqref{qqsup} and \eqref{fatassump}.

Note that \eqref{exponentfat} was phenomenologically justified via the power-law requirement \eqref{fatassump}. Different power laws would lead to different definitions of fat superdiffusion, and change some of the results below. The precise power law \eqref{fatassump}, giving  \eqref{exponentfat}, therefore appears to be a stronger assumption than the simpler finite-variance assumption of normal superdiffusion, giving \eqref{exponent}. However, the fact that in both cases, the full range $u\in(1,2)$ is covered by the superdiffusion range $\alpha\in(1/2,1)$, is an indication that these may indeed be generic.

\section{Hydrodynamic projections for three-point functions}\label{appproj}

Euler-scale correlation functions $\bra \la a_1(x_1,t_1)\la a_2(x_2,t_2)\cdots \la a_x(x_n,t_n)\ket^{\rm eul}$ are connected correlation functions of local observables at large space-time scales,
\[
	\lim_{\lambda\to\infty} \lambda^{-n+1}\bra \overline{\la a_1}(\lambda x_1,\lambda t_1)\overline{\la a_2}(\lambda x_2,\lambda t_2)\cdots\overline{\la a_n}(\lambda x_n,\lambda t_n)\ket^{\rm c},
\]
where $\overline{\la a_i}(\lambda x_i,\lambda t_i)$ are averages over ``fluid cells" of extent $o(\lambda)$ around the space-time points $(\lambda x_i,\lambda t_i)$. Fluid cell averaging may be done in a variety of ways, see for instance the discussion in \cite{Spohn-book} and in \cite{doyoncorrelations,10.21468/SciPostPhys.4.6.045}. Euler-scale correlation functions may be evaluated by the linear-response principles  proposed in \cite{doyoncorrelations}.

Here, I show two fundamental formulae for Euler-scale three-point correlation functions using linear response arguments.  The arguments are formal, and I will consider the classical case for simplicity. I expect the results, being at the Euler scale, to apply in the quantum case as well.

The first result is
\beq\label{first}
	\bra \,(1-\mathbb P_\tau)\la a(x,t) \,\la b(y,0) \,\la c (z,0)\,\ket^{\rm eul} = \bra \,(1-\mathbb P_\tau)\la a(x,t) \,\mathbb P\la b(y,0) \,\mathbb P\la c (z,0)\,\ket^{\rm eul}
\eeq
where here and below $\mathbb P = \mathbb P_\tau$. Thus, upon projecting out conserved densities at $(x,t)$, the only contribution from observables at time 0 come from their projection on the conserved densities. Recall that the projection is with respect to the inner product $\bra \la a,\la b\ket := \bra \la a,\la b\ket_{\whi H} = \int \dd x\,\bra \la a(x)\la b(0)\ket^{\rm c} $, and is written explicitly as \eqref{proj}:
\beq\label{projection}
	\mathbb P_\tau\la a = \sum_{ij} \la q_i \mathsf C^{ij}\bra \la q_j,\la a\ket.
\eeq
The second formula is
\beq\label{second}
	\bra\,(1-\mathbb P_\tau)\la a(x,t) \,\la q_k(y,0) \,\la q_l (z,0)\,\ket^{\rm eul} =
	\bra(1-\mathbb P_\tau)\la a,\la q_{i} ,\la q_{j}\ket \mathsf S^{i}_{~k}(x-y,t)\mathsf S^{j}_{~l}(x-z,t)
\eeq
where the three-point coupling is
\beq\label{3ptcpl}
	\bra\la b,\la q_{i} ,\la q_{j}\ket = 
	\int \dd y\dd z \,\bra \la b(0) \la q_{i}(y)\la q_{j}(z)\ket^{\rm c}.
\eeq
That is, upon projecting out conserved densities at $x,t$,  conserved densities at $y,0$ and $z,0$ can brought to time $t$ by their Euler-scale propagators, and the overlap with the observable at $x,t$ is the three-point coupling. Recall that, at the Euler scale
\beq
	\mathsf S^i_{~j}(x,t)= \big[\mathsf C^{-1}\delta(x-\mathsf At)\mathsf C\big]^{i}_{~j}.
\eeq

\proof A proof for \eqref{first} and \eqref{second} is as follows. Consider the average $\bra \la a(x,t)\ket$ as evolved from an inhomogeneous initial state. By linear response \cite{doyoncorrelations}, one can insert the Euler-cell averages of $\la b(y,0)$ and $\la c(z,0)$ by deformation of their conjugate thermodynamic potential in the initial condition,  $\beta^{\la a}(y)$ and $\beta^{\la b}(z)$, respectively. We evaluate the final result in the homogeneous, stationary state (the latter is not necessary for the general linear response argument, but sufficient for our purposes). This gives:
\beqa
	\bra \la a(x,t) \la b(y,0)\la c(z,0)\ket^{\rm eul}
	&=& \frc{\delta}{\delta\beta^{\la c}(z)}\frc{\delta}{\delta\beta^{\la b}(y)}
	\bra \la a(x,t)\ket\n
	&=& \frc{\delta}{\delta\beta^{\la c}(z)}\Big[\bra \la a,\la q_i\ket_{x,t}\mathsf C^{ij}_{\ x,t}\bra \la q_j(x,t)\la b(y,0)\ket^{\rm eul} \Big].
\eeqa
We evaluate the derivative by applying it to each factor, and then specialising to the stationary and homogeneous state. Applying it to the last factor $\bra \la q_j(x,t)\la b(y,0)\ket^{\rm eul}$, this gives the term
\beq
	\bra \la a,\la q_i\ket\mathsf C^{ij}\bra \la q_j(x,t)\la b(y,0)\la c(z,0)\ket^{\rm eul}
\eeq
as, by definition, the derivative inserts the observable $\la c(z,0)$. The derivative of the first factor $\bra \la a,\la q_i\ket_{x,t}=\int \dd y\,\bra \la a(x,t) \la q_i(y,t)\ket^{\rm c}$, is $\int \dd y\,\bra \la a(x,t) \la c(z,0) \la q_i(y,t)\ket^{\rm c}  = \bra \la a(x,t)\la c(z,0),\la q_i\ket$, so we obtain the term
\beq
	\bra \la a(x,t)\la c(z,0),\la q_i\ket\mathsf C^{ij}\bra \la q_j(x,t)\la b(y,0)\ket^{\rm eul}.
\eeq
On the term $\mathsf C^{ij}_{\ x,t}$, the inverse of the matrix $\mathsf C$, we use the rule of differentiation of inverse matrices and obtain similarly
\beq
	-\bra \la a,\la q_i\ket\mathsf C^{ij} 
	\bra \la q_j(x,t)\la c(z,0),\la q_k\ket\mathsf C^{kl} \bra \la q_l(x,t)\la b(y,0)\ket^{\rm eul}.
\eeq
Using \eqref{projection}, the result is
\beq\label{res0}
	\bra \,(1-\mathbb P_\tau)\la a(x,t) \,\la b(y,0) \,\la c (z,0)\,\ket^{\rm eul} = \bra\ (1-\mathbb P_\tau)\la a(x,t)\,\la c(z,0),\la q_i\,\ket
	\mathsf C^{ij}\bra \la q_j(x,t)\la b(y,0)\ket^{\rm eul}.
\eeq
Specialising to $\la b(y,0) = \la q_k(y,0)$, this gives
\beq\label{res1}
	\bra \,(1-\mathbb P_\tau)\la a(x,t) \,\la c (z,0)\,\la q_k(y,0) \,\ket^{\rm eul} = \bra\ (1-\mathbb P_\tau)\la a(x,t)\,\la c(z,0),\la q_i\,\ket
	\mathsf S^{i}_{~k}(x-y,t).
\eeq
Specialising to $\la c(z,0) = \la q_l(z,0)$, and applying \eqref{res1} to $\bra(1-\mathbb P)\la a(x,t)\la c(z,0),\la q_i\ket
= \int \dd y\bra (1-\mathbb P)\la a(x,t)\la c(z,0)\la q_i(y,0)\ket^{\rm eul}$, we find
\beq
	\bra \,(1-\mathbb P_\tau)\la a(x,t) \,\la q_k(y,0) \,\la q_l (z,0)\,\ket^{\rm eul} = \bra(1-\mathbb P_\tau)\la a(x,t),\la q_i,\la q_j\ket
	\mathsf S^{i}_{~k}(x-y,t) \mathsf S^{j}_{~l}(x-z,t).
\eeq
This shows \eqref{second}. Finally, writing the usual hydrodynamic projection formula $\bra \la q_j(x,t)\la b(y,0)\ket^{\rm eul} = \mathsf S_{jk}(x-y,t) \mathsf C^{kl}\bra \la q_l,\la b\ket$ and applying \eqref{res1} to \eqref{res0}, we obtain
\beqa
	\bra \,(1-\mathbb P_\tau)\la a(x,t) \,\la b(y,0) \,\la c (z,0)\,\ket^{\rm eul} &=& \bra\ (1-\mathbb P_\tau)\la a(x,t)\,\la c(z,0)\la q_i(y,0)\,\ket
	\mathsf C^{ij}\bra \la q_j,\la b\ket \n
	&=& \bra\ (1-\mathbb P_\tau)\la a(x,t)\,\mathbb P\la b (y,0)\,\la c(z,0)\,\ket.
\eeqa
Applying the same formula for $\la c$ instead of $\la b$, we obtain \eqref{first}.
\eproof

\section{Representation of bilinear charges}\label{apprep}

Using the matrix notation, let us introduce the projected three-point coupling symmetric matrix (see \eqref{3ptcpl})
\beq
	\mathsf M^{\la a}_{ij} = \bra (1-\mathbb P_\tau)\la a,\la q_i,\la q_j\ket
\eeq
as well as the Euler-scale correlation matrix
\beq
	\mathsf E_{ij}^{\la a}(x,y,t)=\bra\,(1-\mathbb P_\tau)\la a(0,0) \,\la q_i(x,t) \,\la q_j (y,t)\,\ket^{\rm eul} .
\eeq
Then, from \eqref{second},
\beq\label{ESS}
	\mathsf E^{\la a}(x,y,t) = \mathsf S^{\rm T}(x,t) \mathsf M^{\la a}
	\mathsf S(y,t)
\eeq
where we used space-time reversal symmetry. Diagonalising the flux Jacobian as \eqref{AR} we get
\beqa
	\mathsf E^{\la a}(x,y,t) &=& \mathsf S^{\rm T}(x,t) \mathsf M^{\la a}
	\mathsf S(y,t) \n
	&=& \mathsf R^{-1} \delta(x-v^{\rm eff}t)\mathsf R\mathsf M^{\la a}\mathsf R^{\rm T}\delta(y-v^{\rm eff}t)\mathsf R^{-\rm T} 
\eeqa

Since $\int_\R \dd x\, \mathsf S(x,t)={\bf 1}$, we find from \eqref{ESS} that
\beq
	Q_{ij;\hi H''}(\la a) = \int \dd x \dd y\,\mathsf E_{ij}^{\la a}(x,y,t)
	= \mathsf M^{\la a}_{ij}
\eeq
in agreement with the definition. Further, it is clear that
\beq
	\lim_{t\to\infty} \int_\R \dd x\int_{x-L}^{x+L}\dd y\,
	E^{\la a}(x,y,t) = 0
\eeq
in agreement with the usual two-point hydrodynamic projections.

Most importantly, it is simple to verify that
\beq
	\int_\R \dd t\,\int_\R \dd x \int_{x-L}^{x+L} \dd y\,
	\Big[\mathsf R \mathsf E^{\la a}(x,y,t)\mathsf R^{\rm T}\Big]_{ij}
	= 
	\frc{2L}{|v^{\rm eff}_i-v^{\rm eff}_j|} \Big[\mathsf R \mathsf M^{\la a}\mathsf R^{\rm T}\Big]_{ij}.
\eeq
Therefore, defining the observable
\beq
	\la w_{kl} = \frc1{2L} \int_{-L}^L \dd x\,\big[\mathsf R^{-1}\big]^{~i}_{k}(\mathsf R\la q)_i\iota_x (\mathsf R\la q)_j\,|v^{\rm eff}_i-v^{\rm eff}_j|\mathsf R^{~j}_{l}
\eeq
we see that
\beq
	(\la w_{kl},\la a)_{\hi H''} = \int \dd t\,\bra \tau_t \la w_{kl} ,(1-\mathbb P)\la a\ket = \mathsf M^a_{kl}
	= Q_{kl;\hi H''}(\la a).
\eeq
Thus we have found the element in $\hi H''$ representing the quadratic charges as per \eqref{Qw}.

Naturally, we can consider the linear normal modes
\beq
	{\la q}_I = (\mathsf R \la q)_I
\eeq
and in this basis
\beq
	\la w_{IJ} = 
	\frc1{2L} \int_{-L}^L \dd x\,\la q_I\iota_x \la q_J\,|v^{\rm eff}_I-v^{\rm eff}_J|
\eeq
and
\beq
	(\la w_{IJ},\la a)_{\hi H''} 
	= Q_{IJ;\hi H''}(\la a).
\eeq

\section{Prosen's quadratically extensive charges}\label{sectprosen}

For definiteness I take again the situation of Subsection \ref{ssectbaldif}, with $\iota_x$ and $\tau_t$ the space and time translations; however the construction is rather general.

Prosen introduced in \cite{Prosenquad} charges which grow quadratically with the length of the region on which they are supported, and showed that they led to a lower bound for diffusion. The quadratically extensive charges defined in Subsection \ref{ssectquadext} are not exactly of the same type. Here I discuss Prosen's construction, but now interpreted via the Hilbert space structures I have introduced. I reproduce a diffusion bound that is slightly stronger than that obtained in \cite{Prosenquad}, and that agrees with the bounds one can obtain from the charges of Subsection \ref{ssectquadext}.

Here, the condition of projection at infinity (or clustering) \eqref{req}, used sequentially for $\hi H\stackrel{\iota}\rightarrow\whi H$, and then for $\whi H\stackrel{\iota}\rightarrow\wwhi H$, as it is in Subsection \ref{ssectquadext}, is not quite strong enough. Instead, a stronger condition is needed: a projection at infinity that is uniform with respect to both $\iota_x$ and $\tau_t$
\beq\label{uniclus}
	\bra (1-\mathbb P_\tau)(1 -\mathbb P_\iota)\la a(x,t),\la b\ket_{\hi H}\to 0 \quad \mbox{sufficiently fast and uniformly as $|x|, |t|\to\infty$} \quad (\la a,\la b\in\hi V).
\eeq
Of course, in my original construction, $\mathbb P_\iota$ acts on $\hi H$, but $\mathbb P_\tau$ acts on $\whi H$. Here and below, however, we need to consider both acting on $\hi H$. For this purpose, we select a representative in $\hi H$ of the $\whi H$-equivalence class for each conserved density $\la q_i$ (as discussed in Remark \ref{remaqi} and Section \ref{sectwave}), and apply the basic formula \eqref{proj}. The results below are independent of the choice of representative. By the usual hydrodynamic projection arguments, projecting out the conserved quantities as in \eqref{uniclus} indeed should make the correlator vanish in all of space-time fast enough.

Consider a sequence $Q = (Q_n\in\hi V:n\in\N)$ of elements of $\hi V\subset\hi H$. Assume, without loss of generality, that $Q_n = (1-\mathbb P_\tau)(1-\mathbb P_\iota)Q_n$.

Suppose the sequence satisfies the following conditions:
\bi
\item[$1''$.] There exists $\gamma>0$ such that $||Q_n||_{\hi H}^2 < \gamma n^2$ for all $n\in\N$.
\item[$2''$.] The limit $ Q_{\wwhi H}(\la a):= \lim_{n\to\infty} \bra Q_n,\la a\ket_{\hi H} $ exists for all $\la a\in\hi V$.
\item[$3''$.] There exists $k>0$ and $v_+>v_-\in\R$ such that $\lim_{n\to\infty} \sup_{(x,t),(y,s)\in D_{n,k,v_-,v_+}}|\bra Q_n,\la a(x,t)-\la a(y,s)\ket_{\hi H}| = 0$ for all $\la a\in\hi V$, where
\beq
	D_{n,k,v_-,v_+}=\Big\{(x,t):-kn+v_+t<x<kn+v_-t,\,|t|<\frc{2kn}{v_+-v_-}\Big\}.
\eeq
\ei
Condition $1''$ is the quadratic extensivity. Condition $2''$ is the usual requirement of the existence of the limit action. Condition $3''$ is the requirement of space-time translation invariance. The special (possibly slanted) diamond shape of the domain $D_{n,k,v_-,v_+}$ is from the intuition that $Q_n$ is essentially homogeneous and time-independent on an interval $[-kn,kn]$, but under evolution, the effect of its boundary points moves within the interval; the left (right) boundary point move at velocity $v_+$ ($v_-$). This is the picture of quadratically extensive ``almost" conserved charges considered by Prosen (see \cite[Fig.~1]{Prosenquad}). The terminology ``almost" was used because of the motion of the boundary points. However this motion is generic, and the result (the limit action of Point 2') is truly time-translation invariant; hence I find there is no need for the adverb ``almost".

In quantum spin chains, or whenever a Lieb-Robinson bound exists, typically one can take any
\beq\label{vlr}
	v_+=-v_->v_{\rm LR},\quad \mbox{$v_{\rm LR}$ the Lieb-Robinson velocity.}
\eeq
$v_{\rm LR}$ represents the maximal velocity of propagation of local disturbances as set by the dynamics. The Lieb-Robinson velocity is not state dependent, but for a given state it is generically an over-estimate of the actual propagation velocities. It is natural to assume that it is the spectrum of the flux Jacobian (see Appendix \ref{diagonal}) that describes all relevant velocities, much as I did in Subsection \ref{ssectlinresp}, although this is hard to make rigorous for any given sequence $Q$. Under this hydrodynamic picture, one may take any
\beq\label{vmaxhydro}
	v_- < {\rm \inf}(v^{\rm eff}),\quad 
	v_+ > {\rm \sup}(v^{\rm eff}) \quad \mbox{(hydrodynamic picture),}
\eeq
and this ultimately gives stronger bounds.

Note that with the uniform clustering condition \eqref{uniclus}, the inner product \eqref{hydro2} on $\wwhi H$ can be written as
\beq
	\bra \la a,\la b\ket_{\wwhi H} = \lim_{n\to\infty} \int_{nD} \dd x\dd t\,
	\bra(1-\mathbb P_\tau)(1 -\mathbb P_\iota)\la a(x,t),\la b\ket_{\hi H}
\eeq
for any bounded open domain $D\in \R^2$ containing $0$, and a formula similar to \eqref{intin} holds:
\beq\label{intin2}
	\bra \la a,\la b\ket_{\wwhi H} = 
	 \lim_{n\to\infty} \frc1{|nD|} \int_{nD} \dd x\dd t\int_{nD} \dd y\dd s
	\bra(1-\mathbb P_\tau)(1 -\mathbb P_\iota)\la a(x,t),\la b(y,s)\ket_{\hi H}.
\eeq

We need to prove that $Q_{\wwhi H}$ is bounded with respect to $\wwhi H$. The proof of boundedness is similar to the case of linearly extensive charges.  By Point $3''$, there exists $k>0$, $v_+>v_-\in\R$, and $\delta_n>0$ with $\lim_{n\to\infty}\delta_n=0$, such that
\beq\label{Qnah}
	\bra Q_n,\la a\ket_{\hi H} = \frc{v_+-v_-}{4k^2n^2}  \int_{D_{n,k,v}}\dd x\dd t\, \bra Q_n, (1-\mathbb P_\tau)(1 -\mathbb P_\iota) \la a(x,t)\ket_{\hi H} + \delta_n
\eeq
where I used $|D_{n,k,v_-,v_+}| = 4k^2n^2/(v_+-v_-)$. The first term on the right-hand side is bounded by using the Cauchy-Schwartz inequality:
\beq\label{Qnah2}
	\leq \frc{\sqrt{v_+-v_-}}{2kn}||Q_n||_{\hi H} \;
	\Bigg[\frc{v_+-v_-}{4k^2n^2}\int_{D}\dd x\dd t\int_{D}\dd y\dd s\, \bra (1-\mathbb P_\tau)(1 -\mathbb P_\iota)\la a(x,t), \la a(y,s)\ket_{\hi H}\Bigg]^{1/2}
\eeq
where $D = D_{n,k,v_-,v_+}$.
By using \eqref{intin2}, we obtain
\beq\label{quadbd}
	Q_{\wwhi H}(\la a) \leq \frc{\sqrt{(v_+-v_-)\gamma}}{2k}
	||\la a||_{\wwhi H}.
\eeq
As a consequence, for every $Q_{\wwhi H}$, there exists $\la b\in\wwhi H$ such that
\beq\label{quadhil}
	Q_{\wwhi H}(\la a) = \bra \la b,\la a\ket_{\wwhi H} \quad \forall\;\la a\in\hi V,\qquad ||\la b||_{\wwhi H}\leq 
	\frc{\sqrt{(v_+-v_-)\gamma}}{2k}.
\eeq

Using known space-time uniformity of correlation functions in order to establish Point $3''$, and in particular quantifying the clustering property \eqref{uniclus}, the result \eqref{quadhil} and \eqref{quadbd}, with \eqref{vlr}, can be expressed in a quite rigorous fashion in quantum spin chains; see the arguments in \cite{Prosenquad}.

Therefore we obtain a statement that parallels \eqref{bounhpp},
\beq\label{bdaa}
	\mathfrak L_{\la a,\la a} \geq \frc{4k^2}{(v_+-v_-)\gamma}|Q_{\wwhi H}(\la a)|^2. 
\eeq
With a Lieb-Robinson bound, \eqref{vlr}, in the second line, $v_+=-v_-=v_{\rm LR}$ (taking the limit over all bounds for $\mathfrak L_{\la a,\la a}$). With the hydrodynamic picture \eqref{vmaxhydro}, in the second line we can instead set $v_- = {\rm \inf}(v^{\rm eff})$, $v_+ = {\rm \sup}(v^{\rm eff})$.

Let me make the connection with Prosen's bound (obtained in the context of quantum chains) more explicit. Following Prosen, I assume that there is no ballistic transport, hence $\mathbb P_\tau = 0$, which simplifies the discussion. Prosen requires a sequence $(Q_n)$ such that $\lim_{n\to\infty} \bra Q_n\la a\ket^{\rm c}$ exists, which is quadratically growing, $\bra Q_n^2\ket^{\rm c}\leq \gamma n^2$ for some $\gamma>0$. Without loss of generality, $Q_n$ is hermitian, and $n$ represents the length of the interval on which $Q_n$ is supported. Prosen assumes $Q_n$ to be almost homogeneous (homogeneous except near the boundaries of the support), and almost conserved (the time derivative only involves terms at the boundaries of the support). These requirements can be seen to satisfy the three points $1'$, $2'$ and $3'$ above. With strong enough clustering (for instance, exponential) and $v_+>v_-\in\R$ the maximal and minimal propagation velocities, then in \eqref{bdaa} we take $k=1/2$, and we have
\beq\label{bdpro}
	\mathfrak L_{\la a,\la a} 
	\geq \lim_{n\to\infty} \frc{|\bra Q_n\la a\ket^{\rm c}|^2}{(v_+-v_-)\gamma}.
\eeq
where I use the explicit form $Q_{\wwhi H}(\la a)= \lim_{n\to\infty} \bra Q_n\la a\ket^{\rm c}$.

This is an improvement on the bound found in \cite{Prosenquad} in two ways. First, with the Lieb-Robinson bound, the denominator is $2v_{\rm LR}\gamma$. This is a stronger bound than that with factor $8$ instead of $2$ found in \cite{Prosenquad}. The difference can be traced back to the use of a reduced volume of integration in \cite{Prosenquad}, instead of the full diamond $D_{n,k,v_-,v_+}$. Second, the hydrodynamic velocities are bounded by the Lieb-Robinson velocity $|v_+|<v_{\rm LR}$ and $|v_-|<v_{\rm LR}$. Thus, within the hydrodynamic picture, the result is stronger.

\section{Superdiffusion bound from Prosen-type charges}\label{appsuper}

I show that one may obtain the superdiffusion bound os Subsection \ref{ssectsuper} by using instead the Prosen-type charges of Appendix \ref{sectprosen}.

The form \eqref{qqsup} also suggests that we define an $\alpha'$-sequence $(Q_n\in\hi V)$, for $\alpha'>0$, as one that satisfies Points $1''$ and $2''$ of Appendix \ref{sectprosen}, as well as the following space-time invariance condition:
\bi
\item[$3'''$.] There exists $k>0$ and $\lambda>0$ such that $\lim_{n\to\infty} \sup_{(x,t),(y,s)\in D_{n,k,\lambda}}|\bra Q_n,\la a(x,t)-\la a(y,s)\ket_{\hi H}| = 0$ for all $\la a\in\hi V$, where
\beq
	D_{n,k,\lambda}=\Big\{(x,t):-kn+|\lambda t|^{\alpha'}<x<kn-|\lambda t|^{\alpha'},\,|\lambda t|<(kn)^{1/\alpha'}\Big\}.
\eeq
\ei
The region $D_{n,k,\lambda}$ is an elongated, curvy diamond, whose sides are power-law curves with power $\alpha'$. This is simply the region where we would expect space-time invariance for the Prosen-type charge generated by the sequence
\beq\label{qchf}
	Q_{II;n} = (1-\mathbb P_\tau)(1-\mathbb P_\iota)\Big[\int_{-n}^n \dd x \,\la q_I(x)\Big]^2.
\eeq
Indeed, its boundaries should move at the velocity $v^{\rm eff}_I=0$ (assumed to vanish for simplicity) but extend superdiffusively, and uniformity in space-time is only expected away from this extension. Hence, the sequence $Q_n = Q_{II;n}$ chosen as \eqref{qchf} is expected to be an $\alpha'$-sequence for all $\alpha'>\alpha$. Naturally, with this sequence, the result is
\beq
	\lim_{n\to\infty} \bra Q_{II;n},\la a\ket_{\hi H} =
	Q_{II;\wwhi H}(\la a) = \bra Q_I^2\la a\ket^{\rm c}.
\eeq
Note that the volume of the region is
\beq\label{volgr}
	|D_{n,k,\lambda}| = \frc{4\alpha'}{(\alpha'+1)\lambda} (kn)^{1+1/\alpha'}.
\eeq

An analysis as in \eqref{Qnah} and \eqref{Qnah2} gives
\beq\label{bdal}
	|\bra Q_n,\la a\ket_{\hi H}|\leq
	\sqrt{\frc{||Q_n||^2_{\hi H}}{n^2}}
	\Bigg[\frc{n^2}{|D_{n,k,\lambda}|^2}
	\int_{D_{n,k,\lambda}}\dd x\dd t\int_{D_{n,k,\lambda}}\dd y\dd s\, \bra (1-\mathbb P_\tau)(1 -\mathbb P_\iota)\la a(x,t), \la a(y,s)\ket_{\hi H}
	\Bigg]^{1/2}.
\eeq
Let us apply this on $\la j_I$. Consider \eqref{jjsup}, and the fact that, thanks to \eqref{qqsup}, the two-point function $\bra (1-\mathbb P_\tau)(1-\mathbb P_\iota)\la j_I(x,t),\la j_I(0,0)\ket_{\hi H}$ is supported along a ballistic trajectory of velocity $v^{\rm eff}_I=0$, as per \eqref{jjcorsup}. Thus, at large $n$, we have
\beqa
	\lefteqn{\int_{D_{n,k,\lambda}}\dd x\dd t\int_{D_{n,k,\lambda}}\dd y\dd s\, \bra (1-\mathbb P_\tau)(1 -\mathbb P_\iota)\la j_I(x,t), \la j_I(y,s)\ket_{\hi H}} && \n &\sim &
	c\int_{|s|<\frc{(kn)^{1/\alpha'}}{\lambda}}\dd s
	\int_{|y|<kn-(\lambda s)^{\alpha'}}\dd y
	\int_{|t|<\frc{(kn-y)^{1/\alpha'}}{\lambda}}\dd t\,
	 |t-s|^{2\alpha-2} \quad(n\to\infty)
\eeqa
where $c = \alpha(2\alpha-1)\lambda^{2\alpha} \int \dd u\,u^2 f(u)$. A simple scaling analysis of this integral then gives
\beq
	\int_{D_{n,k,\lambda}}\dd x\dd t\int_{D_{n,k,\lambda}}\dd y\dd s\, \bra (1-\mathbb P_\tau)(1 -\mathbb P_\iota)\la j_I(x,t), \la j_I(y,s)\ket_{\hi H}
	\sim c n^{1+2\alpha/\alpha'}\quad (n\to\infty).
\eeq
Combining this with \eqref{volgr} and \eqref{bdal}, we find that if $Q_{II;\wwhi H}(\la j_I)\neq 0$, then we must have
\beq
	1 + \frc{2\alpha-2}{\alpha'} \geq 0\quad\Rightarrow\quad
	\alpha \geq \frc{2-\alpha'}2.
\eeq
This is valid for all $\alpha'>\alpha$, and the tightest lower bound is obtained by taking $\alpha'$ as near to possible to $\alpha$. Thus we conclude
\beq
	\alpha \geq \frc23.
\eeq

This shows that there is superdiffusion. Furthermore,
the lower bound on $\alpha$ is indeed the expected KPZ superdiffusive exponent. Clearly, if we further impose that the bounding argument above, for $|Q_{II;\wwhi H}(\la j_I)|$, gives a {\em finite} bound, proportional to the finite norm $||j_I||_{\hi H^{(3-2\alpha)}}$, then we must have $\alpha = (2-\alpha')/2$; but as $\alpha\geq 2/3$, we must choose $\alpha'=\alpha$ and we obtain $\alpha = 2/3$. The consideration $\alpha'=\alpha$ is more subtle however; Point $3'''$ must be modified to account for finite corrections. These corrections can be argued to scale appropriately under integration, and the bounding argument can be modified accordingly. A precise analysis may produce bounds on $\lambda$; however this is beyond the scope of this paper.

%bibliography -0000b0


\begin{thebibliography}{99}
\bibitem{Spohn-book}
 H.~Spohn, { {Large Scale Dynamics of Interacting Particles}} (Springer-Verlag, Heidelberg, 1991).
 
 
 \bibitem{SpohnNonlinear} H. Spohn, Nonlinear Fluctuating Hydrodynamics for Anharmonic Chains, { J. Stat. Phys.}~{\bf 155}, 1191 (2014). 
 
 \bibitem{1742-5468-2015-3-P03007}
 C.~B. Mendl and H.~Spohn, {Current Fluctuations for Anharmonic Chains in
   Thermal Equilibrium}, { J. Stat. Mech. Theor. Exp.}~{\bf 2015},
   P03007~(2015). 
 
 \bibitem{KulkarniFluctuating2015}
 M. Kulkarni, D. A. Huse and H. Spohn, Fluctuating hydrodynamics for a discrete Gross-Pitaevskii equation: mapping to Kardar-Parisi-Zhang universality class, { Phys. Rev. A}~{\bf 92}, 043612 (2015). 
 
 \bibitem{PopkovFibonacci2015}
 V. Popkov, A. Schadschneider, J. Schmidt and G. M. Sch\"utz, Fibonacci family of dynamical universality classes,
 { PNAS}~{\bf 112}, 12645-12650  (2015). 
 
 \bibitem{PopkovFibonacci2016}
 V. Popkov, A. Schadschneider, J. Schmidt and G.M. Sch\"utz, Exact scaling solution of the mode coupling equations for non-linear fluctuating hydrodynamics in one dimension, { J. Stat. Mech.}~{\bf 2016}, 093211 (2016).
 
 \bibitem{SchutzFibonacci2018}
 G.M. Sch\"utz, On the Fibonacci Universality Classes in Nonlinear Fluctuating Hydrodynamics, in: { From Particle Systems to Partial Differential Equations}, O. Gon{\c{c}}alves, 
 A. J. Soares (Ed.),  Springer Proceedings in Mathematics \& Statistics (Springer International Publishing), 149 (2018). 
 
 \bibitem{ChenDeGierHirikiSasamotoFluctuHydro18} Z. Chen, J. de Gier, I. Hiki and T. Sasamoto, Exact Confirmation of 1D Nonlinear Fluctuating Hydrodynamics for a Two-Species Exclusion Process, { Phys. Rev. Lett.}~{\bf 120}, 240601 (2018).
 
 \bibitem{BulchandaniKardar2019}
 V. B. Bulchandani, Kardar-Parisi-Zhang universality from soft gauge modes, { Phys. Rev. B}~{\bf 101}, 041411(R) (2020). 
 
 \bibitem{Prosenquad}
 T. Prosen, Lower bounds on high-temperature diffusion constants from quadratically extensive almost conserved operators, { Phys. Rev. E}~{\bf 89}, 012142 (2014). 
 
 \bibitem{PhysRevLett.117.207201}
 B.~Bertini, M.~Collura, J.~De~Nardis, and M.~Fagotti, {Transport in
 out-of-equilibrium $XXZ$ chains: exact profiles of charges and currents},
   { Phys. Rev. Lett.}~{\bf 117}, 207201~(2016). 
 
 \bibitem{PhysRevX.6.041065}
 O.~A. Castro-Alvaredo, B.~Doyon, and T.~Yoshimura, {Emergent hydrodynamics in 
 integrable quantum systems out of equilibrium}, { Phys. Rev. X}~{\bf 6}, 041065~(2016). 
 
 \bibitem{dNBD} J. De Nardis, D. Bernard and B. Doyon, Hydrodynamic Diffusion in Integrable Systems, { Phys. Rev. Lett.}~{\bf 121}, 160603~(2018). 
 
 \bibitem{dNBD2} J. De Nardis, D. Bernard and B. Doyon, Diffusion in generalized hydrodynamics and quasiparticle scattering, { SciPost Phys.}~{\bf 6}, 049 (2019). 
 
 \bibitem{GHKV2018} S. Gopalakrishnan, D. A. Huse, V. Khemani and R. Vasseur, Hydrodynamics of Operator Spreading and Quasiparticle Diffusion in Interacting Integrable Systems, { Phys. Rev. B}~{\bf 98}, 220303 (2018). 
 
 \bibitem{IlievskiSuper2018}
 E. Ilievski, J. De Nardis, M. Medenjak and T. Prosen, Super-diffusion in one-dimensional quantum lattice models, { Phys. Rev. Lett.}~{\bf 121}, 230602 (2018). 
 
 \bibitem{LZP19} M. Ljubotina, M. Znidaric and T. Prosen, Kardar-Parisi-Zhang physics in the quantum Heisenberg magnet, { Phys. Rev. Lett.}~{\bf 122}, 210602 (2019). 
 
 \bibitem{GopalaKinetic2019}
 S. Gopalakrishnan and R. Vasseur, Kinetic theory of spin diffusion and superdiffusion in XXZ spin chains,
 { Phys. Rev. Lett.}~{\bf 122}, 127202 (2019). 
 
 \bibitem{GopalaAnomalous2019}
 S. Gopalakrishnan, R. Vasseur and B. Ware, Anomalous relaxation and the high-temperature structure factor of XXZ spin chains, { PNAS}~{\bf 116}, 16250 (2019). 
 
 \bibitem{deNardisAnomalous2019}
 J. De Nardis, M. Medenjak, C. Karrasch and E. Ilievski, Anomalous spin diffusion in one-dimensional antiferromagnets, { Phys. Rev. Lett.}~{\bf 123}, 186601 (2019).
 
 
 \bibitem{SciPostPhys.3.6.039}
 B.~Doyon and H.~Spohn, {Drude Weight for the Lieb-Liniger Bose Gas}, { SciPost Phys.}~{\bf 3}, 039~(2017). 
 
 \bibitem{Doyon2017}
 B.~Doyon, {Thermalization and pseudolocality in extended quantum systems},  { Commun. Math. Phys.}~{\bf 351}, 155~(2017). 
 
 \bibitem{DoyonProjection} B. Doyon, Hydrodynamic projections and the emergence of linearised Euler equations in one-dimensional isolated systems, preprint {\tt arXiv:2011.00611} (2020).
 
 \bibitem{mazur69}
 P. Mazur, Non-ergodicity of phase functions in certain systems, { Physica}~{\bf 43}, 533 (1969). 
 
 \bibitem{CasZoPre95}
 H. Castella, X. Zotos, and P. Prelov\v{s}ek, Integrability and Ideal Conductance at Finite Temperatures, { Phys. Rev. Lett.}~{\bf 74}, 972 (1995). 
 
 \bibitem{ZoNaPre97}
 X. Zotos, F. Naef and P. Prelov\v{s}ek, Transport and conservation laws, { Phys. Rev. B}~{\bf 55}, 11029 (1997). 
 
 \bibitem{ProsenQuasilocal1} T. Prosen and E. Ilievski, Families of quasilocal conservation laws and quantum spin transport, { Phys. Rev. Lett.}~{\bf 111}, 057203 (2013).
 
 
 \bibitem{ProsenQuasilocal2} E. Ilievski and T. Prosen, Thermodynamic bounds on Drude weights in terms of almost-conserved quantities, { Commun. Math. Phys.}~{\bf 318}, 809 (2013).
 
 
 \bibitem{MedenjakDiffusion2019}
 M. Medenjak, J. De Nardis and T. Yoshimura, Diffusion from Convection, { SciPost Phys.}~{\bf 9}, 075 (2020).
 
 
 \bibitem{shinNormal2018}
 H. K. Shin, B. Choi, P. Talkner and E. K. Lee, Normal versus anomalous self-diffusion in two-dimensional fluids: Memory function approach and generalized asymptotic Einstein relation, { J. Chem. Phys.}~{\bf 141}, 214112 (2014).
 
 
 
 \bibitem{BratelliRobinson12} O. Bratteli and D. W. Robinson, { Operator Algebras and Quantum Statistical Mechanics 1}. (Springer, Berlin, 1987); {\em 2}. (Springer, Berlin, 1987)
 %
 %\doi{10.1007/978-3-662-09089-3}
 
 \bibitem{IlievskietalQuasilocal} E. Ilievski, M. Medenjak, T. Prosen and L. Zadnik, Quasilocal charges in integrable lattice systems, { J. Stat. Mech.}~{\bf 2016}, 064008 (2016). 
 
 \bibitem{RudinFunctional}
 W. Rudin, { Functional Analysis}, International Series in Pure and Applied Mathematics (McGraw-Hill, Singapore, 1991).
 
 \bibitem{Araki} H. Araki, Gibbs States of a One Dimensional Quantum Lattice, { Commun. Math. Phys.}~{\bf 14},120 (1969). 
 
 \bibitem{IsraelConvexity} R. B. Israel, { Convexity in the Theory of Lattice Gases} (Princeton University Press, Princeton, 1979).
 
 \bibitem{DurninTherma2020} J. Durnin, M. J. Bhaseen and B. Doyon, Non-equilibrium dynamics and weakly broken integrability, preprint arXiv:2004.11030 (2020).
 
 \bibitem{ProsenPseudo1} T. Prosen, Quantum invariants of motion in a generic many-body system, { J. Phys. A}~{\bf 31}, L645 (1998). 
 
 \bibitem{ProsenPseudo2} T. Prosen, Ergodic properties of a generic nonintegrable quantum many-body system in the thermodynamic limit, { Phys. Rev. E}~{\bf 60}, 3949 (1999). 
 
 \bibitem{LiebRobinson} E. H. Lieb and D. W. Robinson, The finite group velocity of quantum spin systems, { Commun. Math. Phys.}~{\bf 28}, 251 (1972).
 
 
 \bibitem{BHC06} S. Bravyi, M. B. Hastings and F. Verstraete, Lieb-Robinson bounds and the generation of correlations and the topological quantum order, { Phys. Rev. Lett.}~{\bf 97}, 050401 (2006). 
 
 \bibitem{doyoncorrelations}
 B.~Doyon, {Exact Large-Scale Correlations in Integrable Systems Out of Equilibrium}, { SciPost Phys.}~{\bf 5}, 054~(2018). 
 
 \bibitem{SpohnInteracting} H. Spohn, Interacting and Noninteracting Integrable Systems, { J. Math. Phys.}~{\bf 59}, 091402 (2018).
 
 \bibitem{DoMy19} B. Doyon and J. Myers, Fluctuations in ballistic transport from Euler hydrodynamics, { Ann. Henri Poincar\'e}~{\bf 21}, 255 (2019). 
 
 \bibitem{FagottiLocally} M. Fagotti, Locally quasi-stationary states in noninteracting spin chains, { SciPost Phys.}~{\bf 8}, 048 (2020). 
 
 
 \bibitem{SmirnovBook}
 F.~Smirnov, { Form factors in completely integrable models of quantum field theory},  Adv. Series in Math. Phys. 14 (World Scientific, Singapore, 1992).
 
 \bibitem{CortesPanfil2019}
 A. Cort\'es Cubero and M. Panfil, Thermodynamic bootstrap program for integrable QFT's: Form factors and correlation functions at finite energy density, { JHEP} 104 (2019). 
 
 \bibitem{El-2003}
 G.~A. El, {The Thermodynamic Limit of the Whitham Equations}, { Phys.
   Lett. A}~{\bf 311}, 374~(2003). 
 
 \bibitem{El-Kamchatnov-2005}
 G.~A. El and A.~Kamchatnov, {Kinetic Equation for a Dense Soliton Gas}, { Phys. Rev. Lett.}~{\bf 95}, 204101~(2005). 
 
 \bibitem{El2011}
 G.~A. El, A.~M. Kamchatnov, M.~V. Pavlov, and S.~A. Zykov, {Kinetic Equation
   for a Soliton Gas and its Hydrodynamic Reductions}, { J. Nonlinear
   Sci.}~{\bf 21}, 151~(2011). 
   
 \bibitem{bulint17}
 V. B. Bulchandani, On classical integrability of the hydrodynamics of quantum integrable systems, { J. Phys. A: Math. Theor.}~{\bf 50}, 435203 (2017). 
 
 \bibitem{Kov03}
 P. Kovtun, D. T. Son and A. O. Starinets, Holography and hydrodynamics: diffusion on stretched horizons, {  J. High Energy Phys.} 0310:064 (2003). 
 
 \bibitem{KarevskiCharge2019}
 D. Karevski and G. M. Sch\"utz, Charge-current correlation equalities for quantum systems far from equilibrium, { SciPost Phys.}~{\bf 6}, 068 (2019). 
 
 \bibitem{10.21468/SciPostPhys.4.6.045}
 A.~Bastianello, B.~Doyon, G.~Watts, and T.~Yoshimura, Generalized  Hydrodynamics of Classical Integrable Field Theory: the Sinh-Gordon Model,  { SciPost Phys.}~{\bf 4}, 45~(2018). 
 
 \end{thebibliography}
\end{document}